\documentclass[12pt]{article}
\usepackage{amsfonts,amsthm,amsmath,amssymb,upgreek}
\usepackage{hyperref}
\usepackage[paper=letterpaper,margin=1in]{geometry}
\usepackage{graphicx}
\usepackage{units}
\usepackage{color}
\usepackage[export]{adjustbox}
\usepackage{float}
\usepackage [english]{babel}
\usepackage [autostyle, english = american]{csquotes}
\MakeOuterQuote{"}
\numberwithin{equation}{section}

\def\Tr{{\rm Tr }}
\def\hat{\widehat}
\newcommand{\be}{\begin{equation}}
\newcommand{\ee}{\end{equation}}

\begin{document}
\thispagestyle{empty}

\vspace*{.5cm}
\begin{center}

{\bf {\LARGE Late Time Correlation Functions, Baby Universes, and ETH in JT Gravity}\\
\vspace{1cm}}

\begin{center}

 {\bf Phil Saad}\\
  \bigskip \rm
  
\bigskip
Stanford Institute for Theoretical Physics,\\Stanford University, Stanford, CA 94305

\rm
  \end{center}

\vspace{1.5cm}
{\bf Abstract}
\end{center}
\begin{quotation}
\noindent

Quantum black holes are described by a large number of macroscopically indistinguishable microstates. Correlation functions of fields outside the horizon at long time separation probe this indistinguishability. The simplest of these, the thermal two-point function, oscillates erratically around a nonperturbatively small average ``ramp'' and ``plateau'' after an initial period of decay; these non-decaying averaged features are signatures of the discreteness of the black hole spectrum. For a theory described by an ensemble of Hamiltonians, the two-point function follows this averaged behavior. 

In this paper we study certain correlation functions in Jackiw-Teitelboim (JT) gravity and find precise agreement with the behavior expected for a theory described by an ensemble of Hamiltonians with random matrix statistics- the eigenstates obey the Eigenstate Thermalization Hypothesis (ETH) and the energy levels have random matrix level statistics. 

A central aspect of our analysis is an averaged bulk Hilbert space description of the relevant behavior. The mechanism behind this behavior is topology change due the the emission and absorption of closed "baby universes". These baby universe effects give two complementary pictures of the non-decaying behavior, related by different continuations of a Euclidean geometry. A long Einstein-Rosen bridge can become short by emitting a large baby universe, and baby universes emitted and reabsorbed at points widely separated in space and time creates a "shortcut", allowing particles to leave the interior of the black hole.

\end{quotation}

\setcounter{page}{0}
\setcounter{tocdepth}{2}
\setcounter{footnote}{0}
\newpage

\tableofcontents

\pagebreak

\section{Introduction}\label{SectionIntroduction}
A central issue underlying the black hole information problem is the indistinguishability of black hole microstates to observers outside the horizon. Maldacena \cite{Maldacena:2001kr} posed a particularly sharp version of this problem for simple perturbations of the eternal black hole in AdS. The two-point function of quantum fields outside the eternal black hole, widely separated in time, distinguishes the perturbed black hole from the eternal black hole. Semiclassically, this two-point function decays forever; this decay is described by the quasinormal modes of the black hole \cite{Horowitz:1999jd}. However, using the AdS/CFT duality with a boundary field theory on a compact space with a discrete spectrum, one may see that this decay cannot last forever \cite{Maldacena:2001kr,Goheer:2002vf,Dyson:2002pf,Barbon:2003aq}. The two-point function must eventually begin to fluctuate around its non-zero late time average value. This late time average value is exponentially small in the entropy of the system,\footnote{The late time average is sometimes dominated by the thermal AdS saddle point \cite{Maldacena:2001kr,Barbon:2003aq}. However, we may consider correlators in the microcanonical ensemble to avoid these contributions.} and is invisible in perturbation theory.

The precise behavior of the two-point function at late times is challenging to describe analytically; the fluctuations around the late time average are extremely erratic and sensitive to the details of the energy spectrum \cite{Barbon:2003aq,prange1997spectral}. However, one might imagine studying a similar problem in the context of an ensemble of systems, such as the SYK model \cite{Sachdev:1992fk,KitaevTalks,Polchinski:2016xgd,Maldacena:2016hyu,Jensen:2016pah,Kitaev:2017awl,Cotler:2016fpe,Saad:2018bqo}. In this context we may consider the average behavior of the two-point function over the ensemble.\footnote{One might also consider studying higher moments of the correlation function to probe the full statistics of this noise. We briefly comment on this in Section \ref{SectionDiscussion}.} With the expectation that the Hamiltonians drawn from the ensemble have random matrix statistics, and thus the matrix elements of the fields obey the Eigenstate Thermalization Hypothesis (ETH) \cite{srednicki1994chaos,deutsch1991quantum}, and that nearby energy levels have random matrix statistics, one expects a simple universal form for the two-point function at late times. After decaying for a while, the two-point function begins a period of linear growth, called in this context the ``ramp'' \cite{Cotler:2016fpe}. At at time exponentially long in the entropy, this growth stops and the ``plateau'', and the correlator stays at its long time average value. 

In this paper, we will study this problem in a theory of Jackiw-Teitelboim (JT) gravity \cite{Jackiw:1984je,Teitelboim:1983ux,Almheiri:2014cka,Maldacena:2016upp} coupled to matter.\footnote{Correlation functions in JT gravity have been studied in \cite{Maldacena:2016upp,Yang:2018gdb,Gross:2017aos,Lam:2018pvp,Mertens:2017mtv,Blommaert:2018oro,Blommaert:2019hjr,Bulycheva:2019naf,Iliesiu:2019xuh}.} In \cite{Saad:2019lba}, the authors found that JT gravity is described by an ensemble of Hamiltonians, drawn from a certain random matrix ensemble. A given Hamiltonian drawn from this ensemble has a discrete spectrum. In a theory of JT gravity coupled to matter, we expect that two-point functions of the matter fields will be consistent with two-point functions for an ensemble averaged theory, and the underlying discreteness of the spectrum will be reflected in the non-decaying late time behavior. Through an explicit calculation in JT gravity, we confirm this expectation, matching to precise predictions. 

We also expect that certain out-of-time-ordered correlation functions (OTOCs) in this theory will exhibit a ramp and plateau structure \cite{Cotler:2017jue}. The behavior of these correlators is sensitive to the correlations between energy eigenstates; ETH posits that these correlations are small, giving a simple prediction for these OTOCs. Confirming the expected behavior of these OTOCs in JT gravity then serves as a test of the fine-grained predictions of ETH. In this paper we will calculate the expected ramp and plateau contributions for an out-of-time-ordered four point function in JT gravity, leaving the calculation of higher-point OTOCs to future work.

The mechanism behind the non-decaying behavior of these correlation functions is topology change due to Euclidean wormholes \cite{Lavrelashvili:1987jg,Hawking:1987mz,Giddings:1987cg,Coleman:1988cy,Coleman:1988tj,Giddings:1988wv,Klebanov:1988eh,Maldacena:2004rf,ArkaniHamed:2007js}, which also accounted for the non-decaying behavior of the spectral form factor in \cite{Saad:2018bqo,Saad:2019lba}. These Euclidean wormholes correspond to a type of tunneling process in which a closed ``baby JT universe'' is emitted from or absorbed by a ``parent'' asymptotically AdS universe.\footnote{We may also think of these effects in terms of a topological ambiguity of the state, instead of as a dynamical topology-changing process \cite{Jafferis:2017tiu}. However, we will use the language of baby universe "emission"and "absorption" as we believe it more clearly illustrates our results.} These baby universes can form ``loops'', where they are emitted and then reabsorbed, or they can end in a ``D-brane'' state.

\begin{figure}[H]
\centering
\includegraphics[scale=0.6]{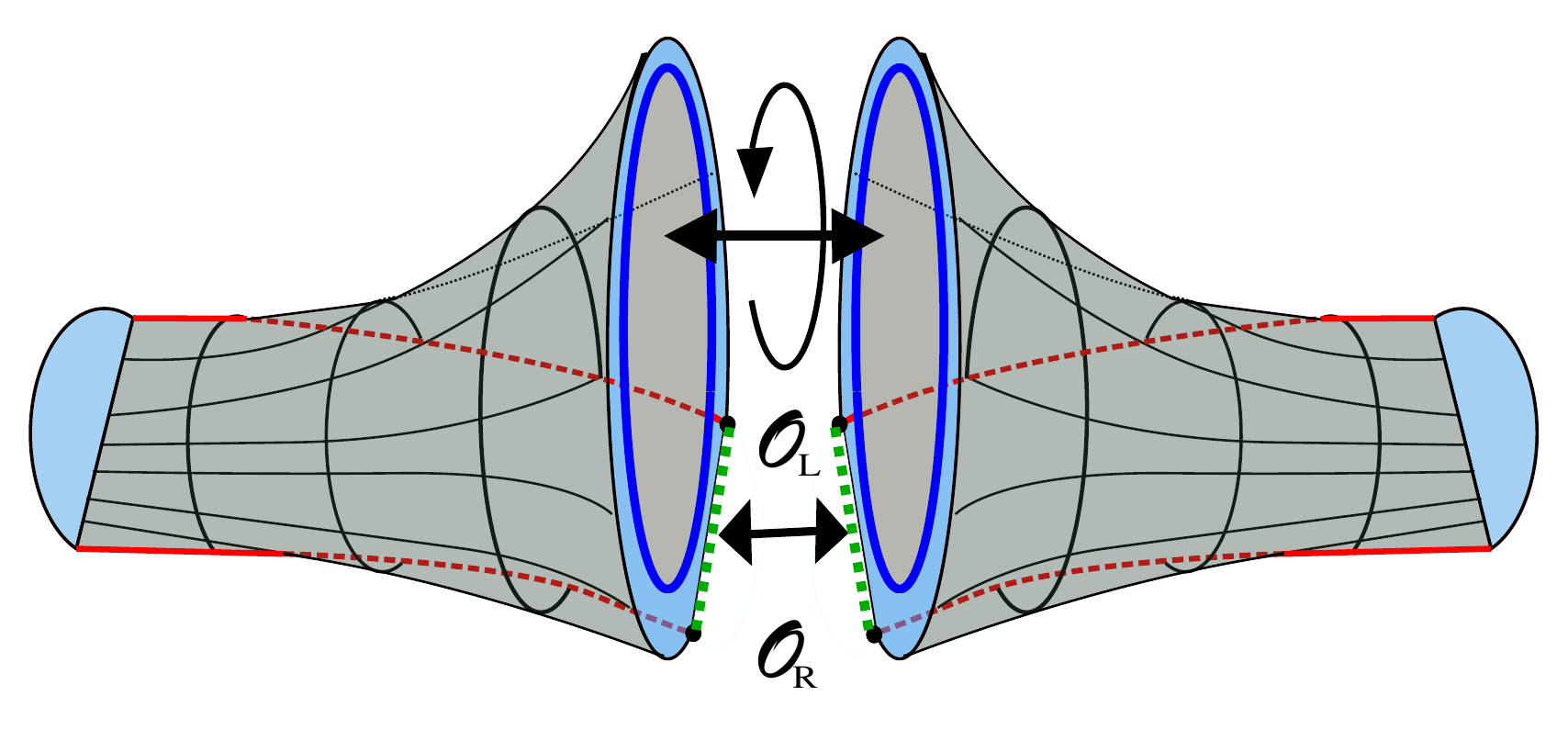}\hspace{20pt}\includegraphics[scale=1.4]{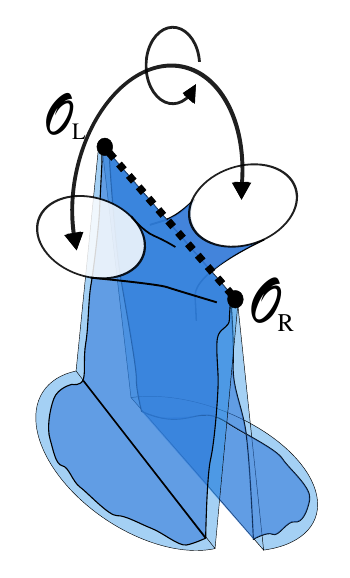}
\caption{\small Here we have pictured the sort of geometry that leads to the ramp in the two-sided two-point function. We can view the correlator as the expectation value of the length of the ERB in the density matrix $\rho(\ell,\ell')$. The ramp comes from a contribution to the density matrix where we trace over the state of a baby universe that has been emitted. At long times $T$, the amplitude to have a short ERB and a large baby universe, with size of order $T$, does not decay, leading to a non-decaying contribution to the correlator. The trace over the baby universes leads to an overall factor of $T$ because of the $\sim T$ ways to match two baby universes of size $T$.}
\label{fig:FigureShortening}
\end{figure}

\begin{figure}[H]
\centering
\includegraphics[scale=1.4]{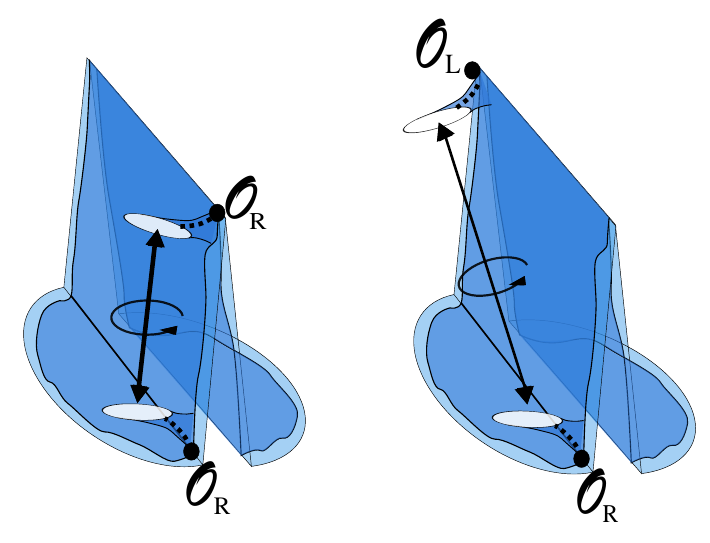}
\caption{\small One may view the two-point function as the transition amplitude between two state, an initial state $|i_R\rangle$ where a particle is near the right boundary at early times, and a state $|f_{L,R}\rangle$, where a particle is near the left or right boundary at late times. Baby universes can facilitate a ``shortcut'' between these two points in spacetime.}
\label{fig:FigureShortcut}
\end{figure}

These topology changing effects have two important consequences for matter correlators. First, a very large parent JT universe may become small by emitting a very large baby universe. Second, matter emitted with a baby universe may be reabsorbed by the parent universe with an amplitude that does not decay with time. We refer to these two effects as ``shortening'' and ``shortcuts'' respectively. We will find that these two effects are complementary descriptions of the non-decaying behavior of the two-point and four-point correlation functions at late times; these descriptions correspond to two different continuations of a Euclidean geometry. We picture these two effects in Figure \ref{fig:FigureShortening} and Figure \ref{fig:FigureShortcut}. The shortening picture is particularly useful for understanding the growth in the ramp region of the correlators. On the other hand, the shortening picture is particularly interesting; in this picture, the particle may have fallen deep into the interior of the black hole, but may escape through a baby universe and be measured at the boundary.

\subsection*{Layout and summary}
We now describe the layout of this paper and give a brief summary of each section.

In \textbf{Section \ref{SectionPredictions}} we describe the expected behavior of correlation functions in ensemble averaged theories and make precise predictions for the late time behavior of the two-point function in the thermofield double state and a class of out-of-time-ordered correlation functions (OTOCs). These predictions are based on the expectations that nearby energy levels have random matrix statistics, and that the matrix elements of the operators under consideration obey the Eigenstate Thermalization Hypotheses (ETH). Our formulas for the late time behavior of these correlation functions involves a model-dependent input, the averaged squared matrix elements of the operator. We may extract these matrix elements from the two-point function at early times; a precise formula for the matrix elements was given in \cite{Yang:2018gdb}. 

In \textbf{Section \ref{SectionJTGravity}} we describe some aspects of the late time behavior of pure JT gravity. JT gravity is a theory of two-dimensional gravity with a metric $g_{\mu \nu}$ and a dilaton $\phi$, defined by the Euclidean action
\be\label{JTaction}
I = -\underbrace{\frac{S_0}{2\pi}\left[\frac{1}{2}\int_{\mathcal{M}}\sqrt{g}R + \int_{\partial\mathcal{M}}\sqrt{h}K\right]}_\text{topological term $= S_0\,\chi(\mathcal{M})$} -\bigg[\underbrace{\frac{1}{2}\int_{\mathcal{M}}\sqrt{g}\phi(R+2)}_\text{sets $R = -2$} +\underbrace{\int_{\partial\mathcal{M}}\sqrt{h}\phi (K-1)}_\text{gives action for boundary}\bigg].
\ee
The integral over the dilaton localizes the path integral onto surfaces of constant negative curvature, dramatically simplifying the theory. In addition to an integral over moduli of these surfaces, we have a nontrivial integral over the shape of the asymptotic boundary. 

These simplifications are reflected in the Hilbert space description of JT gravity on a spatial line with asymptotically AdS boundary conditions. This theory has a single degree of freedom \cite{Harlow:2018tqv,Yang:2018gdb}); in this paper we will describe the Hilbert space in the length basis $|\ell\rangle$, where $\ell$ describes the renormalized length of a spatial slice. Contributions to the path integral from surfaces of higher topology lead us to consider the ``third-quantized'' JT Hilbert space, consisting of states of arbitrarily many spatial universes, which may be closed or have asymptotically AdS boundaries.

We begin Section \ref{SectionJTGravity} with some background about Lorentzian JT gravity, and give a Hilbert space description of the spectral form factor as a transition amplitude in the Hilbert space of two copies of JT gravity. With $|HH_{\beta, L}, HH_{\beta, R}\rangle$ the tensor product of two Hartle-Hawking states at inverse temperature $\beta$,
\be
|Z(\beta+i T)|^2_{JT} = \langle HH_{\beta,L}, HH_{\beta, R}| e^{-i\frac{T}{2} H_{Bulk} } |HH_{\beta, L}, HH_{\beta, R}\rangle
\ee
Time evolution with $H_{Bulk}$ acts by evolving forwards in time on the right boundaries and backwards on the right boundaries. 

As pictured in Figure \ref{fig:SFFFigure}, the spectral form factor decays initially during the ``slope'' region. Eventually, during the ``ramp'' region, the spectral form factor is exponentially small but growing linearly. This linear growth ends at the ``plateau'', where the spectral form factor remains at an exponentially small value. 

Using the spectral form factor as a probe of the length basis wavefunctions of Hartle-Hawking states, along with an exact formula for Hartle-Hawking state from \cite{Yang:2018gdb}, we discuss the long time behavior of the Hartle-Hawking wavefunction. Finally we discuss the Lorentzian interpretation of the ramp in the spectral form factor, which is a result of a topology changing tunneling process in which two JT universes exchange a baby universe \cite{Saad:2018bqo,Saad:2019lba}. At late times, the two JT universes have become very large, but they have a non-decaying amplitude to emit large baby universes and transition to a small universe. This process gives the dominant contribution to the spectral form factor during the ramp region. In this section we also give an exact formula for the amplitude to emit a baby universe. 

In \textbf{Section \ref{SectionRamp2pt}} we calculate the ``ramp'' contribution to the two-point function of a probe scalar field in the Hartle-Hawking state. We begin with setup, describing the calculation of the early time behavior of the correlator from \cite{Yang:2018gdb}, and discussing the geometry responsible for the ramp behavior. The ramp in the two-point function comes from geometries with the topology of the handle on a disk; upon continuation to Lorentzian signature, these describe a process in which a baby universe is emitted and then reabsorbed by a parent JT universe. 

For simplicity, our strategy is to first calculate the relevant contributions to the Euclidean correlator, and then continue the answer. However, we later describe the Lorentzian correlators more directly. 

The calculation of the Euclidean correlator suffers from two complications, and we find an expression for the correlator which appears difficult to evaluate. However, we will find that these two complications end up essentially cancelling each other out, resulting in a simple expression which we can directly match with out predictions. 

The first complication is that the matter two-point function on the relevant geometries is given by a sum over an infinite number of geodesics. Some of these geodesics contribute decaying corrections at late times, but we must still sum up contributions from an infinite number of geodesics to correctly describe the ramp. The second complication is that the integral over the moduli space of handle on a disk geometries is difficult to describe directly. The moduli space of geometries is most simply described in Fenchel-Nielsen coordinates, which describe the length and twist of one of infinitely many circular geodesics on the handle. To assure that we only integrate over distinct geometries, we must integrate over these parameters in a complicated fundamental domain. As we integrate over one fundamental domain, each distinct length and twist is represented by one of the infinitely many circular geodesics once and only once.

Fortunately, the sum over geodesics allows us to simplify the integral over moduli space. We can express the integral over moduli space of the sum over relevant geodesics to an integral over the union of infinitely many fundamental domains of a single term in the sum. The union of these infinitely many fundamental domains is simple, and we can perform the integral.\footnote{This is reminiscent of the method used in \cite{Polchinski:1985zf} to evaluate the one-loop string partition function.} 

The result is a formula for a contribution to the two-point function that precisely matches the prediction from Section \ref{SectionPredictions} before the plateau time. We leave a discussion of the corrections from other geodesics, which we argue give decaying contributions, to Appendix \ref{AppendixGeodesics}.

Finally, we discus the physical interpretation of this contribution to the two-point function, giving a precise Hilbert space description of the ``shortening'' and ``shortcut'' pictures.

In \textbf{Section \ref{SectionRamp4pt}} we perform a similar calculation for the ramps in the out-of-time-ordered four-point function. These come from geometries with the topology of disks with two handles, and we find a formula which precisely matches our predictions from Section \ref{SectionPredictions}. We briefly discuss our expectations for how ramps in $2k$-point OTOCs should come from geometries with $k$ handles, and describe how the calculation might work, but we leave the detailed calculation for future work. 

In \textbf{Section \ref{SectionPlateau}} we describe the plateau contributions to the two-point and four-point functions. We first give a brief overview of the origin of the plateau in the spectral form factor, discussed in \cite{Saad:2019lba}. There the plateau comes from processes in which baby universes are emitted but end in a ``D-brane'' state instead of being reabsorbed. We then find that we may simply adapt that calculation to find the plateaus in the two-point and four-point functions, finding precise agreement with our predictions from Section \ref{SectionPredictions}.

In \textbf{Section \ref{SectionDiscussion}} we discuss some future directions and open questions.

\subsection*{Discussion of previous work}

In \cite{Blommaert:2019hjr}, the authors studied the first order formalism of JT gravity, which is described by a topological BF-type theory, and also found a formula for the two-point function at late times. Our formula (\ref{TwoPointPlateau}) reduces to their formula after approximating the matrix elements as constants. However, in \cite{Blommaert:2019hjr} the authors did not divide out by the mapping class group when performing the integral over connections, so distinct geometries were counted infinitely many times. The authors also considered the contribution of only one of the infinitely many non-self-intersecting geodesics which we find contribute to the late-time behavior. However, as we explain in this paper, these two complications cancel out, and we find the same formula. 

\section{Averaged correlation functions at late times}\label{SectionPredictions}
In this paper we will be interested in the late time behavior of a certain class of correlation functions in a theory of JT gravity coupled to a free scalar field. The simplest of these, which we will study in the most detail, is the two-point function of the scalar field in the Hartle-Hawking state. These correlation functions decay for early times. However, we expect that these correlation functions should behave like an ensemble average of correlation functions in the thermofield double state for a system of finite entropy. For systems of finite entropy, these correlation functions cannot decay forever, so there must be corrections to this behavior.

In this section we give predictions for the late time behavior of these correlation functions in JT gravity. The main inputs for these predictions will be our expectation that JT gravity coupled to matter behaves like an ensemble averaged system, where the Hamiltonians drawn from the ensemble have random matrix statistics. As a consequence of this, we expect that correlations of the density of states for nearby energies behave like those of a random matrix, and that the matrix elements of the fields $\mathcal{O}$ obey the Eigenstate Thermalization Hypothesis (ETH). 

We begin the section with some overview of how two-point functions in in systems with finite entropy behave at late times. We then discuss in more detail our expectations for two-point functions in ensemble-averaged theories. With these expectations, and using input from the early time decaying behavior of two-point function, we give a precise prediction for the behavior of two-point functions at late times in JT gravity. Finally we discuss a class of higher point OTOCs at late times, and make precise predictions for their behavior as well.

\subsection{Two-point functions at late times}\label{Subsection2ptLateTimes}
Consider a chaotic quantum mechanical theory with a discrete spectrum, such as the SYK model with a fixed set of couplings. Two correlation functions of primary interest to us are the thermal two-point function $G_{2,\beta}(T)$ and the two-sided two-point function in the (unnormalized) thermofield double state, $G^{LR}_{2,\beta}(T)$. For an operator $\mathcal{O}$, these are defined as
\be
G_{2,\beta}(T) \equiv \Tr \big[e^{-\beta H} \mathcal{O}(T)\mathcal{O}(0)\big] \label{Gtherm}
\ee
and
\be
G^{LR}_{2,\beta}(T)\equiv \langle TFD_\beta| \mathcal{O}_L(T)\mathcal{O}_R(0)|TFD_\beta \rangle \label{Gtwoside}
\ee
where $|TFD_\beta\rangle \equiv \sum_n e^{-\frac{\beta}{2}E_n} |E_n^*\rangle_L |E_n\rangle_R$ is an entangled state of two copies of our system. 

These correlation functions are related by analytic continuation in $T$, and can both be obtained by analytic continuation of the Euclidean two-point function.

Thermal correlation functions are often defined to be normalized, with normalization factor $Z(\beta)$. In this paper we will work with unnormalized correlation functions, as they are simpler in gravity. We do not expect significant differences in behavior between the normalized and unnormalized correlators.\footnote{In gravity, this would require us to introduce replicas. There would be replica symmetry breaking effects from Euclidean wormholes, but we expect that they would give exponentially small corrections. The non-decaying effects from replicas should provide a multiplicative correction of $1+\mathcal{O}(e^{-2 S(\beta)} )$, so the corrections are unimportant for all times.}

By expanding the trace in (\ref{Gtherm}) or the $TFD$ state in (\ref{Gtwoside}) in terms of a sum over energy eigenstates and inserting a complete set of energy eigenstates in between the two operators, we may express these correlators in the forms
\be
G_{2,\beta}(T) = \sum_{n,m} e^{-\beta E_m} e^{-i T(E_n-E_m)} |\langle E_n |\mathcal{O}|E_m\rangle|^2,
\ee
\be
G^{LR}_{2,\beta}(T) = \sum_{n,m} e^{-\frac{\beta}{2} (E_m+E_m)} e^{-i T(E_n-E_m)} |\langle E_n |\mathcal{O}|E_m\rangle|^2. \label{G2sideenergy}
\ee
In (\ref{G2sideenergy}) we used the property $\langle E_n^*| \mathcal{O}_L | E_m^*\rangle=\langle E_m| \mathcal{O}_R | E_m\rangle\equiv \langle E_m| \mathcal{O}| E_m\rangle$.\footnote{The * operation is a product of time reversal and an exchange of the $L$ and $R$ systems.}

For short times $T$ we may approximate these double sums over energies by integrals weighted by the smooth density of states, $\sum_n f(E_n) \rightarrow \int dE \rho_0(E) f(E)$, where the smooth density of states $\rho_0(E)$ can be defined by an average of the discrete density over energy windows much larger than the level spacing.

If our operators satisfy the ETH, as is expected for simple operators $\mathcal{O}$ in a chaotic system, we expect that the off-diagonal matrix elements $|\langle E_n|\mathcal{O}|E_m \rangle|^2$ may be well approximated by smooth functions of the energies $E_n$ and $E_m$ that are of order $1/\rho_0(\frac{E_n+E_m}{2})$. We will discuss more details of the ETH in the next section.

For early times $T$, we then expect that $G_{2,\beta}(T)$ and $G_{2,\beta}^{LR}(T)$ may be well approximated by integrals of smooth functions of the energies. In particular, the integrals over $E_n-E_m$ give Fourier transforms of smooth functions,\footnote{The ``smooth'' density of states may have sharp edges, as in the SYK model or JT gravity \cite{Maldacena:2016hyu,Cotler:2016fpe,Bagrets:2017pwq,Stanford:2017thb,Mertens:2017mtv,Kitaev:2018wpr,Yang:2018gdb} but to leading order this does not stop the decaying behavior. In general we may work in the microcanonical ensemble to avoid these edge effects.}, which should decay. 

We now focus our attention on systems with a zero-temperature entropy $S_0$ parameter, such as SYK or JT gravity. This will be useful as it will allow us to roughly discuss the size of the correlator and the time $T$ in comparison to the entropy by looking at the dependence on the parameter $S_0$. For example, the smooth density of states is exponential in $S_0$
\be
\rho_0(E)= e^{S_0}\tilde{\rho}_0(E),
\ee
with $\tilde{\rho}$ independent of $S_0$. The norm of off diagonal matrix elements is exponentially small in $S_0$, $|\langle E_n|\mathcal{O}|E_m\rangle|^2\sim e^{-S_0} f(E_n,E_m)$. Since our expressions for the two point functions involve two integrals weighted by the smooth density of states and one factor of the squared matrix elements, we expect that for short times they have a magnitude of order $e^{S_0}$
\be
G_{2,\beta}(T),\; G_{2,\beta}^{LR}(T) \sim e^{S_0}, \hspace{20pt} T\sim 1.
\ee
However after a long period of decay, the correlators will become exponentially smaller than their initial values. On the other hand, the expressions (\ref{Gtherm}) and (\ref{Gtwoside}) require that this decay cannot continue forever \cite{Maldacena:2001kr,Goheer:2002vf,Dyson:2002pf,Barbon:2003aq}. The long time average of the correlators can be obtained by setting $E_n=E_m$ in these expressions. For systems with no degeneracies, we find that only the diagonal part of the sum remains.\footnote{For systems with degeneracies, we have extra contributions, but these will not affect the size with respect to $S_0$ so we will ignore them.}

With only the diagonal sum remaining, we have a sum over energies, of order $e^{S_0}$ in a given window, of a function of size of order $e^{-S_0}$, so the long time average of the correlators is of order one.
\be
\lim_{T\rightarrow \infty} \frac{1}{2T} \int_{-T}^T dT' G_{2,\beta}(T') = \sum_n e^{-\beta E_n} |\langle E_n| \mathcal{O} |E_n\rangle|^2 \sim (e^{-S_0})^0 \label{GLongTime}
\ee

We conclude that at some point the decay of the two-point function must stop, indicating that our approximation of the density of states by a smooth function should break down; at long enough timescales, we need to pay attention to the discreteness of the spectrum. 

The actual behavior of the correlation function after the decaying behavior stops is quite complicated. We expect that the correlator begins to oscillate erratically, with the details of the oscillations depending sensitively on the precise set of energies $E_n$.\footnote{At these timescales the correlator is not ``self-averaging'' \cite{prange1997spectral,Cotler:2016fpe}; the size of the fluctuations about the average is comparable to the size of the correlator, so the correlator cannot be well approximated by its average. We will return to this issue in Section \ref{SectionDiscussion}.}

\subsection{Averaged theories, RMT statistics, and ETH}\label{SubsectionAveraged}
We now switch our attention to theories described not by a fixed Hamiltonian, but by an ensemble of Hamiltonians. An example is the ensemble of SYK models with the couplings $J$ drawn from a Gaussian distribution. In these theories, we can think about correlation functions averaged over the ensemble of Hamiltonians. The simplest examples of such correlators involve operators $\mathcal{O}$ defined independently from the Hamiltonian, such as the operators $\frac{1}{N}\sum_i \psi_i \partial_{\tau}^n \psi_i$ in the SYK model. Using the same notation as for unaveraged correlators, we have for example
\be
G_{2,\beta}(T) = \int dH P(H) \;\Tr\big[ e^{-\beta H} \mathcal{O}(T) \mathcal{O}(0)\big],
\ee
where $P(H)$ is the probability density for a Hamiltonian $H$. Here for simplicity we will assume that the Hilbert space has finite dimension $L$.\footnote{The general lessons of this section will not really depend on this fact. We expect that these considerations should apply to the behavior within a finite energy window.} In the SYK model, we can express this average over Hamiltonians as an average over the couplings $J_{ijkl}$.

We can express a Hamiltonian $H$ in a fixed basis $|i\rangle$ in terms of its eigenvalues $E_n$ and a unitary $U_{in}=\langle i|n\rangle$ defining the change of basis between the fixed basis and the energy eigenbasis $|n\rangle$. We may express the integral over Hamiltonians by separate integrals over $E_n$ and $U$, with a probability measure $P(E_n, U)$ that in general does not factorize. 

With the discrete density of states of a given Hamiltonian $\rho_{d}(E)= \Tr \;\delta(E\hat{I}-H)$, where $\hat{I}$ is the identity matrix, define the pair correlation function $\rho(E,E') =\int dH P(H) \;\rho_d(E) \rho_d(E')$. We expect that we may write the averaged correlation function as an integral over energies, replacing the sums over energies by integrals $\sum_{n,m} \rightarrow \int dE dE' \rho(E,E')$, and the averaged matrix elements as functions of the energies $|\mathcal{O}_{E,E'}|^2$. For example, we expect
\begin{align}
G^{LR}_{2,\beta}(T) =\int dE dE'  \rho(E,E') e^{-\frac{\beta}{2}(E+E')} e^{-i T (E-E')} |\mathcal{O}_{E,E'}|^2. \label{GFactorizedMeasure}
\end{align}

In random matrix theory the measure over matrices factorizes $P(E_n, U)= P_E(E_n) P_U(U)$. The measure for the eigenvalues may be described by an appropriate ``potential'' $V(E)$, and the measure for the unitaries is the Haar measure\footnote{For the GUE symmetry class only. For other symmetry classes, we find a different measure. In this paper we will restrict our attention to JT gravity on orientable surfaces, which is in the GUE symmetry class. JT gravity on non-orientable surfaces has been studied in \cite{Stanford:2019vob}.}. For a particular choice of the potential, the random matrix ensemble in the double scaled limit\footnote{In this limit the dimension of the Hilbert space $L$ is taken to infinity but the averaged density of states near the edge of the spectrum is kept finite.} describes JT gravity \cite{Saad:2019lba}, in the sense that to all orders in $e^{-S_0}$, partition functions of JT gravity are equal to partition functions of the Hamiltonian $H$ averaged over the appropriate ensemble;
\be
Z_{JT}(\beta_1,\dots \beta_n)  \sim \int dH P(H) \Tr \big[e^{-\beta_1 H} \big]\dots \Tr \big[e^{-\beta_n H}\big].
\ee
By taking inverse Laplace transforms of these partition functions, we may extract the averaged density of states $\rho(E) = \int d( E_n) P_E (E_n) \rho_d(E)$ and the pair correlation function $\rho(E,E')$ for JT gravity.

\subsubsection{Averaged densities of states in JT gravity}
The averaged density of states $\rho(E)$ may be obtained from the partition function $Z(\beta)_{JT}$. To leading order this is given by an integral over geometries with the topology of a disk. The topological term in (\ref{JTaction}) gives these geometries a weight of $e^{S_0}$, so that the averaged density of states is proportional to $e^{S_0}$. The resulting density of states is \cite{Maldacena:2016hyu,Cotler:2016fpe,Bagrets:2017pwq,Stanford:2017thb,Mertens:2017mtv,Kitaev:2018wpr,Yang:2018gdb},
\be
\rho(E) = \rho_0(E)(1+O(e^{-S_0}) ),\hspace{20pt} \rho_0(E)\equiv e^{S_0}\frac{\sinh(2\pi \sqrt{2 E})}{2\pi^2}.
\ee
To leading order, the pair correlation function $\rho(E,E')$ factorizes
\be
\rho(E,E') = \rho_0(E) \rho_0(E')( 1+ O(e^{-S_0}) ) \sim e^{2 S_0}. \label{rhofactorized}
\ee
For $|E-E'| \ll 1$, corrections to this factorized form become important. These corrections come from contributions that are subleading in the genus expansion, but singular as $E\rightarrow E'$. Away from the edge of the spectrum \cite{Saad:2019lba},
\begin{align}
\rho(E,E')- \rho_0(E) \rho_0(E') &\approx -\frac{1}{2 \pi^2 (E-E')^2} + \frac{\cos\big(2 \pi \rho_0(E)(E-E') \big)}{2 \pi^2 (E-E')^2}+\rho_0(E) \delta(E-E') 
\cr
&\equiv \rho_{RMT}(E,E'),\hspace{20pt} E,E' >0,\;\; |E-E'| \ll 1. \label{rhormt}
\end{align}
The first term is perturbative in the genus expansion parameter $e^{-S_0}$; it comes from cylindrical geometries in JT gravity. The second term is \textit{nonperturbative} in $e^{-S_0}$, and comes from the effects of single eigenvalues in the matrix integral, and from a sum over an infinite number of disconnected geometries. The third term accounts for value of the late time average of the correlation function (\ref{GLongTime}); its origin is similar to that of the second term. 

We will discuss the JT gravity explanation of these terms in more detail in Section \ref{SectionJTGravity} and Section \ref{SectionPlateau}. For now we will just note some important properties. To do so, we will introduce the spectral form factor $|Z(\beta+i T)|^2_{JT}= \int dE dE' \rho(E, E') e^{-\beta(E+E')} e^{-i T(E-E')}$. This quantity is simply what we get by ignoring the matrix elements in (\ref{GFactorizedMeasure}) and setting $\beta\rightarrow 2\beta$. 

For short times, the spectral form factor is well approximated by using the factorized approximation (\ref{rhofactorized}) for the pair correlation function \cite{Cotler:2016fpe}. The result is the factorized expression
\be
|Z(\beta+i T)|^2_{JT}\approx \bigg(\int dE \rho_0(E) e^{-\beta E} e^{-i E T} \bigg)\bigg(\int dE' \rho_0(E') e^{-\beta E'} e^{i E' T} \bigg) = |Z(\beta+i T)_{JT}|^2
\ee
Each factor in parentheses decays in $T$, as we are Fourier transforming the smooth function $\rho_0(E)e^{-\beta E}$ \footnote{$\rho(E)$ is not really smooth, but has a sharp edge at $E=0$. However, this singularity only causes the decay to transition from exponential to power law.}.

For later times, we must include the corrections (\ref{rhormt}). The first term gives rise to a linearly growing contribution to the spectral form factor, the ``ramp''. This linear growth is a direct consequence of the $1/(E-E')^2$ singular behavior. The second term gives a constant contribution until $T\sim e^{S_0}$, when it transitions to a linear behavior that exactly cancels the linear growth from the first term. This transition at exponentially long times is due to the rapid oscillations $\sim \cos(e^{S_0} (E-E'))$, which reflects the underlying discreteness of the energy spectrum. The third gives a constant contribution to the spectral form factor of size $Z(2\beta)$. This is the size of the ``plateau'' that the spectral form factor limits to at exponentially long times.
\begin{figure}
\centering
\includegraphics[scale=0.6]{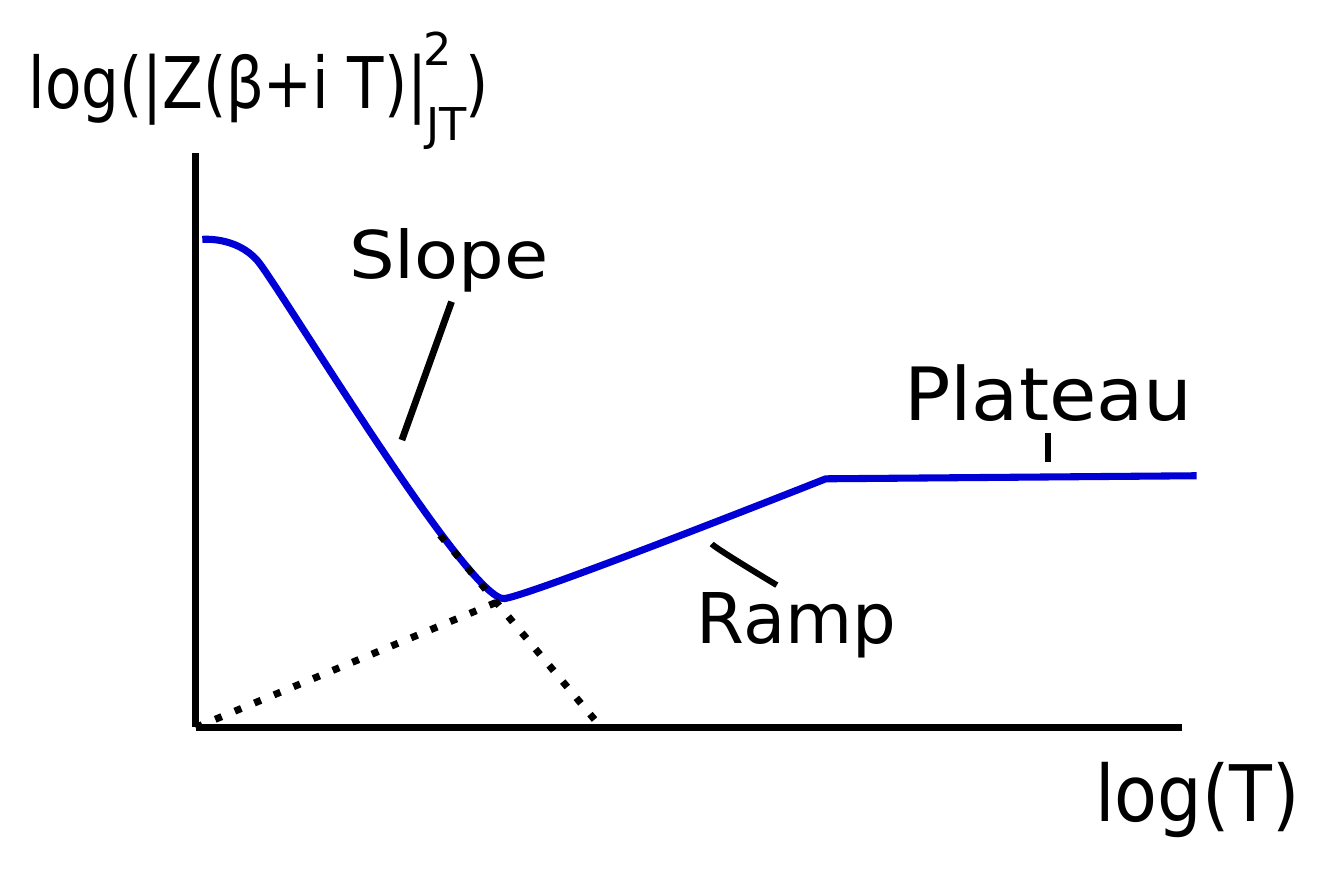}
\caption{\small Above we have pictured the general features in the spectral form factor. The slope transitions to the plateau at a time $T\sim e^{S_0/2}$ \cite{Cotler:2016fpe}, and the ramp transitions to the plateau at a time $T\sim e^{S_0}$. The height of the plateau is smaller than the $T=0$ value of the spectral form factor by a factor of $e^{-S_0}$.}
\label{fig:SFFFigure}
\end{figure}
It is also useful to introduce a fixed energy version of the spectral form factor
\be\label{YET}
Y(E,T) = \frac{1}{\pi i } \int_{\mathcal{C}} d\beta \; e^{2 \beta E} |Z(\beta+i T)|^2_{JT},
\ee
where $\mathcal{C}$ is a vertical contour to the right of the origin. For $E>0$, this function behaves similarly to the spectral form factor at late times.\footnote{The early time behavior of this function is somewhat different than the early time behavior of the spectral form factor. The slow power law decay in the spectral form factor was a consequence of the sharp edge in the energy spectrum at $E=0$, but fixing the energy allows us to avoid this edge. If we fix the energy in a smooth window away from $E=0$, the fixed energy quantity will decay more quickly than the spectral form factor.} $Y(E,T)$ also has a ramp and a plateau,
\be
 Y(E,T) \approx \begin{cases}\frac{T}{2\pi} & 1\ll T< 2 \pi \rho_0(E), \cr \rho_0(E) & T> 2\pi \rho_0(E). \end{cases}\label{YLateTime}
\ee

\subsubsection{Matrix elements and ETH}
Having discussed the expected behavior of the eigenvalue densities in (\ref{GFactorizedMeasure}), we now move on to discussing the matrix elements $|\mathcal{O}_{E,E'}|^2$.

First we should discuss what operators $\mathcal{O}$ we will be interested in. We will imagine coupling a real scalar field $\mathcal{O}$ to JT gravity, and the correlation functions of interest will be of scalar fields at the boundary, appropriately rescaled as in \cite{Yang:2018gdb}. While we know that pure JT gravity is described by an ensemble average over theories with a discrete spectrum, we do not know if this holds true for a putative theory of JT gravity coupled to matter.\footnote{In fact, there are difficulties in coupling JT gravity to matter, which we will discuss in more detail in Section \ref{SectionRamp2pt}. A certain case of JT gravity couple to matter was recently considered in \cite{Iliesiu:2019lfc}.}. However, anticipating our result that correlation functions in such a theory do behave like those of an ensemble averaged theory, we will continue to discuss our expectations for ensemble averaged theories.

Our expectation for the behavior of the averaged matrix elements $|\mathcal{O}_{E,E'}|^2$ with $\mathcal{O}$ a Hermitian operator is that they are described by the \textit{Eigenstate Thermalization Hypothesis} (ETH)\cite{srednicki1994chaos,deutsch1991quantum}. To motivate the ETH ansatz for the averaged matrix elements, we first consider the case of (GUE symmetry class) random matrix theory, where the measure $d U P_U(U)$ is the Haar measure over $U(L)$. In this case, we can do the integral over the unitary matrices exactly. 

We start by first considering the average of the sigle matrix element $\langle E_n|\mathcal{O}|E_m\rangle$, where the dimension of the Hilbert space is $L$. We may write this as
\begin{align}
\int_{Haar} dU\langle E_n|\mathcal{O}|E_m\rangle &=\int_{Haar} dU \sum_{i=1}^L U^\dagger_{ni} U_{i m} \mathcal{O}_i 
\cr
&= \frac{1}{L} \delta_{nm} \sum_i\mathcal{O}_i
\cr
&=\delta_{nm} \frac{\Tr \big[\mathcal{O}\big]}{L} 
\end{align}
This tells us that the off-diagonal matrix elements average to zero, while the diagonal elements are described by the infinite temperature average of $\mathcal{O}$. Let's briefly discuss some expectations of this thermal average. If there is a symmetry that takes $\mathcal{O} \rightarrow -\mathcal{O}$, this average should be zero. For a strictly positive operator like a density, with typical eigenvalues of order one, this average will be of order one.

We may perform a similar calculation to find the average of the squared matrix elements in random matrix theory,
\begin{align}
\int_{Haar} dU \langle E_n|\mathcal{O} |E_m\rangle \langle E_p | \mathcal{O}|E_q\rangle &= \int_{Haar} dU \sum_{i,j=1}^L U^\dagger_{n i }U_{i m} U^\dagger_{pj}U_{jq} \mathcal{O}_i \mathcal{O}_j 
\cr
&\approx \frac{1}{L^2}\delta_{nm} \delta_{pq} \sum_{i,j} \mathcal{O}_i \mathcal{O}_j + \frac{1}{L^2} \delta_{nq} \delta_{mp} \sum_i \mathcal{O}_i^2
\cr
&= \delta_{nm} \delta_{pq} \bigg(\frac{\Tr[\mathcal{O}]}{L}\bigg)^2 + \delta_{nq}\delta_{mp} \frac{\Tr[ \mathcal{O}^2]}{L^2}.
\end{align}
In the second line we dropped terms that are subleading in $1/L$. Our results for these statistics of the matrix elements are reproduced by the ansatz
\be\label{RandomMatrixElements}
\langle E_n|\mathcal{O}|E_m\rangle_{RMT} \approx \begin{cases} \bigg(\frac{\Tr\big[\mathcal{O}\big]}{L}+R_n \sqrt{\frac{\Tr[\mathcal{O}^2]}{L^2}}\bigg) \hspace{20pt} n=m,
\cr
R_{nm} \sqrt{\frac{1}{L}\bigg(\frac{\Tr\big[\mathcal{O}^2\big]}{L}\bigg)}\hspace{40pt} n\neq m,
\end{cases}
\ee
where the $R_{nm}$ are random phases $e^{i \phi_{nm}}$, $\phi_{nm}=-\phi_{mn}$, that are independent and average to zero, and $R_n$ is a random number that averages to zero. This ansatz also reproduces more general statistics of the matrix elements to leading order.

Up to the phase, this form of the matrix elements does not care what the energies $E_n$ and $E_m$ are, it only depends on whether or not $n=m$. The ETH ansatz is a generalization of this form of the matrix elements, essentially describing the eigenfunctions within a small energy window as random. With $\overline{\mathcal{O}}_{E}$ and $\overline{|\mathcal{O}_{E,E}|^2}$ smooth functions of $E$ and $E'$, the ansatz is
\be
\langle E_n|\mathcal{O}|E_m\rangle_{ETH} \approx \delta_{nm} \overline{\mathcal{O}}_{E_n}+ R_{nm}\sqrt{\frac{ \overline{|\mathcal{O}_{E_n,E_m}|^2}}{\rho\big(\frac{E_n+E_m}{2}\big)}}. \label{ETHAnsatz}
\ee
We could also add a small term to the diagonal matrix elements with a random sign to match (\ref{RandomMatrixElements}) more closely. This would be especially appropriate if the average one-point function of $\mathcal{O}$ is zero. We may think of the ETH ansatz for matrix elements as telling us to approximate the energy eigenvector statistics within narrow energy bands as random matrix statistics.

In JT gravity, we will find that the matrix elements are described by this ansatz. The off-diagonal elements will behave like smooth functions of the energies times a random phase. To describe these matrix elements we remove the explicit factor of $e^{S_0}$ from the density of states in (\ref{ETHAnsatz}) and lump the rest of the density of states with the smooth function $\mathcal{O}_{E,E'}$,
\be
\langle E_n|\mathcal{O}|E_m\rangle_{JT} \approx R_{nm} e^{-S_0/2} \sqrt{|\mathcal{O}_{E_n,E_m}|^2},\hspace{20pt} n\neq m.
\ee
We will also find that the diagonal elements average to zero, but the average of their square is nonzero and exponentially small,
\be
\langle E_n|\mathcal{O}|E_n\rangle_{JT} \approx R_n e^{-S_0/2} \sqrt{|\mathcal{O}_{E_n,E_m}|^2}. \label{OnePointPrediction}
\ee
Since the diagonal and off-diagonal matrix elements are described by the same function $|\mathcal{O}_{E_n,E_m}|^2$, the squared matrix elements have a simple uniform behavior,
\be
|\langle E_n| \mathcal{O}|E_n\rangle |^2_{JT} = e^{-S_0}|\mathcal{O}_{E_n,E_m}|^2. \label{JTMatrixElements}
\ee

\subsection{Putting it together: Our prediction for the two-point function}\label{SubsectionPuttingItTogether}

We now put these pieces together to make a prediction for the late time behavior of the boundary two point functions in JT gravity coupled to matter.

First, we write down the expected form of the two-point function at early times, focusing our attention on $G_{2,\beta}^{LR}(T)$ for simplicity. We approximate the pair correlation function in (\ref{GFactorizedMeasure}) as the product of the averaged densities, and input our ETH ansatz (\ref{JTMatrixElements}). Altogether, we have
\be
G^{LR}_{2,\beta}(T) \approx \int dE dE' \rho_0(E) \rho_0(E') e^{-\frac{\beta}{2} (E+E')} e^{-i T(E-E')} e^{-S_0}|\mathcal{O}_{E,E'}|^2, \hspace{20pt} T\sim 1. \label{GPredictionEarlyTime}
\ee
Because of the two factors of $e^{S_0}$ from the densities of states, the total size of the correlator for early times will be of order $e^{S_0}$. 

In \cite{Yang:2018gdb}, up to corrections of order $e^{-S_0}$, the correlator in JT gravity was found to take exactly this form. For boundary operators of scaling dimension $\Delta$, we find the prediction for the off-diagonal matrix elements \cite{Yang:2018gdb}
\be
|\mathcal{O}_{E,E'}|^2= \frac{|\Gamma(\Delta - i(\sqrt{2 E}+\sqrt{2 E'})\Gamma(\Delta - i(\sqrt{2 E}-\sqrt{2 E'})|^2}{2^{2\Delta+1}\Gamma(2\Delta)}\label{OsquaredJT}
\ee
In Section \ref{SectionPlateau} we will explicitly calculate the diagonal matrix elements and their square to confirm our expression (\ref{GPredictionEarlyTime}) with $|\mathcal{O}_{E,E'}|^2$ given by the above formula.

The correlator $G^{LR}_{2,\beta}(T)$ initially decays exponentially in $T$, but eventually begins to decay as $T^{-3}$ \cite{Yang:2018gdb}. This decay continues until the correlator is exponentially small. Eventually, contributions from the $\rho_{RMT}(E,E')$ piece of $\rho(E,E')$ (\ref{rhormt}) should begin to dominate over the decaying contribution from the factorized piece $\rho_0(E)\rho_0(E')$. 

Using our ansatz (\ref{GFactorizedMeasure}) for the two-point function, our expectation that $\rho(E,E')-\rho_0(E)\rho_0(E')\approx \rho_{RMT}(E,E')$ for $|E-E'|\ll 1$, and our prediction (\ref{JTMatrixElements}) and (\ref{OsquaredJT}) for the matrix elements, we can predict that
\be
\boxed{G^{LR}_{2,\beta}(T) \approx e^{-S_0} \int dE dE' \rho(E,E')_{RMT} e^{-\frac{\beta}{2}(E+E')} e^{-i T(E-E')}|\mathcal{O}_{E,E'}|^2, \hspace{20pt} T\gg 1.} \label{GJTPrediction}
\ee
To gain some intuition, let's briefly estimate the behavior of this late time contribution to the correlator. For long times, the integral over $E-E'$ is dominated by small values of $E-E'$. $\mathcal{O}^2(E,E')$ has local maximum at $E-E'$ so we will approximate $\mathcal{O}$ by the value at this maximum. Denoting $|\mathcal{O}_{E,E}|^2= |\mathcal{O}_{E}|^2$, we find
\be
G^{LR}_{2,\beta}(T) \approx e^{-S_0} \int dE \; e^{-\beta E} |\mathcal{O}_{E}|^2 \;Y(E,T),\hspace{20pt} T\gg 1. \label{GLateApprox}
\ee
Using our late time approximation (\ref{YLateTime}) for $Y(E,T)$, we we find an expression for the correlator at late times as a sum over ramps and plateaus, weighted by a Boltzmann factor and the matrix elements $|\mathcal{O}_E|^2$.\footnote{ETH (\ref{ETHAnsatz}) would lead us to expect that $e^{-S_0}\mathcal{O}^2(E)\sim 1/\rho(E)$. For $E\gg \Delta$ we can use Euler's reflection formula for the gamma functions to see that this is indeed the case. Along with the Boltzmann weight, this entropy factor gives a pressure pushing the integral over $E$ in (\ref{GLateApprox}) towards small $E$. In higher dimensional gravity this leads us to expect that the largest contribution at late times is comes from the thermal gas \cite{Maldacena:2001kr,Barbon:2003aq} instead of from black holes; we could remove these contributions by working in the microcanonical ensemble. This entropy factor is absent in the spectral form factor, so there is less pressure towards small $E$.}

\subsection{Higher point OTOCs at Late Times}\label{SubsectionOTOCPredictions}

The averaged two point function does not probe the full structure of the matrix elements $\langle E_n |\mathcal{O} |E_m\rangle$. In particular, the two-point function doesn't depend on the phases of the matrix elements. If we want to probe more of the structure of the matrix elements, we sill have to look at higher point correlators.

Out-of-time-ordered correlators, or OTOCs, turn out to be a particularly useful probe of the structure of these matrix elements. We will focus on OTOCs of the form
\be
G_{2k,\{\beta_i\}}( T) \equiv \Tr\big[ e^{-\beta_1 H}\mathcal{O}(T)e^{-\beta_2 H} \mathcal{O}(0) e^{-\beta_3 H} \mathcal{O}(T) \dots e^{-\beta_{2k-1} H}\mathcal{O}(T) e^{-\beta_{2k} H} \mathcal{O}(0)\big]
\ee

When all of the $\beta_i$ are equal, $\beta_i = \beta/2$, then this is equal to a conventional two-sided OTOC. With $\beta_i=\beta$ for odd $i$ and $\beta_i=0$ for even $i$, this is equal to a conventional one-sided OTOC. However, keeping the $\beta_i$ independent would allow us to probe the matrix elements $\langle E_n |\mathcal{O}|E_m\rangle$ in more detail; for example, it will be useful to fix the intermediate energies between operators with inverse Laplace transforms.

This general correlator can be somewhat unwieldy, so to demonstrate some of the behaviors of these functions we will proceed for now with the special case $k=2$, $\beta_i=\beta/2$. Following our procedure of taking the trace in the energy eigenbasis and inserting complete sets of energy eigenstates in between each pair of $\mathcal{O}$, we find
\begin{align}
G_{4,\beta/2}( T) = \sum_{n,m,p,q} &e^{-\frac{\beta}{2}(E_n+E_m+E_p+E_q)} e^{-i T(E_n-E_m+E_p-E_q)}
\cr
&\times\; \langle E_n|\mathcal{O}|E_m\rangle \langle E_m|\mathcal{O}|E_p\rangle \langle E_p|\mathcal{O}|E_q\rangle \langle E_q|\mathcal{O}|E_n\rangle 
\end{align}

For simplicity, we will assume that $\langle E_n|\mathcal{O}|E_n\rangle=0$. Using our ansatz for the off-diagonal matrix elements, we see that the product of matrix elements includes a product of random phases
\be
R_{nm}R_{mp}R_{pq}R_{qn}
\ee
Unless these phases cancel in pairs, this product of phases to average to zero.\footnote{In order to account for perturbative corrections to these correlation functions, such as those which exhibit Lyapunov behavior, we should relax this assumption \cite{Foini:2018sdb}, allowing non-paired matrix elements to have small correlations. In JT gravity, these effects are captured by the coupling of the matter to the boundary \cite{Maldacena:2016upp,Mertens:2017mtv,Blommaert:2018oro,Lam:2018pvp,Bagrets:2017pwq}. At both early and late times, such effects are subleading and so we will ignore them.}

In order for these phases to cancel in pairs, some of the energies must be equal to each other. In this case, the phases cancel if $n=p$ or $m=q$. In order to get a non-vanishing answer at late times, we must restrict the sum over energies. Since we are summing over one fewer set of energies, we expect that the size of this contribution will be smaller than the early time value of the function by a factor of $e^{-S_0}$ for each pairing of energies. When describing the averaged sum over energies with an integral weighted by the density of states, we should enforce these energy parings with a delta function $\frac{\delta(E-E')}{\rho_0(E)}$.

Replacing the sums over energies by the integrals over energies weighted by the four-density correlation function and inserting our ansatz (\ref{JTMatrixElements}) for the matrix elements, we may write a prediction for the averaged correlator
\begin{align}
G_{4,\beta/2}(T) = 2\;e^{-2 S_0}& \int dE_1 dE_2 dE_4\; \frac{\rho(E_1,E_2,E_1,E_4)}{\rho_0(E_1)} 
\cr
&\times e^{-\frac{\beta}{2}(E_1+E_2+E_1+E_4)} e^{-i T(E_1-E_2+E_1-E_4)} |\mathcal{O}_{E_1, E_2}|^2 |\mathcal{O}_{E_1, E_4}|^2
\end{align}
The factor of $2$ comes from two identical contributions from the two energy pairings. 

At late times, this will be dominated by contributions from when $E_1-E_2+E_1-E_4$ is small. This can happen when $E_2$ and $E_4$ are both close to $E_1$. In this limit, the density correlator, subtracting out the factorized contribution $\rho(E_1)\rho(E_2)\rho(E_4)$ which leads to decay, can be approximated by RMT pair correlators
\be
\rho(E_1,E_2,E_1,E_3) - \rho_0(E_1)^2\rho_0(E_2)\rho_0(E_4) \approx \rho(E_1,E_2)_{RMT} \rho(E_1,E_4)_{RMT},\hspace{6pt} |E_1-E_2|, \; |E_1-E_4|\ll 1.
\ee
Using this approximation for the density correlator, changing variables to $\delta E_{12}=E_1-E_2$, and $\delta E_{14}=E_1-E_4$, and $E= \frac{2E_1+E_2+E_4}{4}$, approximating $\frac{E_1+E_2}{2}\approx \frac{E_1+E_4}{2}\approx E$, $\rho(E_1) \approx \rho(E)$, and $|\mathcal{O}_{E_i,E_j}|^2\approx |\mathcal{O}_{E}|^2$, we find an approximation for the correlator at late times
\be
\boxed{G_{4,\beta/2}(T) \approx 2\; e^{-2 S_0}\int dE\; \frac{e^{-2 \beta E}}{\rho_0(E)} |\mathcal{O}_E|^4 Y(E,T)^2, \hspace{20pt} T\gg 1.} \label{FourPointLateTimePrediction}
\ee
For fixed $E$, $Y(E,T)^2$ has a quadratic ramp, or ``half pipe'' \cite{HalfPipeWiki} which ends at a plateau.

Higher point OTOCs work similarly \cite{Huang:2017fng,Cotler:2017jue}. For $2k$ point functions, we encounter strings of matrix elements of the form
\be
\langle E_{n_1} | \mathcal{O}|E_{n_2}\rangle\langle E_{n_2}|\mathcal{O}|E_{n_3}\rangle \dots \langle E_{n_{2k}}| \mathcal{O}|E_{n_1} \rangle.
\ee
Upon averaging, the phases of these matrix elements average to zero unless they are paired up appropriately. This only happens if we set some of the energies equal to each other. To see which energies must be set equal we can imagine arranging the string of matrix elements in a circle. Below we have pictured this arrangement for $k=4$. The dots along the circle represent matrix elements, in between dots we have denoted the energy in between neighboring operators. For the phases to cancel, we must pair up matrix elements. We denote these pairings by chords connecting the two matrix elements. For each set of pairings, we must set some energies equal to each other. The dominant contributions to the correlator will come from pairings that require the least number of energies set equal to each other, which turns out to leave $k+1$ energies independent. Each integral over independent energies roughly contributes a factor of $e^{S_0}$ to the correlator, so we expect that pairings that leave fewer than $k+1$ energies independent will give exponentially subleading contributions.
\begin{figure}[H]
\centering
\includegraphics[scale=0.5]{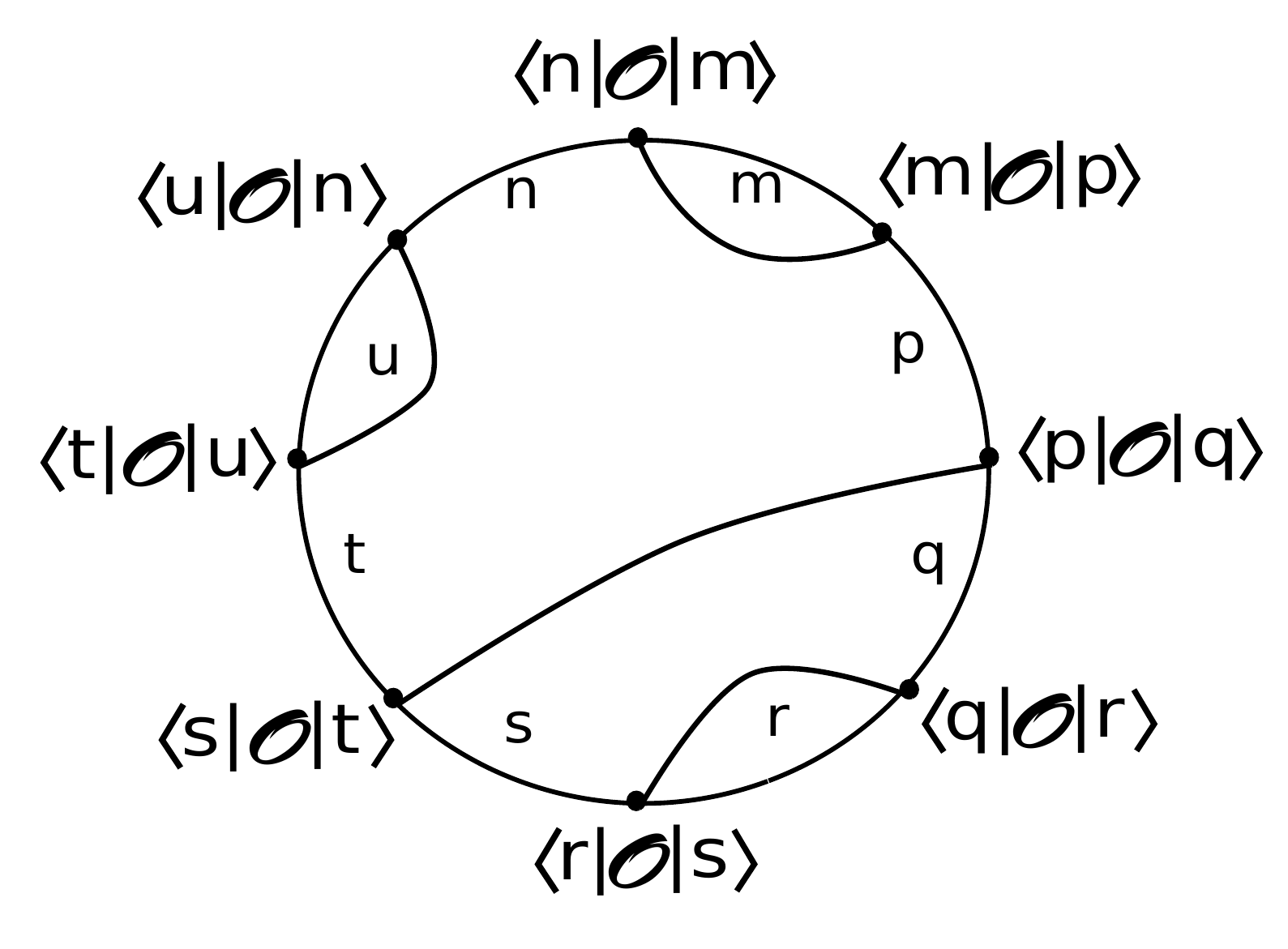} \includegraphics[scale=0.5]{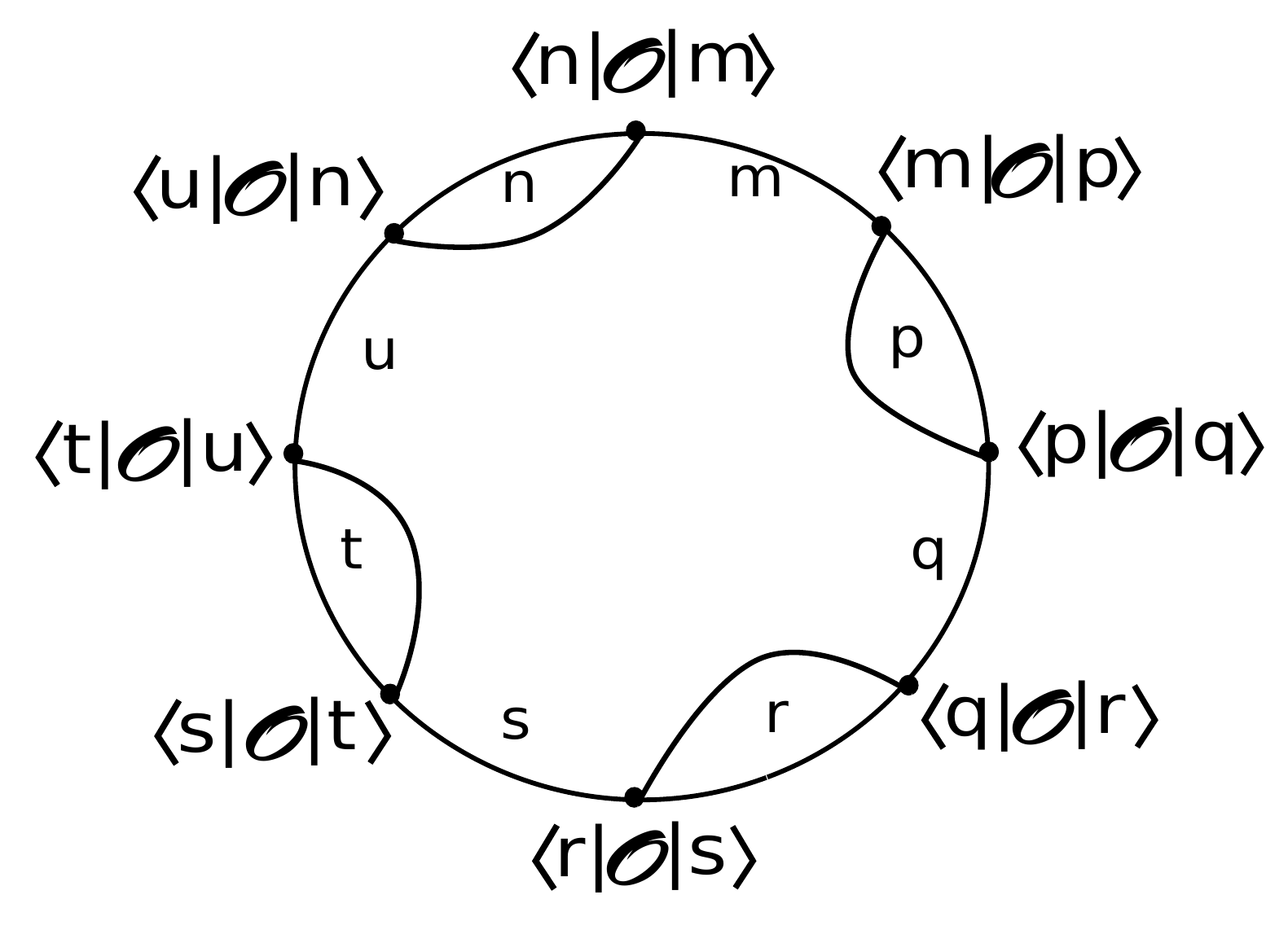}
\end{figure}
The dominant pairings come from \textit{planar} chord diagrams, such as those pictured above. The chords partition the interior of the circle into domains, each domain corresponds to an independent energy. We can see that the above diagrams both have five domains, and thus five independent energies to be integrated over. 

We can imagine writing the integral over the $k+1$ independent energies as an integral over $k$ energy differences $\delta E_i$, and one average energy $E$. At long times, the integral over energies will be dominated by small energy differences, so we can approximate all averages of energies as $E$. 

In the region where the $\delta E_i$ are small, we can approximate the correlation function of the $2k$ densities of states as the product of pair correlation functions, 
\be
\rho(E_{n_1}, \dots, E_{n_{2k}}) \rightarrow \prod_{i=1}^k \rho_{RMT}(E, \delta E_i)
\ee
Following the same procedure that we followed for the four-point function, we may find an approximate expression for the $2k$-point OTOC at late times
\be
\boxed{G_{2k,\beta/2}(T)\approx C_k \;e^{-k S_0} \int dE \;\frac{e^{-k\beta E} }{\rho_0(E)^{k-1}}\; |\mathcal{O}_E|^{2k}\;Y(E,T)^k , \hspace{20pt} T\gg 1.} \label{OTOClatetimeprediction}
\ee
Here $C_k$ is the number of planar chord diagrams with $k$ chords connecting $2k$ distinguishable points on a circle.

\section{Long times and topology change in JT gravity}\label{SectionJTGravity}
The main goal of this paper is to reproduce the formulas (\ref{GJTPrediction}) and (\ref{FourPointLateTimePrediction}) in a theory of JT gravity coupled to matter. The gravitational dynamics will be the key player at late times; we will find that these formulas apply even for a free scalar field. In order to understand the relevant gravitational behavior, in this section we will focus on some aspects of JT gravity without matter at late times. 

Much of this section will be a discussion of results from \cite{Yang:2018gdb,Saad:2018bqo,Saad:2019lba}, rephrased in a Hilbert space language. The natural setting for this discussion is the ``third-quantized'' JT gravity Hilbert space, which consists of states of any number of asymptotically AdS and closed JT gravity universes. In particular, we will focus on the late time dynamics of Hartle-Hawking states. Informed by the results of \cite{Yang:2018gdb,Saad:2018bqo,Saad:2019lba}, we use precise formulas for Hartle-Hawking states and topology-changing transition amplitudes to describe the relevant aspects of the dynamics.

A key player in this section will be the transition amplitude from the state of a single, asymptotically AdS ``parent'' JT gravity universe to a two-universe state of an asymptotically AdS universe and a closed ``baby'' universe.\footnote{These topology changing effects were studied in \cite{Saad:2018bqo,Saad:2019lba,Maldacena:2019cbz,Cotler:2019nbi} and discussed in \cite{Jafferis:2017tiu,Harlow:2018tqv}.} We may calculate this amplitude with a path integral over Euclidean geometries. We refer to these geometries as ``Euclidean wormholes'', and refer to the transition amplitude as the amplitude to ``emit'' or ``absorb'' a baby universe.

We begin the section with some background, discussing some aspects of Lorentzian JT gravity and giving a description of the spectral form factor as a transition amplitude in the third-quantized JT gravity Hilbert space. Next, we describe the results of \cite{Yang:2018gdb} on the late time semiclassical behavior of the Hartle-Hawking state. Finally, we calculate a topology-changing correction to the evolution of the Hartle-Hawking state and use it to describe the physics of the ramp \cite{Saad:2018bqo,Saad:2019lba} in the spectral form factor.

\begin{figure}[H]
\centering
\includegraphics[scale=1]{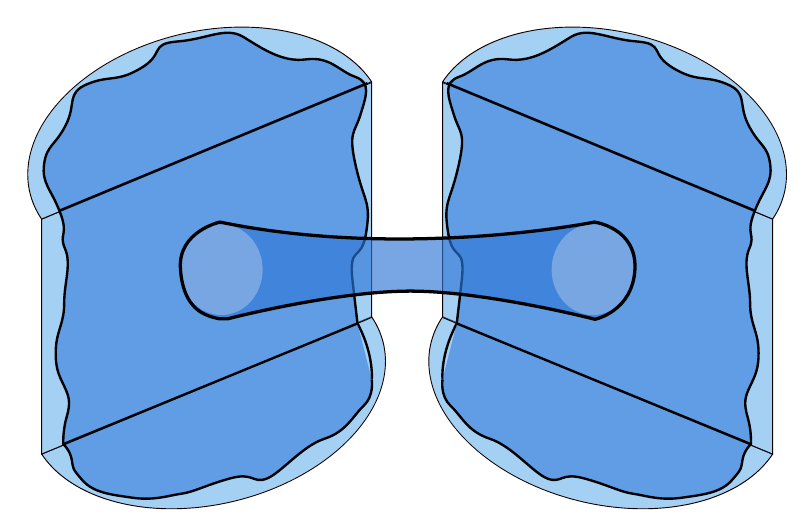}
\caption{\small Here we have pictured a contribution to the spectral form factor which corresponds to a process in which the two systems trade a baby universe.}
\label{fig:DoubleTrumpetTrading}
\end{figure}

\begin{figure}[H]
\centering
\includegraphics[scale=0.65]{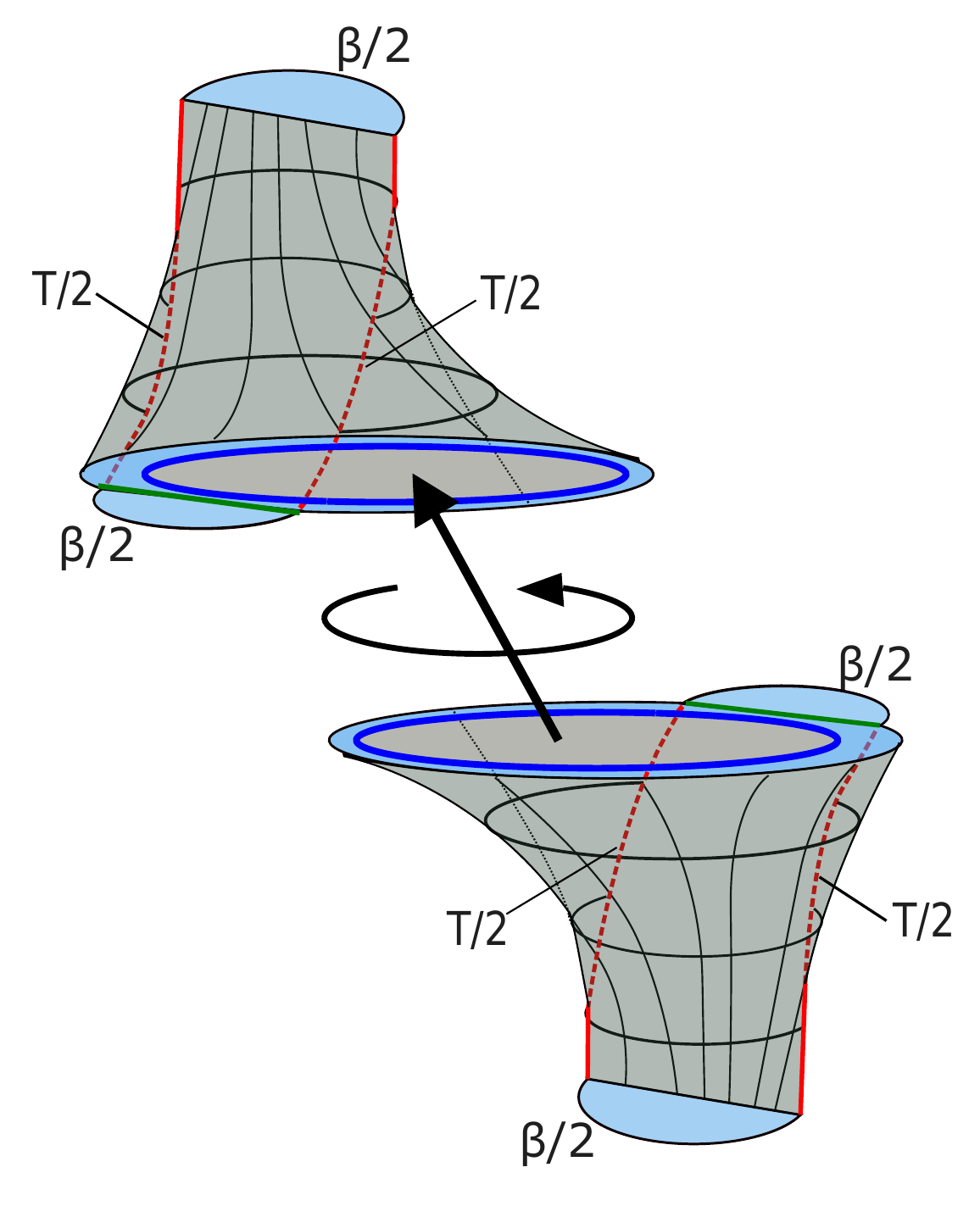}
\caption{\small Here we have pictured the ramp contribution to the spectral form factor, \newline $|Z(\beta+i T)|^2_{JT} = \langle HH_{\beta,L} HH_{\beta,R} | e^{- i \frac{T}{2} H_{Bulk, LR}} | HH_{\beta, L} HH_{\beta, R} \rangle$. The ramp comes from a process in which two Hartle-Hawking states evolve, one forwards in time, the other backwards in time, for a long time $T/2$. After this time, the two Hartle-Hawking states, each with very long ERBs, may trade a large baby universe and transition to the $T=0$ pair of Hartle-Hawking states. The amplitude for the large ERBs to trade a large baby universe at a fixed twist and end up as small ERBs does not decay in time. The $\sim T$ contributions from the different twists lead to a linear growth in $T$.}
\label{fig:RampTradeBig}
\end{figure}

The punchline of this section is as follows: At late times $T$, the Hartle-Hawking state describes a geometry with a very long Einstein-Rosen bridge (ERB), with length of order $T$. Euclidean wormholes allow this state to transition to a state with a short ERB and a large baby universe, with size of order $T$. The amplitude for this process, while exponentially small in the entropy, does not decay as $T$ increases. The ramp in the spectral form factor comes from two copies of the Hartle-Hawking state trading this large baby universe before returning to the $T=0$ Hartle-Hawking state; the $T$ ways in which this baby universe can be ``rotated'' as it is traded leads to a linear factor of $T$. We have pictured this process in Figure \ref{fig:DoubleTrumpetTrading} and Figure \ref{fig:RampTradeBig}.

\subsection{Some aspects of Lorentzian JT gravity}\label{SubsectionLorentzianJT}

In the path integral approach to JT gravity, quantities are defined by the integral over all geometries with a given set of boundary conditions. We are typically interested in quantities that involve imposing asymptotically AdS boundary conditions, but we may also include spacelike boundaries with zero extrinsic curvature and given lengths, or mixed boundaries that have portions of asymptotically AdS boundary conditions and spacelike portions with zero extrinsic curvature.

Partition functions are calculated by integrals over geometries with asymptotically Euclidean AdS boundary conditions. Roughly, this boundary condition is described by fixing the boundary metric as 
\be
ds^2\big|_{\partial M} = \frac{1}{\epsilon^2} d\tau^2,\label{EAdS}
\ee
with $\epsilon$ a holographic renormalization parameter, which should eventually be taken to zero, and $\tau$ periodic as $\tau\sim \tau+\beta$ \cite{Maldacena:2016upp}. One describes these boundaries as Euclidean AdS boundaries with renormalized length $\beta$. 

Spacelike boundaries with zero extrinsic curvature, which we will refer to as ``geodesic boundaries'' may also be described in terms of a renormalized length. With our holographic renormalization parameter $\epsilon$, the renormalized length $\ell$ of a geodesic boundary may be describe in terms of its bare length $\ell_b$ as \cite{Harlow:2018tqv}
\be
\ell \equiv \ell_b - 2 \log\bigg( \frac{2}{\epsilon}\bigg).
\ee

The spectral form factor is an analytically continued partition function, so in JT gravity one may calculate it by first calculating a partition function via an integral over Euclidean geometries with AdS boundaries, then continuing the answer $\beta\rightarrow \beta\pm i T$. However, one may also calculate the spectral form factor more directly by integrating over complex geometries with boundary conditions similar to (\ref{EAdS}), but with $\tau\sim \tau+\beta\pm i T$. One must choose a contour in the $\tau$ plane to fix these boundary conditions, as well as a contour of integration over metrics that satisfy our boundary conditions\footnote{In \cite{Saad:2018bqo} only the saddle point geometry was considered, so this was not important.}. A simple choice of time contour, used in \cite{Saad:2018bqo}, is to choose $\tau= (\beta\pm i T) x$, with $x\sim x+1$ real. We then integrate over geometries with boundary conditions
\be
ds^2\big|_{\partial M} = \frac{(\beta\pm i T)^2}{\epsilon^2} dx^2,\label{CAdS}
\ee
In holography, the partition function $Z(\beta)$ is interpreted as $\Tr\big[e^{-\beta H}\big]$ for a boundary Hamiltonian $H$. This choice of contour for $Z(\beta\pm i T)$ corresponds to slicing
\be
e^{-(\beta\pm i T) H} = \lim_{N\rightarrow \infty} e^{- \frac{(\beta\pm i T)}{N} H}
\ee
Another choice, which will be more useful to us, will be to choose the boundary metric to alternate in signature, but remain real. The time contour consists of a Euclidean segment of length $\beta/2$, followed by a Lorentzian segment of time $\pm T/2$, followed by another Euclidean segment of length $\beta/2$ and finally by a Lorentzian segment of time $\pm T/2$. This corresponds to the decomposition
\be
e^{-(\beta\pm i T) H}  = e^{-\frac{\beta}{2} H} e^{\mp i\frac{T}{2} H} e^{-\frac{\beta}{2} H} e^{\mp i\frac{T}{2} H} \label{TimeDecomp}
\ee
In a quantum system, different choices of time contours, corresponding to different decompositions and slicings of $e^{- (\beta\pm i T) H}$, give the same answer. However, in JT gravity, which we define with a path integral, this property is subtle.

First, the path integral prescription for calculating partition functions in gravity does not obviously give an answer that corresponds to something like $\Tr \big[e^{-(\beta\pm i T) H}\big]$. Of course in AdS/CFT this property is true, but if we do not assume the existence of a holographic description of our system, we should not assume that different boundary time contours correspond to different slicings of $e^{-(\beta \pm i T) H}$ for some $H$ and are thus equivalent.

Second, for a generic choice of time contour, one must be careful in defining the integration contour over metrics with the corresponding boundary conditions. For example, with the contour (\ref{CAdS}), one must choose an integration contour over complex metrics. 

In pure JT gravity, the first issue is not a problem, as \cite{Saad:2019lba} shows that partition functions in JT gravity are disorder averages of ordinary quantum mechanical partition functions. However, it will be useful to see directly in JT gravity how different boundary time contours produce the same answer. To see how this works we will translate a Hilbert space description of the spectral form factor into an expected formula in gravity.

To deal with the second issue, we will restrict our attention to time contours for which the corresponding contour of integration is more clear. The integrals with boundary time contours corresponding to the decomposition (\ref{TimeDecomp}) may be performed over piecewise Euclidean and Lorentzian metrics. We may think of the geometries being integrated over as being glued together along slices with zero extrinsic curvature, which are spacelike geodesics.

Let's start with a single analytically continued partition function $Z(\beta + i T)$. For an ordinary quantum system, we may express this as the \textit{return amplitude} \cite{Papadodimas:2015xma,Numasawa:2019gnl} of the (unnormalized) thermofield double state.
\be
\Tr \big[e^{-(\beta+i T) H} \big]= \langle TFD_\beta| e^{- i \frac{T}{2} (H_L+ H_R)} | TFD_\beta \rangle.
\ee
The spectral form factor is then the \textit{return probability} of the thermofield double state.

We expect that the thermofield double $|TFD_\beta\rangle$ is dual to the Hartle-Hawking state $|HH_\beta\rangle$, in the sense that we expect that by taking formulas for a (possibly averaged) quantum system and replacing
\be
|TFD_\beta\rangle\rightarrow |HH_\beta\rangle,
\ee
as well as making other appropriate replacements, we find matching behavior between the quantum mechanics formula and the gravity formulas. We may define $|HH_\beta\rangle$ by its matrix elements in the length basis \cite{Yang:2018gdb, Harlow:2018tqv}, with $\langle \ell |HH_\beta\rangle \equiv \psi_{D,\beta/2}(\ell)$ equal to the integral over all Euclidean geometries with the topology of a disk and boundary consisting of an asympotically AdS portion of renormalized length $\beta/2$ and a geodesic portion of renormalized length $\ell$.
\begin{figure}[H]
\centering
\includegraphics[scale=0.6]{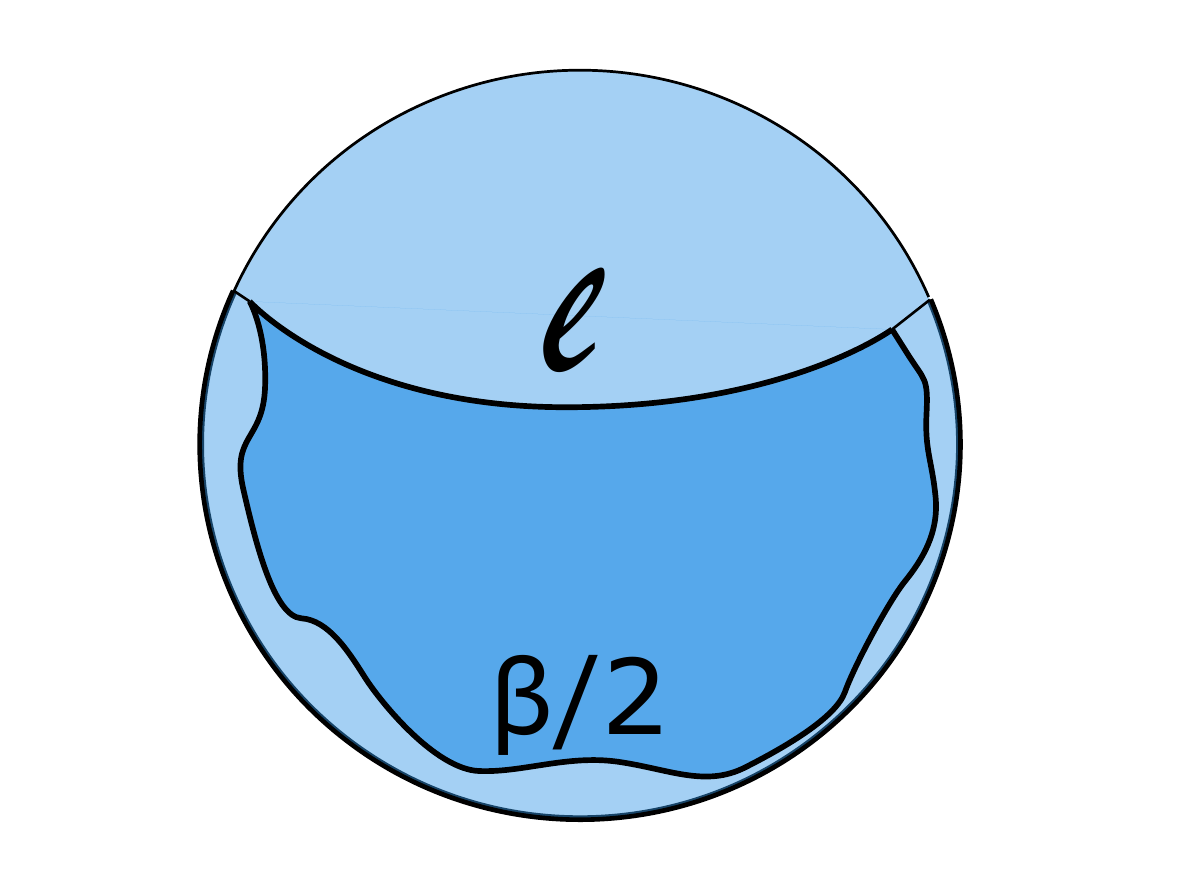}
\caption{\small Here we have pictured an example geometry that contributes to the Hartle-Hawking wavefunction $\psi_{D,\beta/2}(\ell)$. The light blue circle represents the hyperbolic disk, while the dark blue represents the piece of the disk that we take as our Euclidean spacetime. The wiggly asymptotic boundary has renormalized length $\beta/2$, while the geodesic piece of the boundary has renormalized length $\ell$.}
\end{figure}
Notice that with our definition the Hartle-Hawking state is not normalized, with a wavefunction exponential in $S_0$, $\psi_{D,\beta/2}(\ell)\sim e^{S_0}$. With the appropriate inner product, which we describe later, the norm is equal to the disk contribution to the partition function.

We also expect that evolution with $e^{- i \frac{T}{2} (H_L+ H_R)}$ is dual to evolution with an appropriate bulk Hamiltonian $H_{Bulk}$. We define $e^{-i \frac{T}{2} H_{Bulk}}$ by the path integral formula for its matrix elements in the length basis; $\langle \ell | e^{-i \frac{T}{2} H_{Bulk}}|\ell'\rangle$ is equal to the JT path integral over appropriate geometries\footnote{For contributions with nontrivial topology, the question of which geometries we should integrate over is subtle, and will address this question in more detail later in this section. However, we will find that the natural answer gives matrix elements that are equal to the analytic continuation of the matrix elements of the Euclidean evolution $e^{-\tau H_{Bulk}}$, which are naturally calculated via integrals over all real geometries with the appropriate boundary conditions} with a single boundary consisting of two spatial geodesic slices of renormalized length $\ell$ and $\ell'$ connected by two asympotically AdS boundaries of time $T/2$.
\begin{figure}[H]
\centering
\includegraphics[scale=0.9]{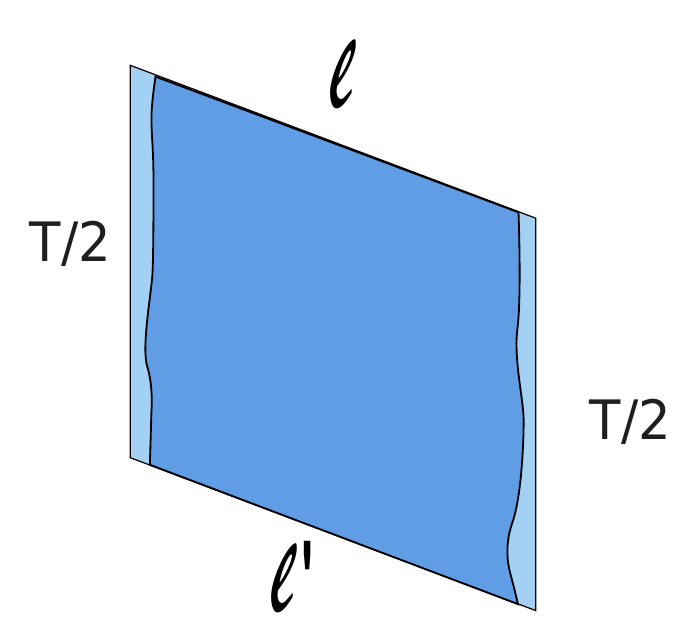}
\caption{\small Matrix elements of the time evolution operator $\langle \ell | e^{-i \frac{T}{2} H_{Bulk}}|\ell'\rangle$ are calculated by the path integral over all geometries with the appropriate boundary conditions.}
\label{fig:Propagatorfigure}
\end{figure}
Altogether, we expect that the analytically continued partition function may be described in gravity as
\begin{align}\label{ZAmplitude}
Z(\beta+ i T)_{JT} &= \langle HH_\beta| e^{-i \frac{T}{2} H_{Bulk}} | HH_\beta\rangle
\cr
&=\int \mu(\ell) d\ell \int \mu(\ell') d\ell' \; \langle HH_\beta| \ell\rangle \langle \ell|e^{-i \frac{T}{2} H_{Bulk}} |\ell'\rangle\langle \ell'| HH_\beta\rangle.
\end{align}
We define the left hand side as the analytic continuation of the partition function $Z(\beta)$, and we define the right hand side via our definitions of the Hartle-Hawking state and time evolution operator in the length basis, and an inner product of length basis states described by the measure $\mu(\ell)$ which we will discuss in the next section.
\begin{figure}[H]
\centering
\includegraphics[scale=0.6]{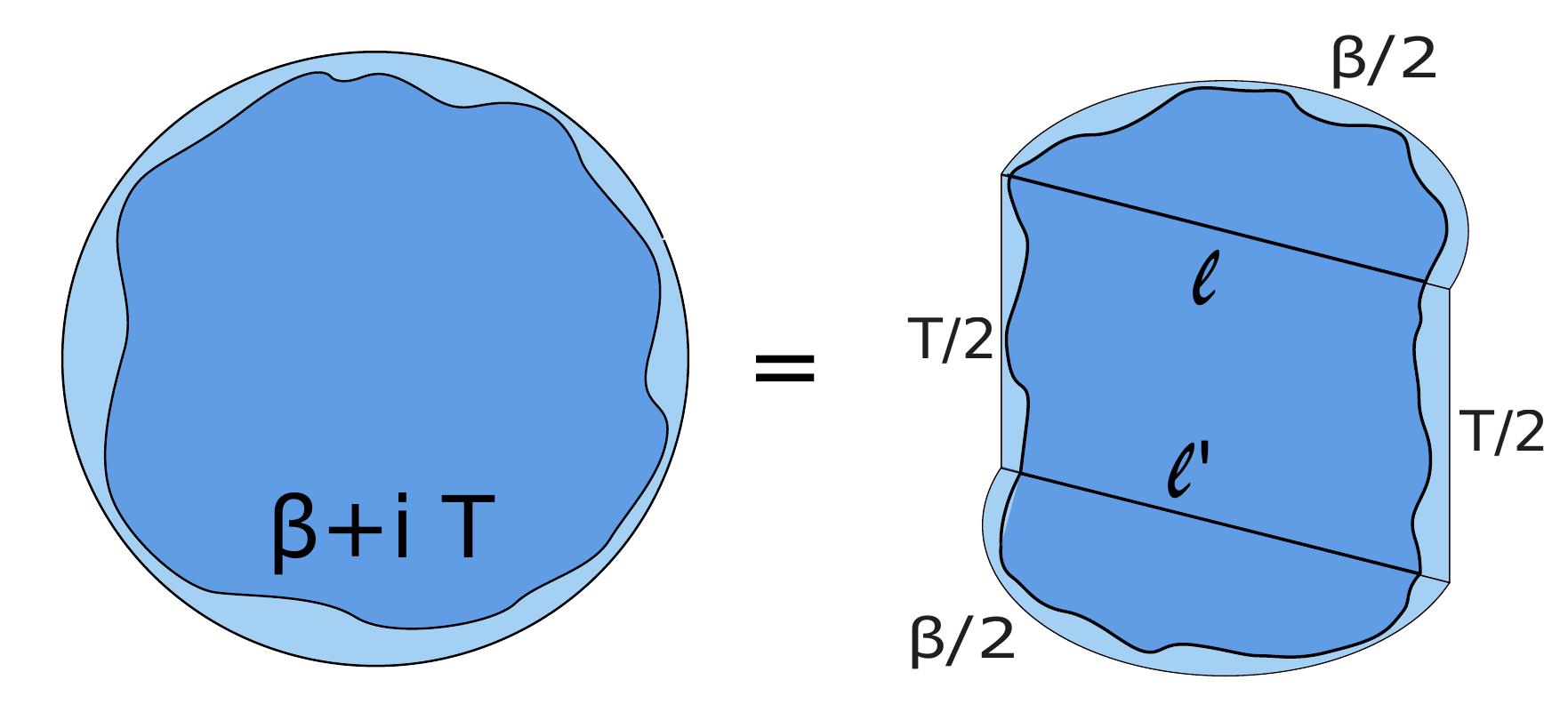}
\caption{\small Here we have represented equation (\ref{ZAmplitude}) to leading order in $e^{-S_0}$. On the left, we show a piece of complexified hyperbolic space with the topology of a disk and a boundary of renormalized length $\beta+i T$. On the right we have shown another piece of complexified hyperbolic space with the topology of a disk and a boundary of length $\beta+i T$. However, this geometry is piecewise Euclidean and Lorentzian, with the two half-disks representing the Euclidean geometries which prepare the Hartle-Hawking state, and the vertical, rectangular piece representing the Lorentzian evolution between these two states. These pieces are joined along slices of zero extrinsic curvature of renormalized lengths $\ell$ and $\ell'$.}
\end{figure}
We may then interpret the spectral form factor $|Z(\beta+i T)|^2_{JT}$ as the return probability of the Hartle-Hawking state, or the return amplitude for two copies of the Hartle-Hawking state, with one of the copies evolved backwards in time.
\be
\boxed{|Z(\beta+i T)|^2_{JT} = \langle HH_{\beta,L} HH_{\beta,R} | e^{- i \frac{T}{2} H_{Bulk, LR}} | HH_{\beta, L} HH_{\beta, R} \rangle}
\ee
Here $H_{bulk, LR}$ evolves the $R$ system forwards in time and the $L$ system backwards. We do not write $H_{bulk, LR}= H_{Bulk, R}- H_{Bulk, L}$, as Euclidean wormholes couple the $L$ and $R$ systems. For example, during the time evolution of the full system, the $R$ system may emit a baby universe which is absorbed by the $L$ system, as pictured below.
\begin{figure}[H]\label{FigureDoubleTrumpetRamp1}
\centering
\includegraphics[scale=1]{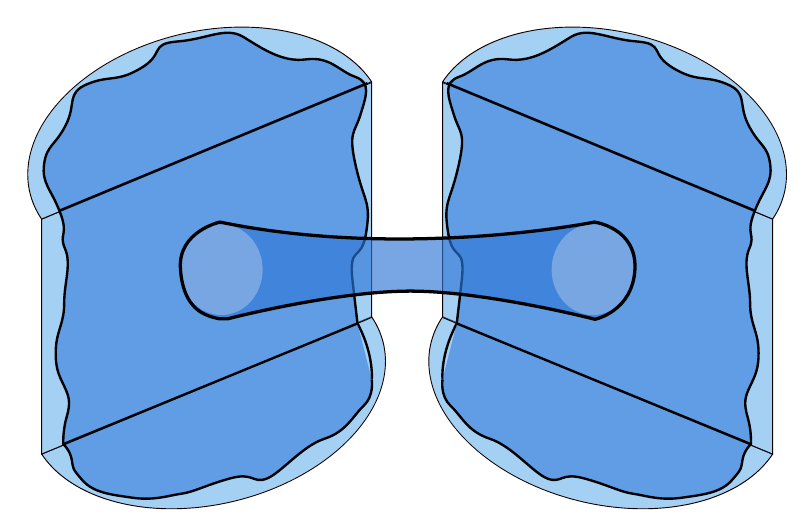}
\caption{\small Here we have pictured a contribution to the spectral form factor, viewed as the return probability of a pair of Hartle-Hawking states. Contributions such as these couple the two systems, so that  $H_{bulk, LR}\neq H_{Bulk, R}- H_{Bulk, L}$.}
\end{figure}
\subsection{The spectral form factor in JT gravity}\label{SubsectionSFFJT}
In Figure \ref{fig:SFFFigure}, we have pictured the main features in the spectral form factor: the slope, ramp and plateau.

To leading order, the slope comes from two disconnected disk geometries, each corresponding to the leading order contribution to $Z(\beta\pm i T)_{JT}$, so we may understand the decay by looking at just one copy of the partition function. 
\be
|Z(\beta+i T)|^2_{JT} \approx  Z(\beta+i T)_{JT} Z(\beta-i T)_{JT},\hspace{20pt} T\ll e^{S_0/2}.
\ee
Up to nonperturbative corrections in $e^{-S_0}$, but including all contributions from more complicated topologies, $Z(\beta \pm i T)$ decays forever.\footnote{Higher genus contributions change the sharp edge of the spectrum at $E=0$, which can affect the late-time behavior, but we can easily remove these effects by working in the microcanonical ensemble.} We may give a useful physical picture for this decay by relating this quantity to the return amplitude of the Hartle-Hawking state.

Let's think about the Hartle-Hawking wavefunction in the length basis. In \cite{Yang:2018gdb}, and exact formula for this wavefunction was given,
\be
\psi_{D,\beta/2}(\ell)\equiv\langle \ell | HH_\beta \rangle= \int_0^\infty dE \rho_0(E) e^{-\frac{\beta}{2} E} \big(4 e^{-\ell/2} K_{i \sqrt{8 E}} ( 4 e^{-\ell/2})\big).
\ee
The ``D'' denotes the disk topology of the geometries that produce this wavefunction. $\rho_0(E)$ is the leading order density of states of JT gravity
\be
\rho_0(E)= e^{S_0} \frac{\sinh(2\pi \sqrt{2E})}{2\pi^2}.
\ee
This wavefunction has a peak at $\ell \sim \log \beta$ with a width of order $\sqrt{\beta}$ \cite{Yang:2018gdb, Harlow:2018tqv}. This expression for the wavefunction is reminiscent of the thermofield double state, with energy eigenstates $|E_L E_R\rangle$ replaced by a bulk ``energy eigenstate'' $|E\rangle$. Defining 
\be
\psi_E(\ell)\equiv\langle \ell |E \rangle \equiv 4 e^{-\ell/2} K_{i \sqrt{8E}} ( 4 e^{-\ell/2}), \label{EnergyWavefunction}
\ee
we may write
\be
\psi_{D,\beta/2}(\ell)= \int_0^\infty  dE \rho_0(E) e^{-\frac{\beta}{2} E} \psi_E(\ell) \label{HHWavefunctionFormula}
\ee
The overlap of different Hartle-Hawking wavefunctions is defined as 
\be
\langle HH_\beta| HH_{\beta'}\rangle \equiv\\e^{-S_0} \int_{-\infty}^\infty e^\ell d\ell \; \psi_{D,\beta/2}^*(\ell) \psi_{D,\beta'/2}(\ell),
\ee
which may be justified by using the boundary particle formalism from \cite{Kitaev:2018wpr,Yang:2018gdb}. Our normalizations differ from those of \cite{Yang:2018gdb}, as we are choosing to include the weighting $e^{S_0}$ from the topological term in the JT action in our definition of the Hartle-Hawking wavefunction. The factor of $e^{-S_0}$ in the inner product accounts for this normalization. One may see that this inner product reproduces the expectation that the norm of the Hartle-Hawking state is the partition function. The integral
\be
e^{-S_0}\int_{-\infty}^\infty e^\ell d\ell\; \psi_E^*(\ell) \psi_{E'} (\ell) = \frac{\delta(E-E')}{ \rho_0(E)} \label{BesselOrthogonal}
\ee
is key for showing this, and will be very useful in this paper.

Now let's find the wavefunction of the Hartle-Hawking state after evolving it for a time $T/2$. This is defined as the integral over geometries with a disk topology with a boundary that consists of an asymptotic portion of renormalized length $\beta+ i T$ and a geodesic portion of renormalized length $\ell$. To leading order in $e^{-S_0}$ we may simply obtain this wavefunction by continuing $\beta\rightarrow \beta+i T$ in our formulas for the Hartle-Hawking wavefunction. 
\be
\psi_{D,\beta/2+ i T}(\ell) = \langle \ell| e^{-i \frac{T}{2} H_{Bulk}} |HH_\beta\rangle (1+\mathcal{O}(e^{-S_0})).
\ee
However, it is useful to check our expectation that we may also find this wavefunction via an integral over piecewise Euclidean and Lorentzian geometries,

Precisely, our expectation is that to leading order in $e^{-S_0}$ the Hartle-Hawking wavefunction evolved for time $T$ may be calculated via an integral of the form
\be\label{HHWavefunctionTimeEvolved}
\psi_{D,\beta/2+i T}(\ell)= e^{-S_0}\int e^{\ell'} d\ell' \; P_{\chi=1}(T/2, \ell , \ell') \psi_{D,\beta}(\ell'),
\ee
where $P_{\chi=1}(T/2, \ell ,\ell')$ is the leading order propagator between states with renormalized length $\ell$ and length $\ell'$, defined via
\be
\langle \ell| e^{-i \frac{T}{2} H_{Bulk} } |\ell'\rangle \sim \sum_{n=0}^\infty P_{\chi=1-2n}(T/2,\ell, \ell') e^{-2 n S_0}.
\ee
where the tilde indicates that the right hand side is a divergent series asymptotic to the left hand side. $n$ labels the topology of the contributions to the propagator, with $\chi=1-2n$. An example geometry that contributes to $P_{\chi=1}(T/2,\ell,\ell')$ is pictured in Figure \ref{fig:Propagatorfigure}. $P_{\chi=1}(T/2,\ell,\ell')$ is calculated by the JT path integral over Lorentzian geometries with the topology of a disk and a boundary that consists of two geodesic segments of renormalized lengths $\ell$ and $\ell'$ connected by two asymptotically AdS segments of times $T/2$.

The right hand side of this formula (\ref{HHWavefunctionTimeEvolved}) corresponds to the picture
\begin{figure}[H]
\centering
\includegraphics[scale=1.3]{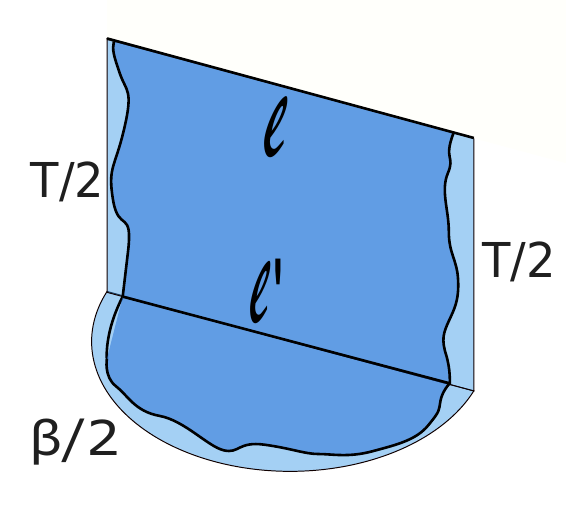}
\end{figure}
Using the boundary particle formalism of \cite{Kitaev:2018wpr,Yang:2018gdb}, one may directly calculate the propagator and find
\be
P_{\chi=1}(T/2,\ell,\ell') = \int dE \rho_0(E) e^{-i \frac{T}{2}E } \psi_E(\ell) \psi_E(\ell'). 
\ee
Along with the orthogonality formula (\ref{BesselOrthogonal}), we confirm our formula (\ref{HHWavefunctionTimeEvolved}).

With the time-evolved Hartle-Hawking wavefunction in hand, we may discuss its long-time behavior. At short times, the wavefunction is peaked around a short length. For $T\gg \beta$, the wavefunction remains peaked around a certain value but we find that the peak of this wavefunction moves towards larger $\ell$. In \cite{Yang:2018gdb}, the expectation value of $\ell$ was found to increase linearly in time,
\be\label{WormholeSize}
\frac{e^{-S_0} \int e^\ell d\ell \; |\psi_{D,\beta/2+i T}(\ell)|^2 \;\ell }{e^{-S_0} \int e^\ell d\ell \; |\psi_{D,\beta}(\ell)|^2}\approx \frac{2\pi T}{\beta}, \hspace{20pt} T\gg \beta.
\ee
$\ell$ is essentially the length of the Einstein-Rosen Bridge (ERB) in the geometry described by the wavefunction. Classically, the length of the ERB increases according to this formula.\cite{Susskind:2014rva,Brown:2018bms} The results of \cite{Yang:2018gdb} show that semiclassical quantum effects do not stop this growth. The main contribution of quantum effects is to add large tails to the wavefunction around this classical peak.

Now we may see the physical origin of the slope: as time increases, the amplitude for the time evolved Hartle-Hawking state to have a small ERB decreases, with most of the support moving to large ERBs. The initial state is mostly supported at small ERB lengths, and so the return amplitude, and thus the spectral form factor, then decreases.

We also see that to all orders in $e^{-S_0}$, this decaying behavior does not change. Subleading contributions in $e^{-S_0}$ correspond to geometries with handles. The corrections to the density of states calculated in \cite{Saad:2019lba} are smooth away from the edge at $E=0$. Working in the microcanonical ensemble to avoid effects from the sharp edge, these smooth deformations in the density lead to decaying behavior in microcanonical versions of $Z(\beta+i T)$.

We may interpret these corrections as follows. First consider the correction to the time evolution operator from one handle with $\chi=-1$, which we have defined as $P_{\chi=-1}(T/2,\ell, \ell') e^{-2 S_0}$. For now we may define this via an integral over Euclidean geometries with $\chi=-1$ that gives the Euclidean transition amplitude $P_{\chi=-1}(- i \tau ,\ell, \ell')$, which we then continue $-i\tau\rightarrow T/2$, but the results of the next section will give us a more physical and direct way to calculate this amplitude.

We can picture $P_{\chi=-1}(T/2,\ell,\ell')$ as accounting for the effects of emitting and reabsorbing a baby universe while evolving in time.
\begin{figure}[H]
\centering
\includegraphics[scale=0.8]{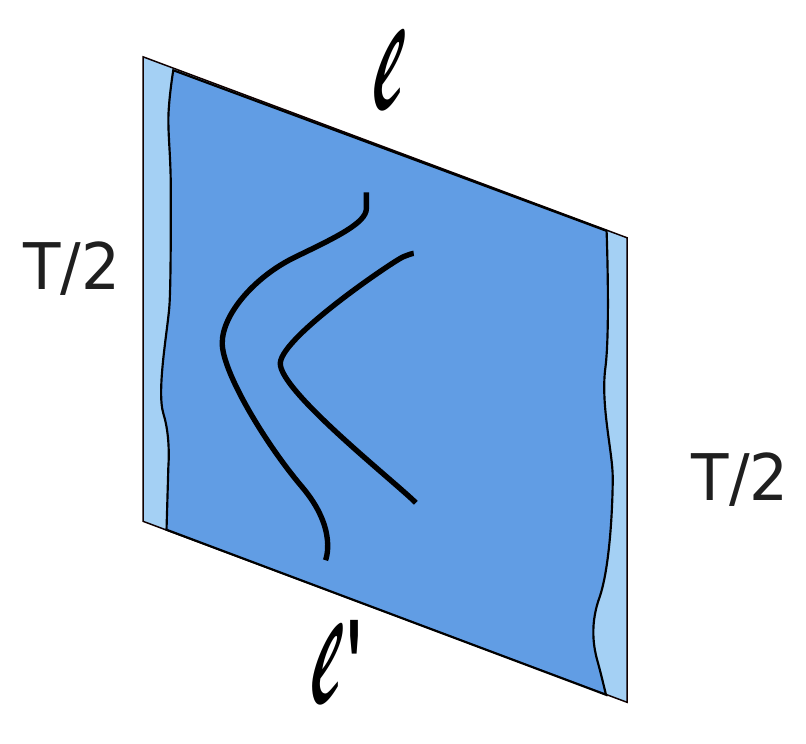}
\end{figure}
The fact that this contribution decays tells us that emitting and reabsorbing a baby universe does not stop the ERB from growing. The decay of contributions to $Z(\beta+i T)_{JT}$ from higher topologies tell us that emitting and reabsorbing any number of baby universes also does not stop this growth.

However, the non-decaying ramp and plateau behavior of the spectral form factor suggests that there are effects that stop the growth of the ERB. As these non-decaying behaviors are present in the spectral form factor, but not the return amplitude, they must rely in some way on the interplay between two copies of the Hartle-Hawking state. These effects should have a small amplitude, but dominate the short $\ell$ behavior of the pair of Hartle-Hawking wavefunctions at long times. 

As described in \cite{Saad:2018bqo,Saad:2019lba}, the ramp comes from geometries with the topology of a cylinder. By choosing the appropriate boundary time contours, we may view these geometries as describing the evolution of two copies of the Hartle-Hawking state, one forwards in time, the other backwards in time, during which the two systems exchange a baby universe.
The $R$ system, which ``absorbs'' the baby universe emitted by the $L$ system, is evolving backwards in time. This means we may also view it as emitting a baby universe. Together, in this process both systems emit the same baby universe and end up back in the original pair of Hartle-Hawking states. While exponentially small, the amplitude for this to happen is non-decaying. This tells us that while emitting and reabsorbing a baby universe cannot stop the growth of the ERB, just emitting a baby universe can.

In the following section, we will describe the physics of this process in detail and find a simple picture of the linear growth of the ramp in the spectral form factor.

\subsection{The ramp from trading a baby universe}\label{SubsectionRampSFF}
The ramp contribution to the spectral form factor comes the continuation of the ``double-trumpet'' \cite{Saad:2019lba}, which is the integral over all Euclidean geometries with the topology of a cylinder and two asymptotically Euclidean AdS boundaries. As the name suggests, the double-trumpet may be decomposed into two ``trumpets''; for a given geometry, we cut along the minimal length geodesic homotopic to either of the two asymptotic boundaries. Labeling the length of this geodesic $b$, the double-trumpet can be written in terms of the trumpet partition function $Z_{Tr}(\beta,b)$, which is an integral over all cylindrical geometries with an asymptotically Euclidean AdS boundary of renormalized length $\beta$ and a geodesic boundary of length $b$.
\be
\text{Double-trumpet}(\beta,\beta') = \int_0^\infty b db\; Z_{Tr}(\beta,b) Z_{Tr}(\beta',b). \label{DoubleTrumpet}
\ee
\begin{figure}[H]
\centering
\includegraphics[scale=0.8]{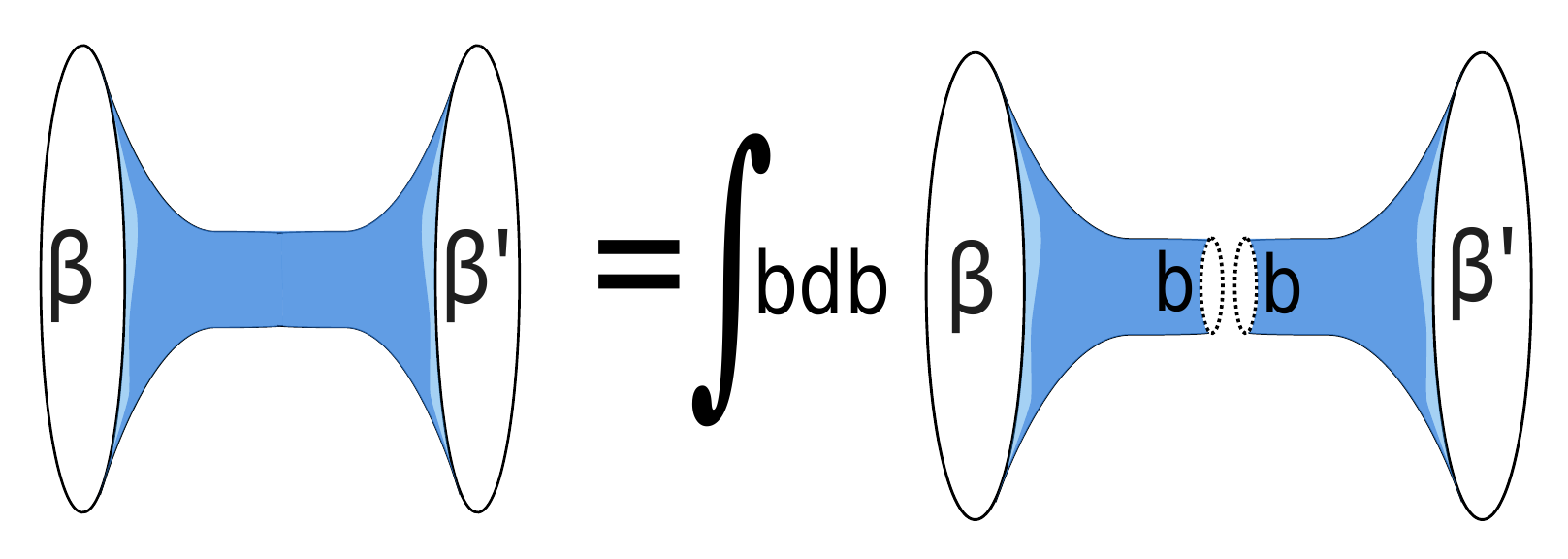}
\end{figure}
The simple physical explanation of the measure is that the factor of $b$ accounts for the different angles at which the two trumpets may be glued. Through a careful calculation \cite{Saad:2019lba} one may find that the correct measure for gluing circular geodesic boundaries in JT gravity is the Weil-Petersson measure \cite{wolpert1985weil}
\be
d(WP)=db\wedge d\tau.
\ee
$b$ and $\tau$ are Fenchel-Nielsen coordinates. $\tau$ is the twist \textit{length} describing the proper distance the two circular boundaries are rotated relative to each other before they are glued. With $\theta$ a relative angle describing the gluing, $\tau=b \theta/2\pi $. $b$ and $\tau$ should be integrated over a region that does not overcount geometries.

For the double-trumpet geometries, we integrate over $\tau$ from $0$ to $b$ and $b$ from $0$ to $\infty$. The integrand, the product of trumpet partition functions, is independent of the twist length $\tau$, so the integral simplifies to (\ref{DoubleTrumpet}).

We can give a rough interpretation of the analytically continued trumpet partition function $Z_{Tr}(\beta+i T)$ as the leading order amplitude for the Hartle-Hawking state to evolve into a the state $|HH_\beta\rangle \otimes |b\rangle\equiv |HH_\beta, b\rangle$, where $|b\rangle$ is the state of JT gravity on a circle of size $b$. By writing the final Hartle-Hawking state in the length basis, we may express this as
\begin{align}
Z_{Tr}(\beta+i T,b) &= e^{-S_0}\int e^\ell d\ell \;\psi^*_{D,\beta/2}(\ell) \psi_{Tr,\beta/2+i T}(\ell,b) 
\cr
&=e^{-S_0}\int e^\ell d\ell\; \langle HH_\beta |\ell \rangle \langle \ell,b | e^{-i\frac{T}{2} H_{Bulk}} |HH_\beta \rangle(1+\mathcal{O}(e^{-S_0}) ).
\end{align}
The corrections in the second line come from removing the contributions of extra Euclidean wormholes.

For $T=0$, the ``trumpet wavefunction'' $\psi_{Tr,\beta/2}(\ell,b)$ is given by the JT path integral over all Euclidean geometries with the topology of a cylinder with a geodesic boundary of length $b$ and a boundary consisting of an asymptotically AdS boundary segment of renormalized length $\beta/2$ and a geodesic segment of renormalized length $\ell$. This is justified using the boundary particle formalism of \cite{Kitaev:2018wpr,Yang:2018gdb} in Appendix \ref{AppendixTrumpetWavefunction}. Explicitly,
\be
\psi_{Tr,\beta/2}(\ell,b) = \int_0^\infty dE \frac{\cos(b \sqrt{2 E})}{\pi \sqrt{2 E}} e^{-\frac{\beta}{2}E} \psi_E(\ell). \label{TrumpetWavefunctionFormula}
\ee
Upon analytic continuation $\beta/2\rightarrow \beta/2+i T$, we may roughly think of this as the amplitude for the Hartle-Hawking state to evolve into the state $|\ell, b\rangle$. 

As pictured below, we may think of this trumpet wavefunction as being calculated by an initial Euclidean evolution that prepares the Hartle-Hawking state, which is then evolved for time $T/2$. During this evolution, a baby universe of size $b$ is emitted.
\begin{figure}[H]
\centering
\includegraphics[scale=1.3]{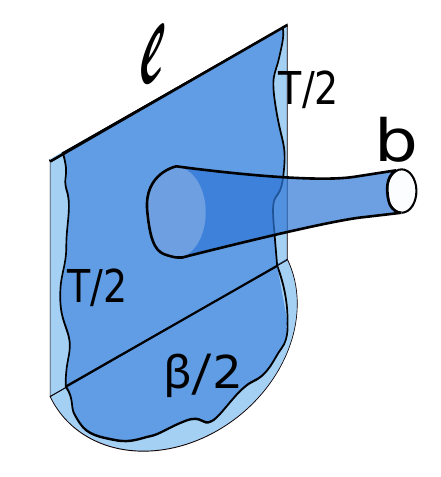}
\caption{\small Here we have pictured a contribution to the trumpet wavefunction $\psi_{Tr,\beta/2+i T}(\ell,b)$, the amplitude for the Hartle-Hawking state to evolve into the state $|\ell,b\rangle$ after a time $T/2$. The Euclidean piece of the geometry at the bottom prepares the Hartle-Hawking state. We then evolve this state for time $T/2$ and include geometries with a circular geodesic boundary of length $b$ in addition to a geodesic boundary of renormalized length $\ell$.}
\end{figure}
This prompts us to think about the propagator $P_{Tr}(T/2, b, \ell, \ell') \sim \langle \ell, b| e^{-i\frac{T}{2} H_{Bulk}} |\ell'\rangle$. We may simply extract this propagator from the trumpet wavefunction by using the orthogonality formula (\ref{BesselOrthogonal})
\be\label{TrumpetPropagator}
P_{Tr}(T/2, b, \ell, \ell') = \int_0^\infty dE \;\frac{\cos(b \sqrt{2E})}{\pi\sqrt{2E}} e^{-i T E} \psi_E(\ell) \psi_E(\ell'). 
\ee
We may also try to calculate this propagator directly in JT gravity. However, we cannot obtain this quantity via an integral over nonsingular Lorentzian metrics. Instead, it will be useful to decompose this propagator into pieces that we may calculate with integrals over purely Lorentzian and Euclidean geometries. An example is the decomposition into an ordinary propagator $P(T/2, \ell, \ell')$ and the Euclidean amplitude $\langle \ell, b| \ell'\rangle$
\be
P_{Tr}(T/2, b, \ell, \ell') = e^{-S_0}\int e^{\ell''} d\ell'' \;  \langle \ell, b| \ell''\rangle P_{\chi=1}(T/2, \ell'', \ell'). \label{TrumpetPropagatorDecomposition}
\ee
Pictorially,
\begin{figure}[H]
\centering
\includegraphics[scale=0.7]{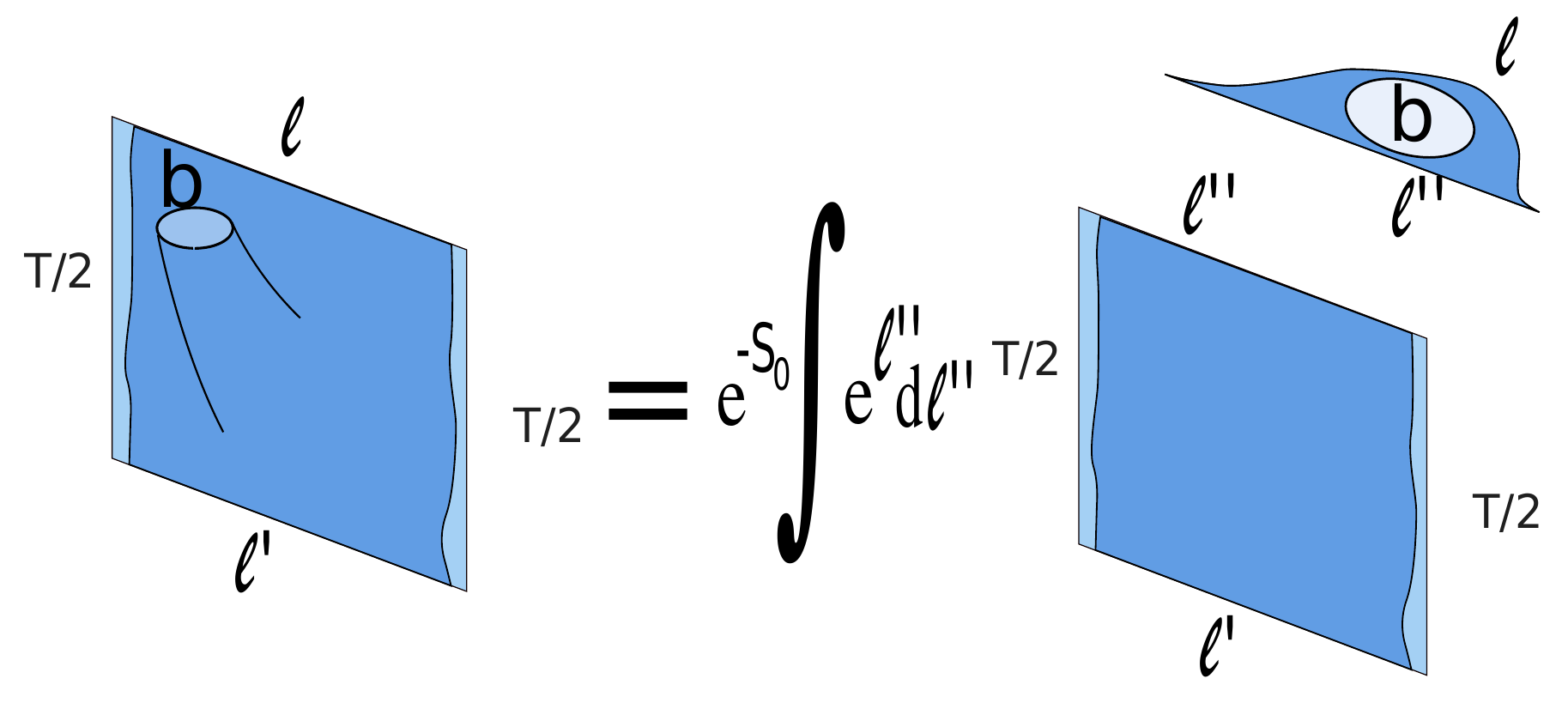}
\end{figure}
This decomposition is part of a family of decompositions that correspond to different time slicings. For any $\delta T$ such that $0<\delta T <T$, 
\be
P_{Tr}(T,b,\ell,\ell') = e^{-2 S_0} \int e^{\ell_1} d\ell_1 \int e^{\ell_2} d\ell_2 P_{\chi=1}(T-\delta T, \ell, \ell_1) \langle \ell_1,b| \ell_2\rangle P_{\chi=1}(\delta T, \ell_2, \ell'). \label{ambiguity}
\ee
The amplitude $\langle \ell, b| \ell'\rangle$ can be calculated directly in JT gravity via an integral over geometries with the topology of a cylinder, with a geodesic boundary of length $b$ and a boundary consisting of two geodesic segments of renormalized length $\ell$ and $\ell'$. These two geodesic segments are joined at infinity. 
\begin{figure}[H]
\centering
\includegraphics[scale=0.5]{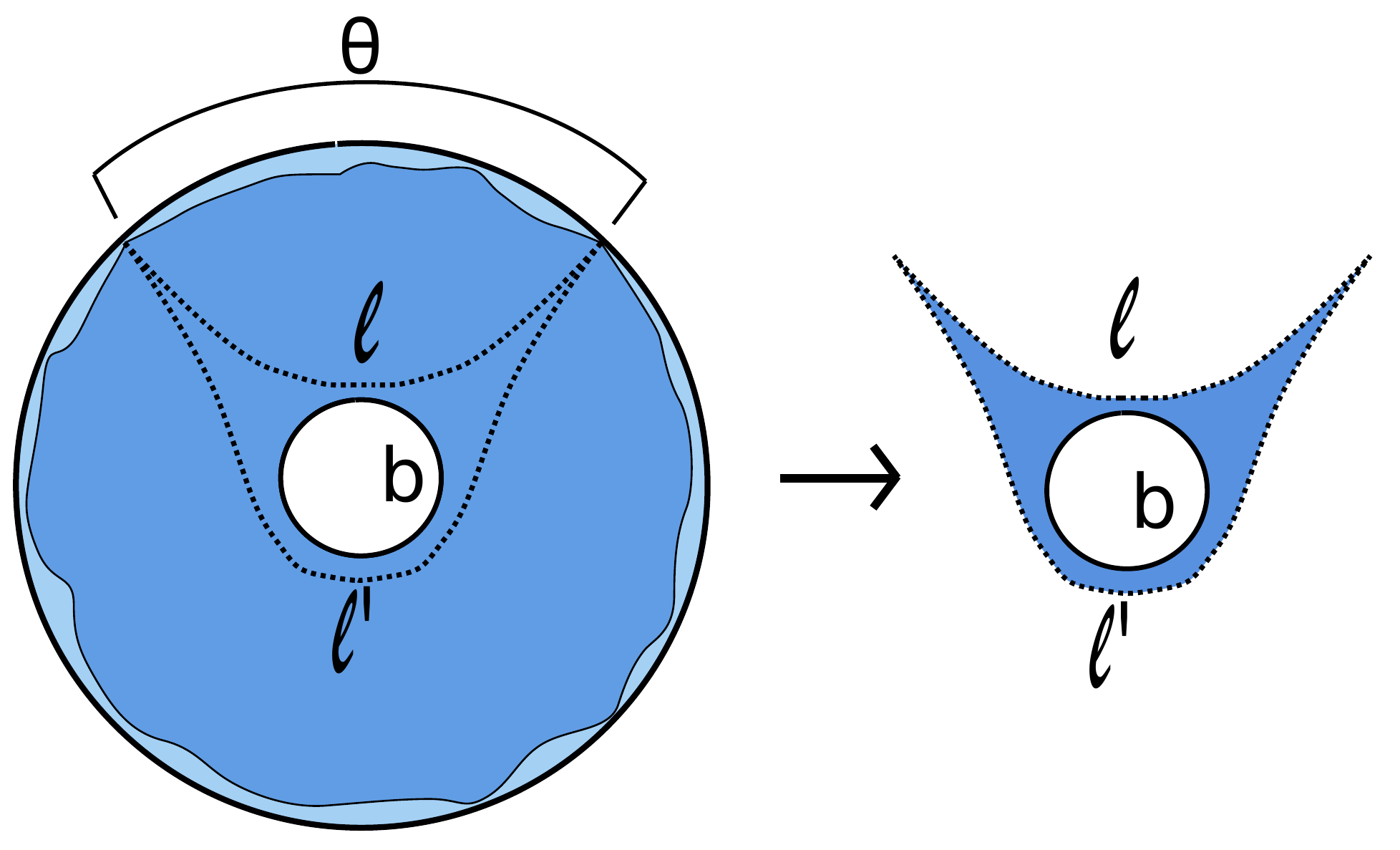}
\caption{\small Above we have pictured a ``top down'' view of the trumpet, with a piece cut out. This piece contributes to the overlap $\langle \ell,b|\ell'\rangle$.}
\label{fig:topdowntrumpet}
\end{figure}
We may label the geodesics by the angle $\theta$ between the two endpoints, defined such that for $\theta> \pi$, the geodesic ``wraps around'' the $b$ circle. The JT action for such a geometry is equal to a topological term plus corner terms from the joining regions at infinity. We can think of the corner terms as being a sum of two terms, one accounting for the angle between each geodesic and a constant trumpet radius circle. Each of these angles just depends on the angle $\theta$ subtended by the geodesics. Since we must have the sum of the two angles equal to $2\pi$ we may write the integral over all such geometries as an integral over the two angles with a delta function restricting their sum,
\be
\langle \ell, b| \ell'\rangle \sim \int d\theta d\theta' \delta(\theta+\theta'- 2\pi) e^{-I_{Corner}(\theta)- I_{Corner}(\theta')}. \label{AngleIntegral}
\ee
Here we have not been careful about the measure for $\theta$ and $\theta'$. We would like to see that our formula from the boundary particle formalism reproduces this formula. By using (\ref{TrumpetPropagator}) and (\ref{TrumpetPropagatorDecomposition}), along with the orthogonality formula (\ref{BesselOrthogonal}), or just by setting $T=0$ in (\ref{TrumpetPropagator}), we find
\be
\langle \ell, b| \ell'\rangle = \int_0^\infty dE \frac{\cos(b \sqrt{2E})}{\pi\sqrt{2E}} \psi_E(\ell) \psi_E(\ell').
\ee
Using our definition (\ref{EnergyWavefunction}) for the wavefunction $\psi_E(\ell)$ and the integral formula for the Bessel function
\be
K_{i \sqrt{8 E} } (4 e^{-\ell/2}) = \frac{1}{2}\int_{-\infty}^\infty dx \; e^{- 4 e^{-\ell/2} \cosh(x)} e^{-i \sqrt{8 E} x},
\ee
and changing variables to $s=\sqrt{2E}$, we find
\be
\langle \ell, b| \ell'\rangle =  \frac{2}{\pi} e^{-(\ell+\ell')/2} \int_{-\infty}^\infty dx \int_{-\infty}^\infty dy \;e^{-4 e^{-\ell/2}\cosh(x)-4 e^{-\ell'/2}\cosh(y)} \int_{-\infty}^\infty ds\; e^{-i s (2 x + 2 y-b)}.
\ee
The $s$ integral is gives a delta function, so
\be
\langle \ell, b| \ell'\rangle=4 e^{-(\ell+\ell')/2}\int_{-\infty}^\infty dx \int_{-\infty}^\infty dy \;\delta (2x+ 2y -b) e^{-4 e^{-\ell/2}\cosh(x)-4 e^{-\ell'/2}\cosh(y)}.
\ee
With an appropriate change of variables, we reproduce the formula (\ref{AngleIntegral}).

The geometrical picture for the transition amplitude $\langle \ell, b| \ell'\rangle$ will be useful. The limit with large $\ell'$ and $b$, both scaling together, is particularly important. In this limit the amplitude is essentially independent of $b$ and $\ell'$ for $\ell$ of order one. Physically this corresponds to the limit in which the angle subtended by the $\ell$ geodesic is very small. Most of the $\ell'$ geodesic wraps around the $b$ circle. As $b$ and $\ell'$ increase together, the behavior of the endpoints of the $\ell'$ geodesic doesn't change, so the action of such geometries are independent of $b$ for large $b$. 
\begin{figure}[H]
\centering
\includegraphics[scale=0.4]{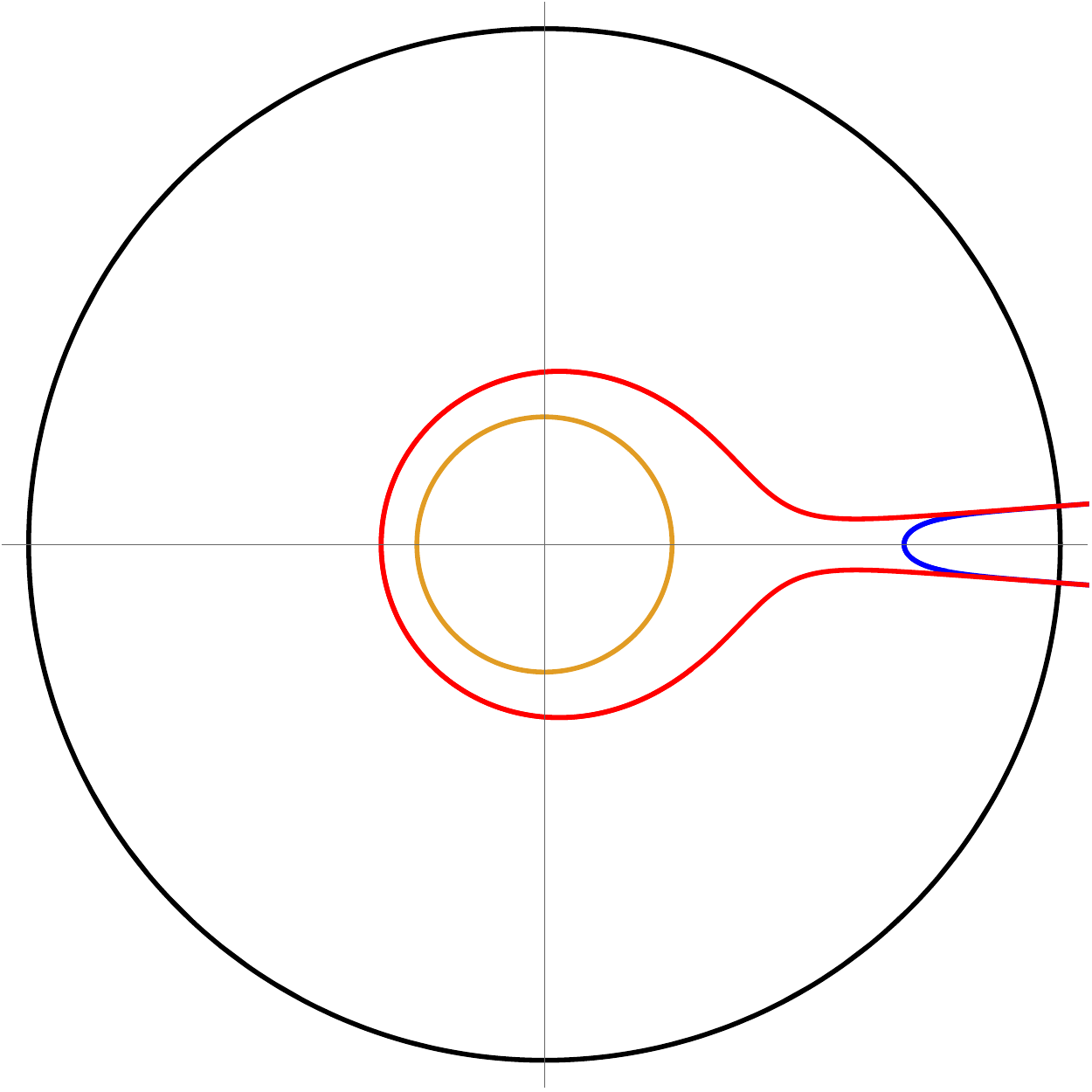}\hspace{40pt}\includegraphics[scale=0.4]{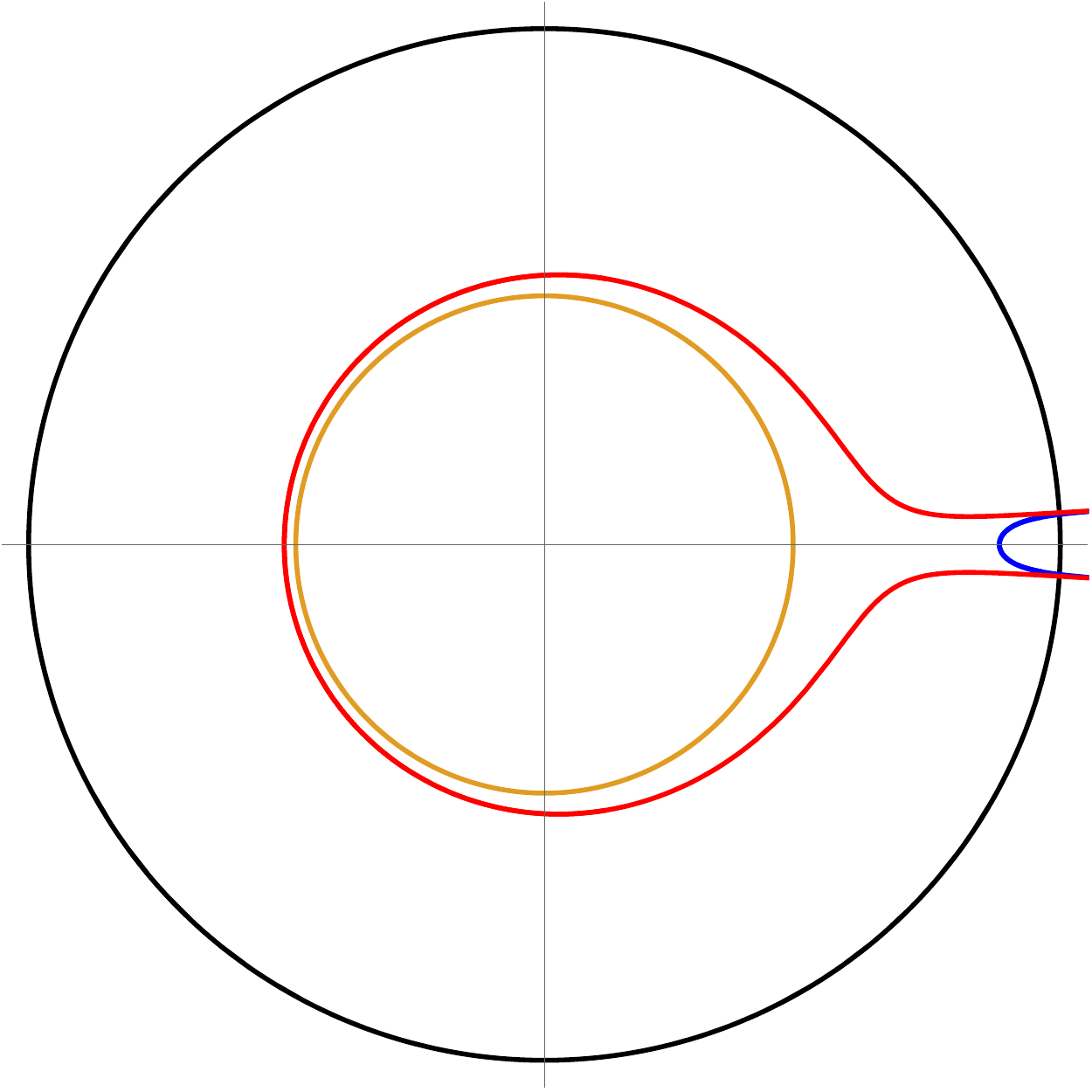}
\caption{\small As in Figure \ref{fig:topdowntrumpet}, we have pictured a ``top down'' view of the trumpet, along with the geodesics on the trumpet that define $\langle \ell, b | \ell'\rangle$. The orange circle represents the $r=0$ circle in trumpet coordinates, described by $ds^2= dr^2 + \frac{b^2}{(2\pi)^2} \cosh^2(r) d\theta^2$. Constant $r$ slices, such as the black circle, are circles centered around the origin, a distance $r$ from the $r=0$ circle. The red and blue lines are geodesics which meet at infinity; the red geodesic has renormalized length $\ell$, and the blue geodesic has renormalized length $\ell'$. Here we can see that for small $\ell'$, as $b$ and $\ell'$ are increased together the red geodesic mostly hugs the $r=0$ circle.}
\end{figure}
The takeaway is that even for a state with a very large ERB, there is a nonperturbatively small amplitude to transition to a state with a small ERB, such as the Hartle-Hawking state, by emitting a baby universe with a size comparable to the original length of the ERB. For large ERB, the action to transition to a small ERB goes to a constant.
\begin{figure}[H]
\centering
\includegraphics[scale=0.7]{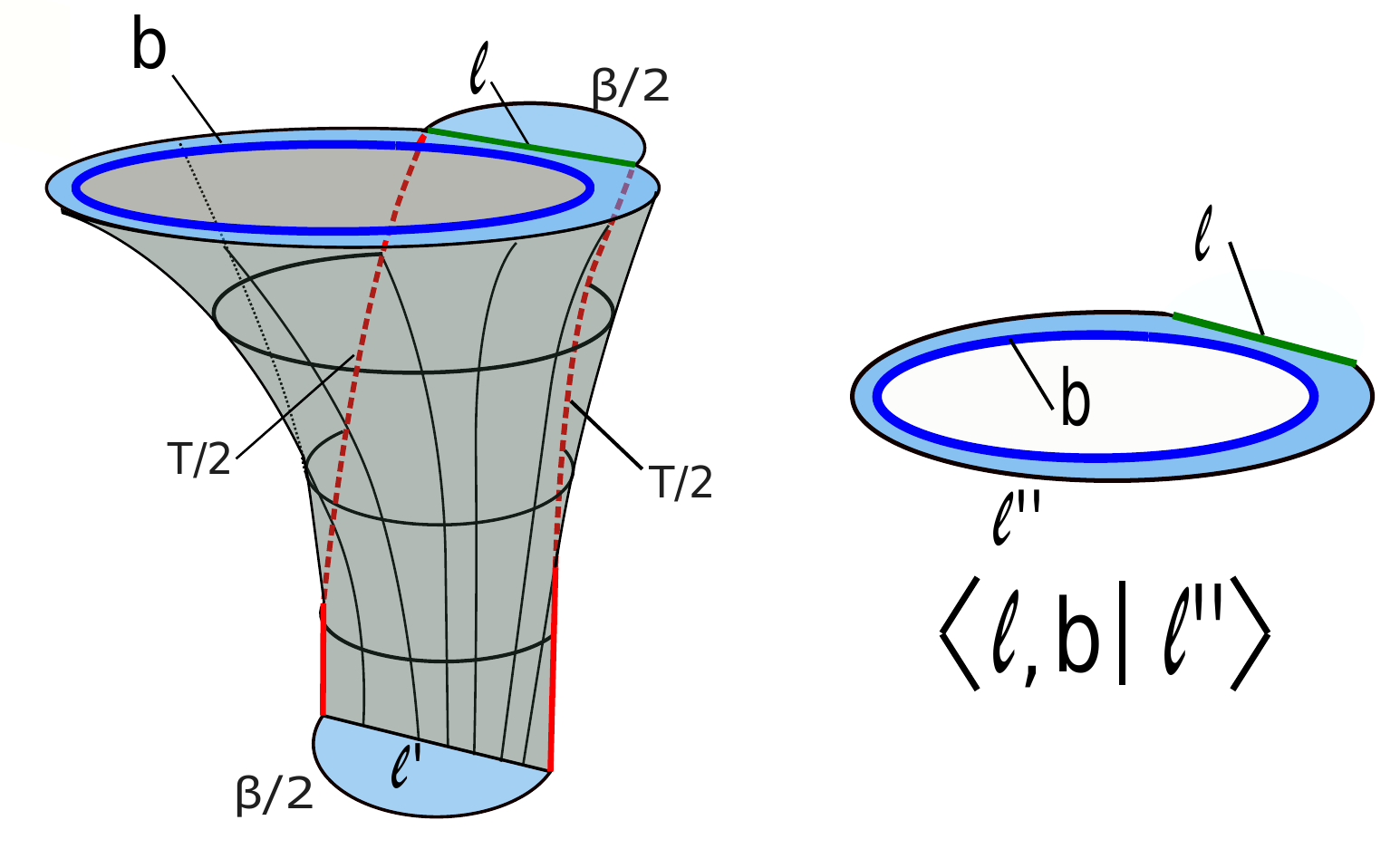}
\caption{\small Above we have pictured the dominant contribution to the process $\langle HH_\beta, b| e^{-i\frac{T}{2} H_{Bulk}} |HH_\beta\rangle$, with $b\sim T$. The grey piece is the Lorentzian portion of the geometry, with time increasing upwards, and the red lines denote asymptotic boundary of the Lorentzian piece. The dark blue circle denotes the geodesic boundary of length $b$, and the green line denotes the geodesic slice of renormalized length $\ell$. On the right we have pictured just the portion $\langle \ell, b |\ell''\rangle$ which causes the transition from a large Einstein Rosen Bridge (ERB) to a small ERB and a large baby universe.}
\label{fig:TrumpetStateLongTime}
\end{figure}
In this paper we will describe the overlap $\langle \ell, b |\ell'\rangle$ as the amplitude to emit a baby universe. However, one may instead think of this overlap as describing a topological ambiguity of a state \cite{Jafferis:2017tiu} instead of a dynamical process. This viewpoint is illustrated in particular by (\ref{ambiguity}). In this equation, we are not summing over different times at which this baby universe emission happens, as might be appropriate for a usual tunneling process. Instead, we choose one time slice on which the baby universe is emitted; due to diffeomorphism invariance, different choices of time slice give the same result. On the other hand, we find that the language of baby universe "emission" and "absorption" illustrates our results more clearly, so we will continue to use this description. As the prescription for including these effects is unambiguously determined by the Euclidean path integral, our choice of terminology will not affect our results.

Now let's use this idea to understand the ramp in the spectral form factor. We start by calculating the double-trumpet (\ref{DoubleTrumpet}) in a particularly useful way.\footnote{We thank Douglas Stanford for bringing this to our attention.} First, we fix the average energy of the two boundaries of the double-trumpet to find a contribution to $Y(E,T)$,
\begin{align}
Y(E,T)& \supset \frac{1}{\pi i}\int_{\mathcal{C}} d\beta \; e^{2\beta E} \int_0^\infty b db \; Z_{Tr}(\beta+i T,b) Z_{Tr}(\beta-i T, b)
\cr
&=  \frac{1}{\pi i}\int_{\mathcal{C}} d\beta \; e^{2\beta E}\int_0^\infty b db \; \frac{-\beta \frac{b^2}{\beta^2+ T^2}}{2\pi \sqrt{\beta^2+T^2}}.
\end{align}
$\mathcal{C}$ is a vertical contour to the right of the origin. Now we take the limit $T\gg \beta$. Exchanging the order of the integrals, the $\beta$ integral simplifies into a delta function,
\begin{align}
Y(E,T) &\supset \int_0^\infty b db\; \frac{1}{2\pi T} \bigg(\frac{1}{2\pi i} \int_{\mathcal{C}} d(2\beta) \; e^{2 \beta (E-\frac{b^2}{2T^2}) }\bigg),\hspace{20pt} T\gg \beta
\cr
& = \int_0^\infty b db\; \frac{\delta(E - b^2/2T^2)}{2\pi T}
\cr
&= \frac{1}{2\pi \sqrt{2E}} \int_0^\infty b db \;\delta( b- T\sqrt{2E}) = \frac{T}{2\pi}.
\end{align}
As this is independent of the energy $E$, Laplace transforming to obtain the double-trumpet just multiplies this by $1/2\beta$, matching the formula for the ramp from \cite{Saad:2019lba} for $T\gg \beta$. Here we see that at long times and at a fixed average energy, the size of the $b$ circle is fixed to be of order $T$, and the linear dependence on $T$ comes from the measure $b db$. In other words, at long times, the dominant contribution to the double-trumpet comes from geometries in which a baby universe of size of order $T$ is emitted, and the linear growth comes from the $T$ different ways in which the baby universe can be rotated before being absorbed.

Our discussions of the semiclassical growth of the ERB (\ref{WormholeSize}) and the overlap $\langle \ell,b |\ell'\rangle$ let us give a more detailed physical picture. After a long time $T$, a pair of Hartle-Hawking states will evolve into a state with two long ERBs, each of size of order T. The spectral form factor measures the overlap of this state with a pair of Hartle-Hawking states, both peaked at small ERB lengths. In order to get a non-decaying overlap, the long ERBs must both emit large baby universe, with sizes of order $T$. The only way for this to happen is for the two ERBs to trade a baby universe with a size of order $T$. The different ways in which the baby universe can be ``rotated'' before being absorbed gives a factor proportional to its size, leading to an overall behavior linear in $T$.
We can also consider contributions from trading multiple baby universes. These geometries give decaying contributions to the spectral form factor. This happens as a result of large cancellations between terms, and is related to the GUE statistics of the ensemble; in theories with GOE or GSE statistics we expect that growing terms at higher genus do not completely cancel.\footnote{In certain generalizations of JT gravity, described by other random matrix ensembles \cite{Stanford:2019vob}, these corrections are related to contributions from non-orientable geometries.}

\section{The ramp in the two-point function}\label{SectionRamp2pt}

The gravitational effects described in the previous section will have signatures in correlation functions; in particular, when operators are sufficiently widely separated in time effects like these dominate the behavior of correlation functions and give rise to the ramp and plateau behavior. Two important ramifications of these topology changing processes on the behavior of bulk matter are roughly as follows:

First, Euclidean wormholes may provide ``shortcuts''. For example, a Euclidean wormhole may connect regions near the boundaries that are widely separated in time, but the length of a geodesic connecting these widely time-separated boundary points may be short if it goes through the wormhole. Physically, this corresponds to the fact that when a baby universe is emitted, it may be reabsorbed near any other point in space with a probability that doesn't decay in time, so matter emitted with the baby universe may then reabsorbed at any faraway point in space with a non-decaying probability.

Second, emitting a large baby universe can cause the parent JT universe to shrink. A two-sided boundary two point function in the thermofield double state with operators separated by time $T$ is approximately given by $e^{-\Delta \ell}$, where $\ell$ is the renormalized length of the maximal length spatial slice between the two operators, which due to the symmetry of the thermofield double is equal to the length of the wormhole at time $T/2$. For large times $T$, this length will be large, of order $T$, unless a baby universe with size of order $T$ is emitted, in which case $\ell$ may be of order one. The shrinking of the parent JT universe also causes single sided two point functions to stop decaying; matter which has fallen deep into the geometry may become close to the boundary again after the baby universe is emitted.

We may think of either of these two behaviors as complementary descriptions of the ramp; the natural description depends on how we continue the Euclidean geometry to find a piecewise Euclidean and Lorentzian geometry. We will discuss this point in more detail later. However, when doing the calculation it is helpful to keep the ``shortening'' picture in mind. 

We will begin this section by describing the setup of the calculation. We review the results from \cite{Yang:2018gdb} for the contribution to the two-point function in the absence of Euclidean wormholes, which gives a prediction for the matrix elements $|\mathcal{O}_{E,E'}|^2$. We then proceed with the calculation of the contribution to the two point function from a single euclidean wormhole. Together with the contribution from \cite{Yang:2018gdb}, we find a formula that exactly agrees with our expectation (\ref{GJTPrediction}) for $T\ll e^{S_0}$. We then argue that at late times, but before the plateau time, any corrections are small. The plateau is left to Section \ref{SectionPlateau}. Finally, we discuss the physical interpretation of this calculation.

\subsection{Geometry and setup}\label{SubsectionGeometrySetup}

In this paper we will work in the probe approximation for the matter fields, ignoring the effects of any backreaction. We expect this to be a good approximation at long times.\footnote{Though minimally coupled matter does not backreact locally in JT gravity, matter loops around cycles contribute Casimir forces. For example, a free scalar field strongly weights the contributions of geometries with small cycles, giving a divergence for vanishingly small cycles. We expect that the late time behavior of the correlator is dominated by large cycles, and would be insensitive to any cutoff on the integral that removes the contributions of small cycles. As an aside, we note that \cite{Maldacena:2018lmt} provides a possible hint about how such a cutoff might be naturally implemented in the SYK model.} We choose our matter to be a single free scalar field $\mathcal{O}$.\footnote{We expect that similar results will hold for fields with spin and with weak interactions.} In this approximation, correlation functions are simply given by free QFT correlators at given boundary times on a fixed geometry, integrated over geometries.
\subsubsection{The disk contribution}
For example, the leading contribution to the Euclidean thermal two-point function
\be
G_{2,\beta}(T=-i \tau)_{JT}\sim  \Tr \big[e^{-\beta H}  e^{\tau H} \mathcal{O} e^{-\tau H} \mathcal{O}\big]
\ee
is given by an integral over pieces of the hyperbolic disk with renormalized boundary length $\beta$ weighted by the JT action and the free propagator between boundary points a distance $\tau$ away from each other on the boundary. For boundary operators $\mathcal{O}$ of conformal weight $\Delta$ rescaled as in \cite{Yang:2018gdb}, the free propagator $\langle\mathcal{O}(x)\mathcal{O}(x')\rangle_{Disk}$ is simply given by
\be
\langle\mathcal{O}(x)\mathcal{O}(x')\rangle_{Disk}= e^{-\Delta \ell(x,x')},
\ee
where $\ell(x,x')$ is the renormalized geodesic distance between boundary points $x$ and $x'$. This expression looks similar to the saddle point expression for the propagator in the worldline formalism, with the mass of the particle replaced by $\Delta$. In that case we would only trust the saddle point expression for large mass. In our case the expression is exact for finite $\Delta$. $\Delta$ has an infinite power series in $m$; keeping only the leading term of this series results in the large mass saddle point expression.

As explained in \cite{Yang:2018gdb}, we may calculate the correlator by integrating from one segment of the boundary of renormalized length $\beta-\tau$ up to a geodesic slice of length $\ell$ to produce the Hartle-Hawking wavefunction $\psi_{D,\beta-\tau}(\ell)$, then integrating from this slice to the rest of the asymptotic boundary to produce the wavefunction $\psi_{D,\tau}(\ell)$, and finally integrating over all possible lengths $\ell$ weighted by the matter correlator $e^{-\Delta \ell}$.
\be
G_{2,\beta}(T=-i \tau)_{JT} \supset e^{-S_0} \int e^\ell d\ell \; \psi_{D,\beta-\tau} (\ell) \psi_{D, \tau}(\ell) e^{-\Delta \ell} \equiv G_{2,\beta,\chi=1}(-i\tau)_{JT}. \label{DiskCorrelatorImaginaryTime}
\ee
We can represent this formula with the diagram
\begin{figure}[H]
\centering
\includegraphics[scale=0.75]{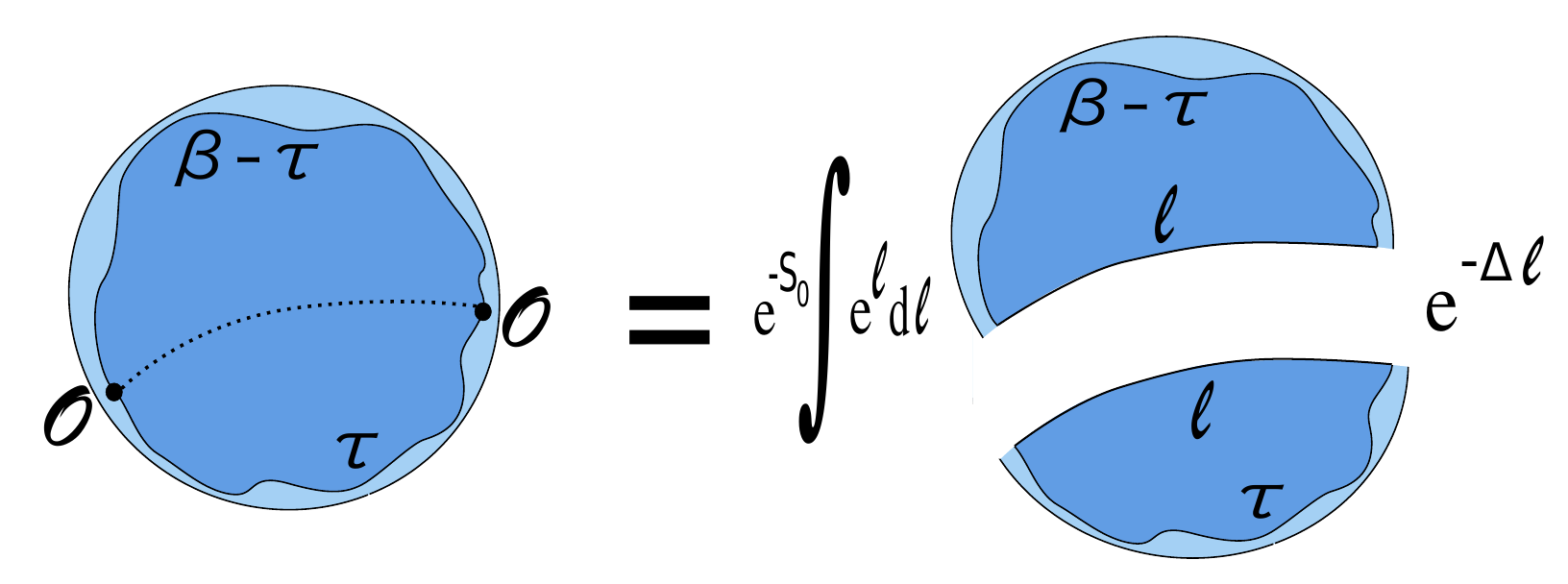}
\label{fig:DiskCorrelatorEuclidean}
\end{figure}
To obtain the Lorentzian correlators $G^{LR}_{2,\beta,\chi=1}(T)$ and $G_{2,\beta,\chi=1}(T)$ we simply continue $\tau\rightarrow \beta/2+ i T$ in the case of the two sided correlator and $\tau\rightarrow i T$ in the case of the thermal correlator.

We may also give a more direct formula for the two sided correlator in terms of the expectation value of $e^{-\Delta \ell}$ in the Hartle-Hawking state evolved for time $T/2$
\be\label{DiskCorrelatorRealTime}
G^{LR}_{2,\beta,\chi=1}(T)= e^{-S_0} \int e^{\ell} d\ell \; |\psi_{D,\beta/2+ i T}(\ell)|^2 e^{-\Delta \ell}.
\ee
\begin{figure}[H]
\centering
\includegraphics[scale=0.9]{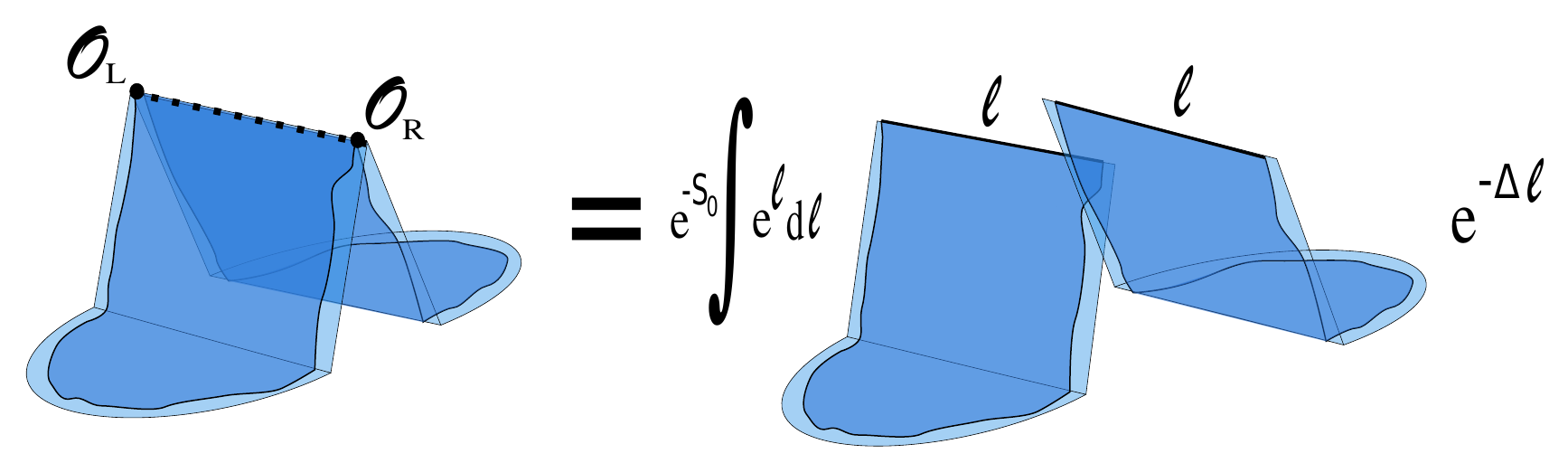}
\caption{\small We may view the two-sided two-point function as the expectation value of the operator $\mathcal{O}_L \mathcal{O}_R$ in the thermofield double state evolved for time $T/2$ on both sides. In JT gravity, the two-point function of the two boundary operators is given by $e^{-\Delta \ell}$, with $\ell$ the geodesic length $\ell$ between the two operators. We can thus express the correlator in gravity as the expectation value of $e^{-\Delta \ell}$ in the Hartle-Hawking state evolved for time $T/2$.}
\end{figure}
Now, following \cite{Yang:2018gdb}, we can massage our formula (\ref{DiskCorrelatorImaginaryTime}) into a form similar to (\ref{GPredictionEarlyTime}). Inputting our integral expression for the Hartle-Hawking wavefunction (\ref{HHWavefunctionFormula}) into our formula (\ref{DiskCorrelatorRealTime}), we find
\be
G_{2,\beta,\chi=1}(-i\tau)_{JT} = e^{-S_0} \int_{-\infty}^\infty e^\ell d\ell \int_0^\infty dE \int_0^\infty dE' \;\rho_0(E)\rho_0(E') e^{-\beta E'} e^{-\tau (E-E')} \psi_E(\ell) \psi_{E'}(\ell) e^{-\Delta \ell}.\label{DiskCorrelatorInProgress}
\ee

We first perform the integral over $\ell$. We write out the relevant integral below, as we will use it again later.
\begin{align}
\int_{-\infty}^\infty d\ell \;\psi_E(\ell)\psi_{E'}(\ell) e^{-\Delta \ell} &= \frac{|\Gamma(\Delta+ i (\sqrt{2E}+\sqrt{2E'})) \Gamma(\Delta+  i (\sqrt{2E}-\sqrt{2E'}))|^2}{2^{2\Delta+1}\Gamma(2\Delta) }, \hspace{20pt} \Delta>0
\cr
&\equiv |\mathcal{O}_{E,E'}|^2.  \label{JTMatrixElementsCalculation}
\end{align}
Here we have defined the quantity $|\mathcal{O}_{E,E'}|^2$. Inputting this into the formula (\ref{DiskCorrelatorInProgress}), we find
\be
G_{2,\beta,\chi=1}(-i\tau)_{JT}  = \int_0^\infty dE dE' \rho_0(E) \rho_0(E') e^{-\beta E'} e^{-\tau (E-E')} e^{-S_0}|\mathcal{O}_{E,E'}|^2.
\ee
This formula matches our prediction for the early time behavior of the correlator (\ref{GPredictionEarlyTime}), justifying our interpretation of $e^{-S_0}|\mathcal{O}_{E,E'}|^2$, defined through (\ref{JTMatrixElementsCalculation}), as the averaged squared matrix elements of the operator $\mathcal{O}$.

\subsubsection{Geometry for the ramp contribution}
We now discuss the setup for the calculation of the correction to the two-point function from the contributions of geometries with a Euclidean wormhole.\footnote{This geometry has previously been suggested to account for the ramp in the two-point function \cite{Saad:2018bqo,Blommaert:2019hjr}.} These are constant negative curvature geometries with Euler character $\chi=-1$ and a single circular asymptotic boundary. We will focus our attention on purely Euclidean geometries for now; we will obtain a Euclidean correlator via integration over these geometries, then obtain Lorentzian correlators via analytic continuation.

We begin by describing the geometries in two different ways. First we discuss the description of these geometries used in \cite{Saad:2019lba}. Take such a geometry and cut along the minimal length circular geodesic homotopic to the asymptotic boundary. Label the length of this geodesic $b_{Tr}$. This decomposition is pictured below.
\begin{figure}[H]
\centering
\includegraphics[scale=0.6]{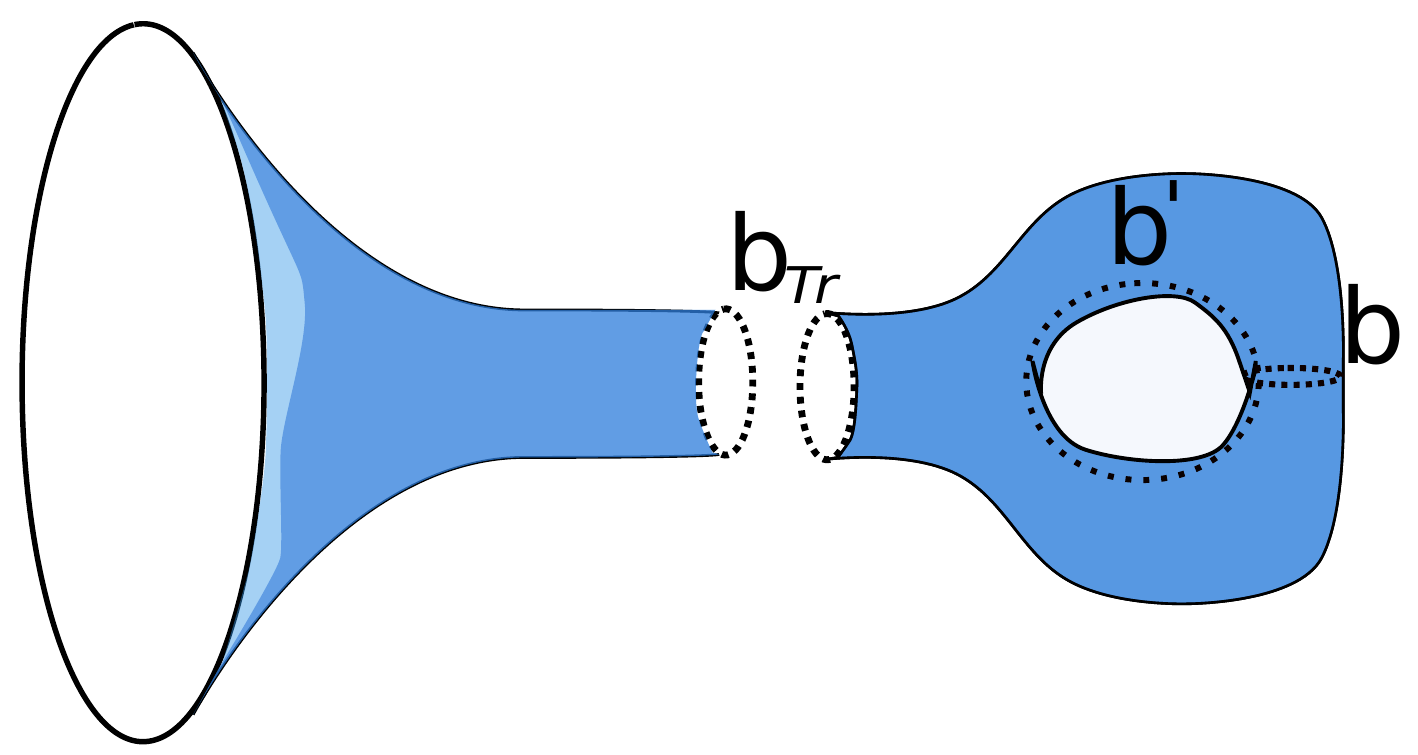}
\caption{\small On the left, we have a constant negative curvature geometry with an asymptotic boundary and a geodesic boundary; this is the trumpet geometry. On the right we have a constant negative curvature geometry with the topology of a handle on a disk, with a geodesic boundary. There are an infinite number of simple closed geodesics on the handle geometry on the right, we have pictured a basis of such cycles as examples. }
\label{fig:HandleOnDiskDecomp}
\end{figure}
We need to understand the moduli space of geometries on the right; the ``handle on a disk'' with a geodesic boundary. For a given boundary length $b_{Tr}$, the geometry may be completely specified by two the length and ``twist'' of any one of the infinitely many simple closed (circular) geodesics on the geometry. To see this, we imagine cutting along such a geodesic, with length $b_i$ and twist $\tau_i$, to produce a ``pair of pants''. 
\begin{figure}[H]
\centering
\includegraphics[scale=0.9]{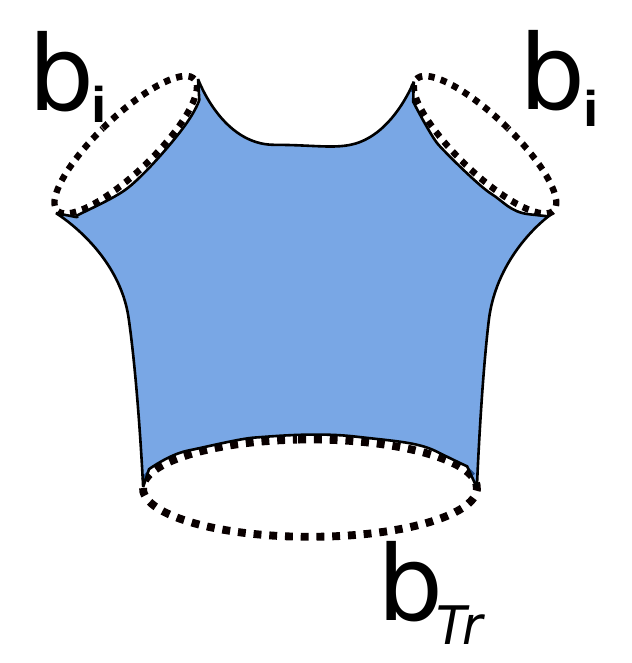}
\caption{\small Here we have pictured a pair of pants geometry with a geodesic boundary of length $b_{Tr}$ and two geodesic boundaries of length $b_i$.}
\end{figure}
This pair of pants geometry has no moduli; given three lengths, there is a single geometry with this topology that has geodesic boundaries with these lengths. 

The original handle on a disk geometry may be obtained by ``gluing'' the two $b_i$ boundaries together with a twist $\tau_i$. As there is a single pair of pants geometry for a given set of boundary lengths, the moduli space of handle on a disk geometries with a geodesic boundary of length $b_{Tr}$ may then be labeled by coordinates $b_i$ ad $\tau_i$, corresponding to the length and twist of the circular geodesic used to define in the pair of pants construction. However, this construction is not unique; we may choose any of the infinitely many circular geodesics on the handle on a disk to label the geometry. This issue is analogous to the issue of modular invariance in defining a torus. As the geometry is completely fixed by the length and twist of just one of the circular geodesics, two different choices of length and twist parameters may define the same geometry; the geometry may have two cycles, each with one of the given pairs of length and twists. The moduli space of geometries is then described by a fundamental domain in the $b_i$-$\tau_i$ plane, such that for a geometry with a a geodesic of length and twist $b_i$ and $\tau_i$ in this domain, none of the other circular geodesics on the geometry have lengths and twists in this domain. There are infinitely many choices of fundamental domain.

We will need to integrate over handle on a disk geometries. As discussed in \cite{Saad:2019lba} and Section \ref{SubsectionRampSFF}, in JT gravity the measure for integration over the moduli space of these geometries is the Weil-Petersson measure, which may be described in the $b_i$-$\tau_i$ coordinates as \cite{wolpert1985weil}
\be
d(WP) =d b_i \wedge d\tau_i. \label{WPMeasure}
\ee
This measure is invariant under different pair of pants decompositions, though it is difficult to see in these coordinates. For different choices of cycle $i$ and $j$, we have $d b_i\wedge d\tau_i= db_j\wedge d\tau_j$. $b_{i}$ and $\tau_{i}$ should be integrated over a fundamental domain $\mathcal{F}_i(b_{Tr})$ \cite{Kim:2015qoa}.

We now describe a different construction of the handle on a disk geometry, which will also prove useful. In this construction we will ignore the boundary wiggles; it should be clear how they are incorporated. The idea behind this construction is to view the handle on a disk geometry as a quotient of the hyperbolic disk by the action of two isometries. There is a family of quotients which covers the whole moduli space of handle on a disk geometries.
\begin{figure}[H]
\centering
\includegraphics[scale=0.5]{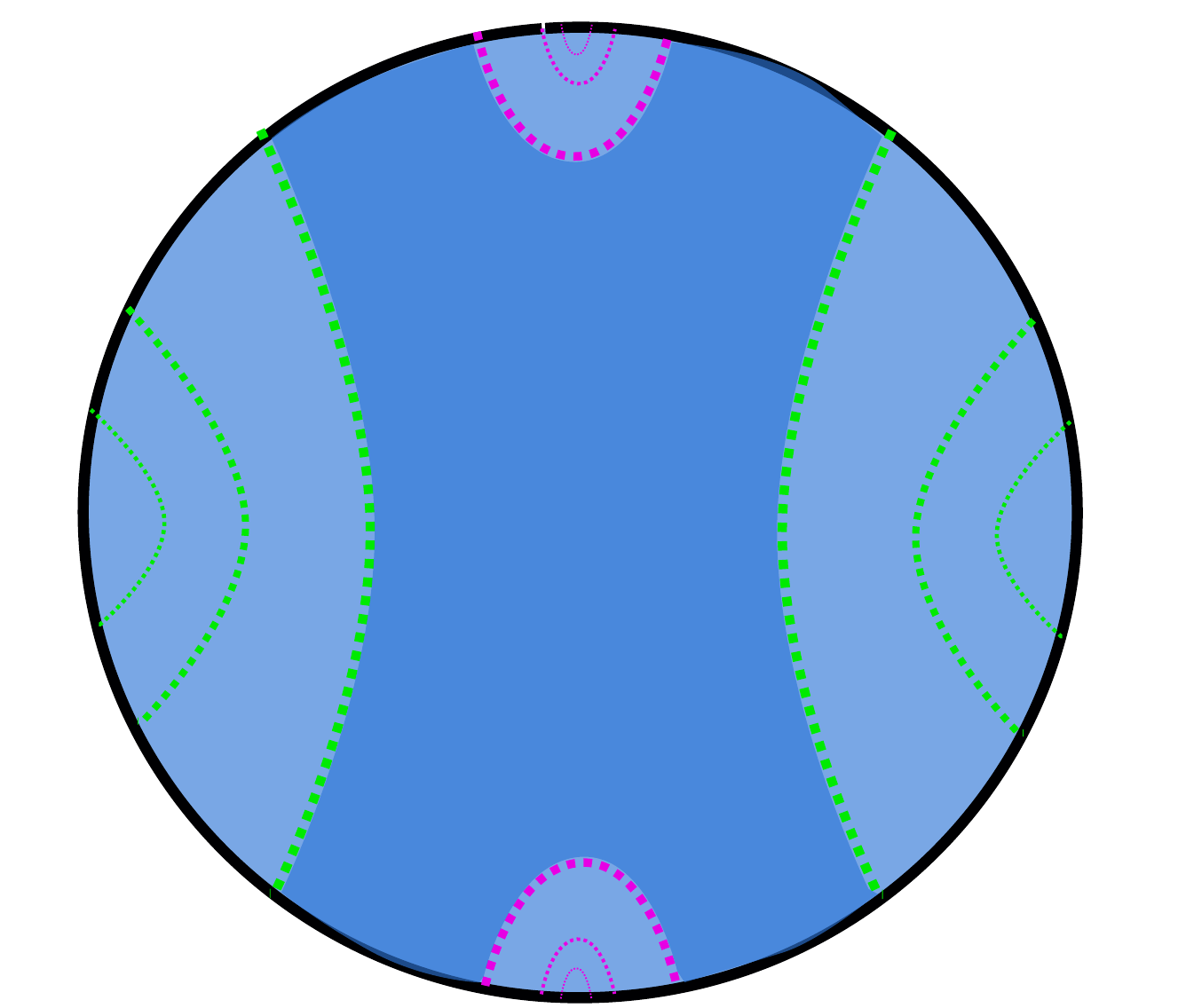}
\caption{\small Above we have pictured two geodesics and some of their images under the actions of these isometries
The green geodesics are images of one of the geodesics under one identification and the purple are images of another geodesic under the other identification. A fundamental domain of this identification is shaded. }
\end{figure}
The geometry resulting from the quotient may be obtained by cutting and gluing the pairs of geodesics that bound this fundamental domain. First cutting and gluing the two green geodesics produces a geometry with the topology of a cylinder; the double-trumpet. Cutting and gluing along the purple geodesics on this double-trumpet geometry produces the handle on a disk geometry. 
\begin{figure}[H]\label{HandelQuotientCircles}
\centering
\includegraphics[scale=0.5]{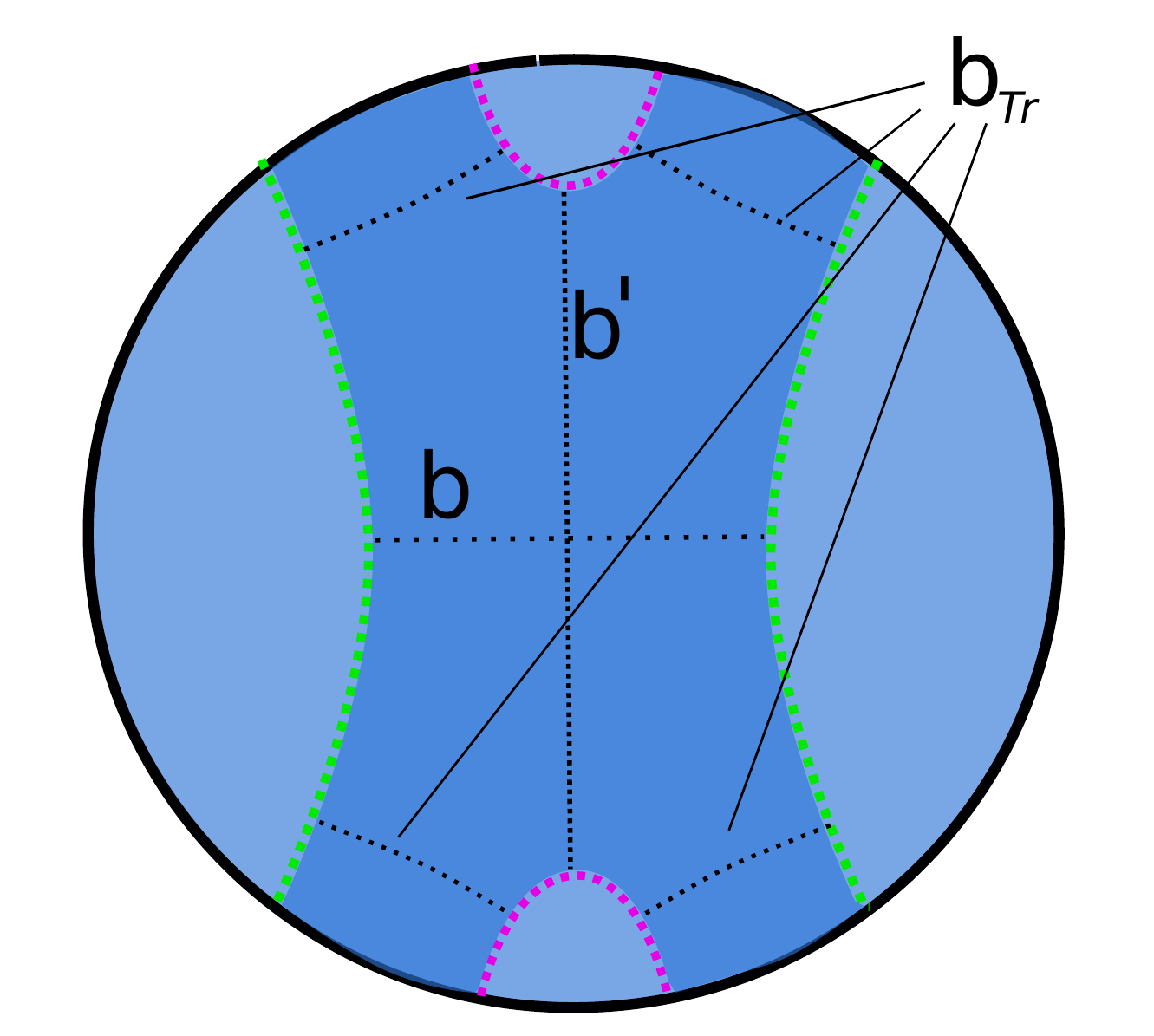} 
\caption{\small To make contact with our previous construction, we give an image of the fundamental domain with the geodesics $b_{Tr}$, $b$, and $b'$ from Figure \ref{fig:HandleOnDiskDecomp} pictured.}
\end{figure}

In \cite{Blommaert:2019hjr}, the authors studied the BF description of JT gravity on this geometry and found a linear ramp behavior in the two-point function. However, their method amounts to choosing a pair of pants decomposition with length and twist $b_i$ and $\tau_i$, and integrating over all $b_i>0$, $0< \tau_i < b_i$. This region covers infinitely many fundamental domains of the mapping class group, so distinct geometries are counted infinitely many times. While in this work we take the mapping class group into account, we will find the same formula for the two-point function. In the rest of this section we will explain why this is the case.

\subsubsection{Matter correlator on the ramp geometry}
We now turn to the matter two-point function on a fixed geometry. We again focus our attention on the case where the operators are on the boundary, where the correlator simplifies. The free two-point function $\langle \mathcal{O}(x)\mathcal{O}(x')\rangle_g$ is obtained by solving the equation $(-\nabla_g+m^2)\langle \mathcal{O}(x)\mathcal{O}(x')\rangle_g=\delta^{(2)}(x-x')$ with the boundary conditions that this falls off as $x$ and $x'$ are widely separated, and where $\nabla_g$ is the Laplacian on the geometry. Fortunately, as our geometries are all quotients of the hyperbolic disk, we may express the solution $\langle \mathcal{O}(x)\mathcal{O}(x')\rangle_g$ as a sum over images of the correlator on the disk. 

As the correlator of boundary points on the disk is simply equal to $e^{-\Delta \ell(x,x')}$ (after rescaling), the correlator on the handle on the disk geometry is equal to a sum over images of terms involving the geodesic distances between boundary points on the disk and their images. These geodesics between boundary points and images correspond to geodesics on the quotient geometry.

The rescaled matter correlator can then be expressed as
\be
\langle \mathcal{O}(x)\mathcal{O}(x')\rangle_g=\sum_{\gamma} e^{-\Delta \ell_\gamma}
\ee
where $\gamma$ labels geodesics between boundary points $x$ and $x'$ and $\ell_\gamma$ is the renormalized length of these geodesics.
\begin{figure}[H]
\centering
\includegraphics[scale=0.75]{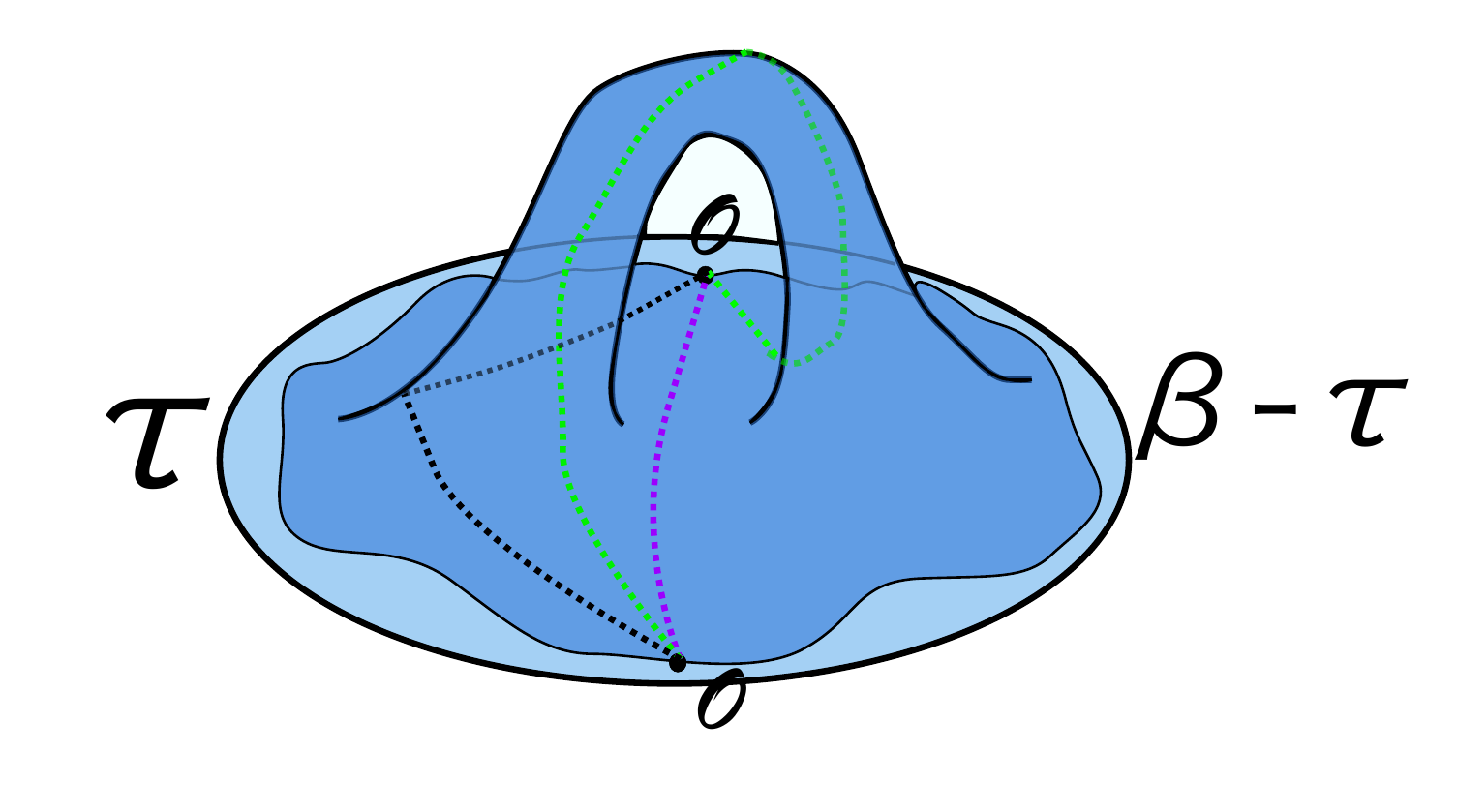}
\caption{\small Above we have pictured three example geodesics that contribute to the correlation function. The colors of the geodesics roughly correspond to the colors labeling geodesics in Figure (\ref{HandelQuotientCircles}); the green geodesic intersects the cycle $b$, wile the purple geodesic intersects the cycle $b'$. The black geodesic cuts through neither of the cycles on the handle.}
\label{fig:HandleDiskGeodesics}
\end{figure}
Now we put our pieces together  and write a schematic expression for the Euclidean correlator of free matter on the $\chi=-1$ geometry, $G_{2,\beta,\chi=-1}(T=-i\tau)_{JT}$ . Denoting the renormalized distance between boundary points $x$ and $x'$ along the boundary as $\ell(x,x')|_{\partial M}$, we write
\be
G_{2,\beta,\chi=-1}(-i\tau)_{JT}= \int\mathcal{D}g \; e^{-I_{JT}(g)}\langle \mathcal{O}(x)\mathcal{O}(x')\rangle_g \bigg|_{\ell(x,x')|_{\partial M}= \tau}
\ee
The integral is over all geometries with $\chi=-1$ and a single asymptotically AdS boundary of renormalized length $\beta$. The action $I_{JT}(g)$ only depends on the Euler character $\chi=-1$ and the boundary curve, not the moduli $b_i$ and $\tau_i$ of the handle. However, the geometry on which the boundary curve lives depends on the length $b_{Tr}$. Denoting the integral over the boundary wiggles by $\int \mathcal{D}(\mathcal{W})$ and the action of the boundary wiggles for fixed $b_{Tr}$ as $I_{\partial}(b_{Tr}, \mathcal{W})$, we schematically write
\begin{align}
G_{2,\beta,\chi=-1}(-i\tau)_{JT}&= e^{-S_0}  \int_0^\infty db_{Tr} \int_0^{b_{Tr}}d\tau_{Tr}
\cr
&  \int \mathcal{D}(\mathcal{W}) \;e^{-I_\partial(b_{Tr}, \mathcal{W})} \int_{\mathcal{F}_i(b_{Tr})} d b_i d\tau_i\langle \mathcal{O}(x)\mathcal{O}(x')\rangle_g \bigg|_{\ell(x,x')|_{\partial M}= \tau}. \label{HandleCorrelatorFull}
\end{align}
\subsection{Calculating the correlator}\label{Subsection2ptCalculation}
Now we move on to calculating the $\chi=-1$ contribution to the correlator. We will organize the calculation by doing a separate calculation for each geodesic between our operators. In this section we will only consider non-self-intersecting geodesics which go through the handle, meaning that they intersect at least one of the circular geodesics on the handle. There are infinitely many of these geodesics. We label these geodesics as $\gamma_i$. In Appendix \ref{AppendixGeodesics}, we will argue that all other geodesics give contributions which decay in time.

Consider any non-self-intersecting geodesic $\gamma_i$ between the boundary points which intersects a circular geodesic on the handle, for example the purple or green geodesics in Figure \ref{fig:HandleDiskGeodesics}. There is a single circular geodesic on the handle which \textit{does not} intersect $\gamma_i$.\footnote{A simple way to see this is to imagine cutting along the geodesic $\gamma_i$. This results in a geometry like the double-trumpet, but where each boundary has asymptotically AdS and geodesic components. The circular geodesic which does not intersect $\gamma_i$ is the circular geodesic homotopic to either boundary, which splits the double-trumpet into two trumpets.} Conversely, every circular geodesic on the handle corresponds to a geodesic $\gamma_i$. Label the length and twist of this circular geodesic as $b_i$ and $\tau_i$. 

Before calculating the full contribution of $\gamma_i$ to the correlator, we will look at the contribution from fixed $b_i$ and $\tau_i$. We may view the integral with this cycle fixed as an integral over boundary wiggles with $\ell_{\gamma_i}\equiv \ell_i$ fixed, then integrated over $\ell_i$, with the usual measure. We justify this by explicit calculation of the Jacobian in Appendix \ref{AppendixTrumpetWavefunction}.
\begin{figure}[H]
\centering
\includegraphics[scale=0.8]{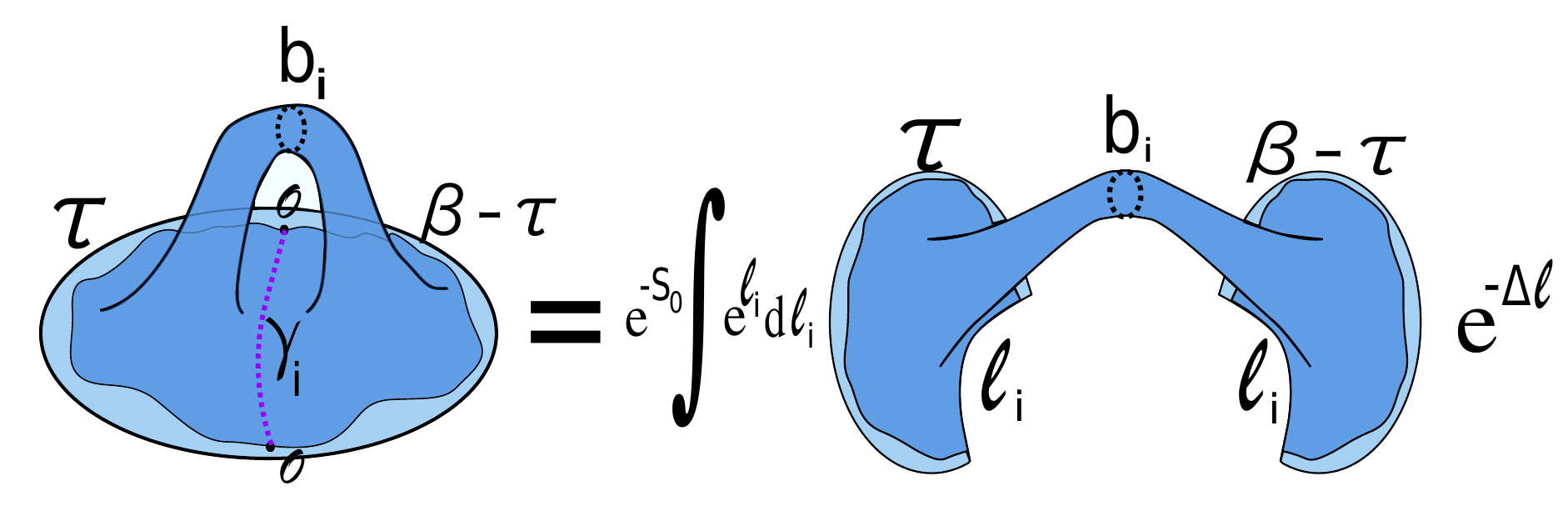}
\caption{\small Here we have pictured how cutting along $\gamma_i$ leads to double-trumpet geometries. The integral over geometries on the left is equal to an integral over boundary wiggles on the right for fixed $\ell_i$, giving a pair of trumpet wavefunctions $\psi_{Tr}(\ell_i,b_i)$, integrated over $\ell_i$ with a weighting factor $e^{-\Delta \ell_i}$. }
\label{fig:CutHandleGeodesic}
\end{figure}
As pictured above, the result can be written as
\be
e^{-S_0}\int e^{\ell_i} d\ell_i \;\psi_{Tr,\tau}( \ell_i, b_i) \;\psi_{Tr,\beta-\tau}(\ell_i,b_i) e^{-\Delta \ell_i} \label{GNaiveContribution}.
\ee

To find a contribution to the correlator, we must integrate over $b_i$ and $\tau_i$ with the measure (\ref{WPMeasure}). However, we must only integrate over a fundamental domain $\mathcal{F}_i(b_{Tr})$. The fundamental domain depends on $b_{Tr}$ \cite{Kim:2015qoa}, so we cannot simply integrate equation (\ref{GNaiveContribution}). 

In order to obtain a formula for $G_{2,\beta,\chi=-1}(-i\tau)_{JT}$ as an integral of (\ref{GNaiveContribution}), we must include every geodesic $\gamma_i$. Starting with our full expression (\ref{HandleCorrelatorFull}) and restricting the matter correlator $\langle \mathcal{O}(x)\mathcal{O}(x')\rangle_g$ to the contributions from the geodesics $\gamma_i$. We find
\be
G_{2,\beta,\chi=-1}(-i\tau)_{JT}\supset e^{-S_0} \int_0^\infty db_{Tr} \int_0^{b_{Tr}}d\tau_{Tr} \int \mathcal{D}(\mathcal{W}) \;e^{-I_\partial(b_{Tr}, \mathcal{W})} \int_{\mathcal{F}_i(b_{Tr})} db_i d\tau_i \sum_{j} e^{-\Delta \ell_j}. \label{Gsimple}
\ee
Let's look more closely at the integral over the moduli $b_i$ and $\tau_i$. We may move the sum over geodesics outside the integral,
\be
\sum_j \int_{\mathcal{F}_i(b_{Tr})} db_i d\tau_i \;e^{-\Delta \ell_j}.
\ee
Here we have made an arbitrary choice of cycle to integrate over. Every geodesic $\gamma_j$ except for $\gamma_i$ intersects the $b_i$ cycle. We may instead choose to integrate over the $b_j$ cycles which do not intersect $\gamma_j$. However, we must integrate these over the fundamental domain $\mathcal{F}_j(b_{Tr})$. This leaves us with the sum of integrals
\be
 \sum_j \int_{\mathcal{F}_j(b_{Tr})} db_j d\tau_j\;e^{-\Delta \ell_j}. 
\ee
The union of the fundamental domains $\cup_j\mathcal{F}_j(b_{Tr})$ is equal to the whole region $0<\tau_j < b_j$, $0<b_j<\infty$; If we are integrating over every distinct handle on a disk geometry, every single combination of length and twist must appear as the length and twist of one of the infinitely many circular geodesics $b_i$ on the geometry. Since each one of these circular geodesics $b_i$ is represented as the circular geodesic on the double-trumpet geometries on the right of Figure \ref{fig:CutHandleGeodesic}, then every physically distinct $b_i$ and $\tau_i$ ($b_i>0$, $0<\tau_i<b_i$ ) must appear in a geometry on the right of Figure \ref{fig:CutHandleGeodesic} for one and only one of the geodesics $\gamma_i$.

Each term in this sum in this sum describes the integral over the length and twist of a cycle of a function of the geodesic which does not intersect it. We re-interpret this sum by thinking of each cycle $b_j$ as the same cycle $b$, where the geodesic $\gamma_j\rightarrow \gamma$ is defined to be the geodesic which does not intersect the cycle $b$. $b$ and $\tau$ are integrated over all physically distinct lengths and twists.
\be
\sum_j \int_{\mathcal{F}_j(b_{Tr})} db_j d\tau_j \;e^{-\Delta \ell_j} \rightarrow \int_0^\infty db \int_0^{b} d\tau \; e^{-\Delta \ell}.\label{ChangeOfVariables}
\ee
Inserting (\ref{ChangeOfVariables}) back into (\ref{Gsimple}), and using the fact that the domain of integration for $b$ and $\tau$ is independent of any of the other integration variables, we find an expression for the contribution to the correlator from the geodesics $\gamma_i$,\footnote{This method of exchanging the sum over geodesics for a sum over fundamental domains is reminiscent of the calculation of the one-loop string partition function in \cite{Polchinski:1985zf}.}
\be
G_{2,\beta,\chi=-1}(-i\tau)_{JT}\supset e^{-S_0}\int_0^\infty db \int_0^{b} d\tau \int_0^\infty db_{Tr} \int_0^{b_{Tr}} d\tau_{Tr}  \int \mathcal{D}(\mathcal{W}) \;e^{-I_\partial(b_{Tr}, \mathcal{W})} \; e^{-\Delta \ell}.
\ee
For fixed $b$ and $\tau$, this integral is exactly the integral we considered earlier corresponding to the contribution from a single geodesic $\gamma_i$ for fixed $b_i$ and $\tau_i$. Inserting our result for this contribution (\ref{GNaiveContribution}), and performing the integral over $\tau$, we find the simple formula
\be
\boxed{G_{2,\beta,\chi=-1}(-i\tau)_{JT}\supset e^{-S_0} \int_0^\infty b db \int_{-\infty}^\infty e^\ell d\ell \;\psi_{Tr,\tau}(\ell,b) \psi_{Tr,\beta-\tau}(\ell,b) e^{-\Delta \ell}.}\label{GHandleCombined}
\ee
Plugging in our expressions (\ref{TrumpetWavefunctionFormula}) for the trumpet wavefunctions, we find
\begin{align}
G_{2,\beta,\chi=-1}(-i\tau)_{JT}\supset e^{-S_0} \int_0^\infty b db \int_{-\infty}^\infty e^{\ell} d\ell &\int_0^\infty dE \int_0^\infty dE'\frac{\cos( b \sqrt{2 E}) \cos( b \sqrt{2 E'})}{2\pi^2 \sqrt{E E'}}.
\cr
&\times e^{-\beta E'} e^{-\tau(E-E')} \psi_E(\ell) \psi_{E'}(\ell) e^{-\Delta \ell}
\end{align}
We may recognize the integral over $\ell$ as the integral (\ref{JTMatrixElementsCalculation}) which we identified as the square of the matrix elements of $\mathcal{O}$. We may also recognize the integral over $b$ 
\begin{align}
\int_0^\infty b db \frac{\cos( b \sqrt{2 E}) \cos( b \sqrt{2 E'})}{2\pi^2 \sqrt{E E'}} =&\frac{1}{(2\pi i)^2}\int_{\mathcal{C}} d\beta d\beta' e^{\beta E+ \beta' E'} \big[Z(\beta) Z(\beta')\big]_{Double-Trumpet}
\cr
=&\rho(E,E')_{Ramp}.
\end{align}
Altogether, we can write our formula (\ref{GHandleCombined}) as 
\be\label{GJTRamp}
\boxed{G_{2,\beta,\chi=-1}(-i\tau)_{JT}\supset \int_0^\infty dE dE' \rho(E,E')_{Ramp} \;e^{-\beta E'} e^{-\tau (E-E')} e^{-S_0}|\mathcal{O}_{E,E'}|^2} 
\ee
After continuing $\tau \rightarrow i T$ or $\tau\rightarrow \beta/2 + i T$ we find that this formula agrees precisely with our expectation (\ref{GJTPrediction}) up until the plateau time.

Since the subset of geodesics that we accounted for gave us the expected behavior for the correlator at late times, we should find that all of the other geodesics give small corrections. We will address these corrections briefly in Appendix \ref{AppendixGeodesics} and find these geodesics give decaying contributions, but we will leave a systematic study of the corrections to future work. We also expect that similarly to the spectral form factor, contributions from higher topology do not affect the late time behavior; we also briefly comment on these corrections in Appendix \ref{AppendixGeodesics}.

\subsection{Physical Interpretation}\label{SubsectionPhysicalInterpretation}
There are many ways to continue the Euclidean handle on a disk geometries in order to find contributions to the Lorentzian correlators. We will focus on the two most natural continuations and their physical interpretations. 

\subsubsection{Shortening the Einstein-Rosen Bridge}
The first continuation is the most useful for thinking about the two-sided correlator, and is most easily connected to the calculation of the Euclidean correlator. We can think about this continuation as follows. First, we take the formula (\ref{GHandleCombined}) for the ramp contribution to the Euclidean correlator, continue $\tau \rightarrow \beta/2+i T$, and write the trumpet wavefunctions as
\be
\psi_{Tr,\beta/2 \pm i T}(\ell, b) = e^{-2 S_0} \int e^{\ell'} d\ell' \int e^{\ell''} \ell''\; \langle \ell,b|\ell'\rangle P_{\chi=1}(\pm T/2,\ell', \ell'') \psi_{D,\beta/2}(\ell'').
\ee
This expression is interpreted as describing the evolution of the Hartle-Hawking state for time $T/2$, after which a baby universe is emitted. This defines a state with a parent universe and a baby universe, where there is a non-decaying amplitude for the parent universe to be small and the baby universe to be of size $b\sim T$. The geometries contributing to this wavefunction are described by a Euclidean half-disk, which defines the Hartle-Hawking state, glued to a Lorentzian geometry that has evolved for a long time, which ends on a Euclidean wormhole geometry. These geometries are similar to those pictured in Figure \ref{fig:TrumpetStateLongTime}, with the final Hartle-Hawking half-disk removed.

The ramp contribution to the correlator is then equal to the expectation value of $e^{-\Delta \ell}$ in a density matrix defined by this state. This density matrix is obtained by tracing over the state of the baby universe using the inner product $\langle \psi_1 |\psi_2\rangle \equiv \int_0^\infty b db \;\psi_1^*(b)\psi_2(b)$.
\be \label{RampDensityMatrix}
G^{LR}_{2,\beta}(T)\supset e^{-S_0}\int_{-\infty}^\infty e^\ell d\ell \; \rho_{Trumpet}(\ell, \ell) \; e^{-\Delta \ell},
\ee
\be
\rho_{Trumpet}(\ell,\ell') \equiv \int_0^\infty b db \; \psi_{Tr,\beta/2+i T} (\ell,b) \psi_{Tr,\beta/2-i T}(\ell',b).
\ee
The geometries contributing to this piece of the correlator consist of two copies of the geometries which describe the trumpet wavefunction, glued together along the final spatial slice and baby universe. Below, we have pictured this in two ways; the first emphasizes the growth of the parent universe before the emission of a large baby universe, while the second is more schematic and is more similar to previous figures.
Let's compare our formula (\ref{RampDensityMatrix}) with the interpretation of the two-sided correlator in a conventional quantum system. We can express such a correlator as the expectation value of the operator $\mathcal{O}_L \mathcal{O}_R$ in the pure state $|TFD_\beta(T/2)\rangle \equiv e^{-i\frac{T}{2} (H_L+H_R)} |TFD_\beta\rangle$,
\be
G_{2,\beta}^{LR}(T) = \langle TFD_\beta(T/2)| \mathcal{O}_L \mathcal{O}_R|TFD_\beta(T/2)\rangle.
\ee
At early times, we have a similar formula for the correlator as the expectation value of $e^{-\Delta \ell}$ in the pure state disk wavefunction $\psi_{D,\beta/2+i T}$. Combining the early time and ramp contributions, we have a formula for the two-sided correlator before the plateau time as the expectation value of $e^{-\Delta \ell}$ in the mixed density matrix
\be
\rho(\ell, \ell') \equiv \psi_{D,\beta/2+ i T}(\ell) \psi_{D,\beta/2-i T}(\ell')+ \rho_{Trumpet}(\ell,\ell').
\ee
In Section \ref{SectionPlateau}, we will find that the plateau can be incorporated by a modification of this density matrix.

This formula seems natural from the point of view of third-quantized JT gravity. Measuring the two-sided correlator is similar to measuring a particle that has been prepared in a known state in a conventional second-quantized theory; if the particle of interest has emitted another particle before it is measured, entanglement between the two particles will lead to the density matrix of the particle of interest being mixed. 

However, we expect that for a non-averaged system there should be modifications which allow us to describe the correlator as an expectation value in a pure state. One is tempted to speculate that in a non-averaged system, the states of the baby universes should not be traced over. We will return to the topic of non-averaged systems in Section \ref{SectionDiscussion}. 

Here we take a moment to emphasize the somewhat surprising fact that during the ramp region, the correlator is dominated by contributions from large baby universes. One might have thought that the contributions from large baby universes are heavily suppressed relative to contributions from tiny baby universes. However, this is not the case; once you have paid the cost from the topological term in the action, the baby universes are allowed to be large enough to take away the whole interior of the black hole.

\subsubsection{Shortcuts in spacetime}

We may also continue our Euclidean geometries so that they describe transition amplitudes. We schematically describe these transition amplitudes as
\begin{gather}\label{OverlapFormula}
G_{2,\beta}(T) \supset \langle f_R| i_R\rangle_{\chi=-1}, \hspace{20pt} G^{LR}_{2,\beta}(T) \supset \langle f_L| i_R\rangle_{\chi=-1},
\cr
|f_{L,R}\rangle \equiv e^{-2S_0} \int e^\ell d\ell \int e^{\ell'}d\ell' \;\mathcal{O}_{L,R} |\ell_2\rangle \big( P_{\chi=1}(T;\ell_2,\ell_1) \langle \ell_1|HH_\beta\rangle).
\cr
|i_R\rangle_{\chi=-1} \equiv  e^{-2 S_0}  \int e^\ell d\ell \int e^{\ell'}d\ell' \; |\ell\rangle \big(P_{\chi=-1}(T;\ell,\ell')\langle \ell'| \mathcal{O}_R|HH_\beta\rangle\big),
\end{gather}

$|f_{L,R}\rangle$ is a state where we let the Hartle-Hawking state evolve without emitting any baby universes for time $T$, then create a particle with $\mathcal{O}_L$ or $\mathcal{O}_R$.

$|i_R\rangle_{\chi=-1}$ is a state where we create a particle on top of the Hartle-Hawking state with $\mathcal{O}_R$, a baby universe is emitted, the parent universe evolves for time $T$, then the baby universe is reabsorbed. We are free to choose different time slicings corresponding to different times at which the baby universe is emitted and absorbed. We will comment on this further at the end of this section; however, for now it is useful to keep in mind a description of $P_{\chi=-1}$ where the baby universe is emitted at time zero and reabsorbed at time $T$.

We also may express the sum over geodesics and integral over geometries such that we include only one geodesic, which goes through the baby universe, and we integrate freely over the length and twist of the baby universe.

With these choices, we may schematically think of the propagator $P_{\chi=-1}$ as
\be
P_{\chi=-1}(T,\ell,\ell') \sim \int db d\tau \int \bigg[\prod_{i=1}^2 e^{-S_0} e^{\ell_i} d\ell_i \bigg]\langle \ell | \ell_2,b\rangle P_{\chi=1}(T,\ell_2,\ell_1) \langle \ell_1,b| \ell'\rangle,
\ee
and on this geometry, we only include contributions to the matter propagator from a single geodesic which goes through the baby universe.

The geometries corresponding to this continuation are pictured below.
In contrast to the "shortening" picture, this continuation is useful for thinking about the single-sided two-point function, as well as the two-sided two-point function. For simplicity, we will focus on the single-sided two-point function $G_{2,\beta}(T)$. 

At early times, the single-sided two-point function may be described by a similar formula, where we replace the state $|i_R\rangle_{\chi=-1}$ with a state where no baby universes are emitted and absorbed. Together with the ramp contribution, 
\begin{gather}
G_{2,\beta}(T) \approx \langle f_R|i_R\rangle,\hspace{20pt} T\ll e^{S_0},
\cr
|i_R\rangle\equiv |i_R\rangle_{\chi=1}+|i_R\rangle_{\chi=-1},
\cr
|i_R\rangle_{\chi=1} \equiv \int \bigg[\prod_{i=1}^2 e^{-S_0} e^{\ell_i}d\ell_i\bigg]\; |\ell_2\rangle \big[P_{\chi=1}(T,\ell_2,\ell_1) \langle \ell_1 |\mathcal{O}_R| HH_\beta\rangle.
\end{gather}
For long times, $|i_R\rangle_{\chi=1}$ descibes a state where the particle created by $\mathcal{O}$ has been falling into a black hole for a long time. The amplitude to stay outside of the black hole decays forever, so the overlap with $|f_R\rangle$ decays forever.

On the other hand, $|i_R\rangle_{\chi=-1}$ describes a state with a non-decaying amplitude for this particle to be outside the black hole. Immediately after the particle is created, a baby universe is emitted. The particle has an amplitude to end up in this baby universe. The parent universe then evolves for a long time before the baby universe is reabsorbed in the same state that it was emitted in. After being reabsorbed into the parent universe, the particle has an amplitude to be out near the boundary, where it can then be measured.

The linear growth of the amplitude is not as straightforwardly visible in this picture. The circular geodesic whose twist leads to the linear growth in the previous picture becomes complexified in this continuation; part of this geodesic goes through the long Lorentzian region of these spacetimes. %The rough picture for this factor of $T$ is that the amplitude for a small baby universe to be reabsorbed into the parent universe grows with the size of the parent universe; there are more places for the universe to be absorbed, each giving an amplitude for the particle to appear near the boundary. As the size of the parent universe is growing linearly with $T$, this amplitude grows linearly with $T$.

We can also describe the state $|i_R\rangle_{\chi=-1}$ in many ways, by changing the times at which the baby universe is emitted and absorbed. For any $T_2 \geq T_1\geq 0$ such that $T-T_1-T_2>0$,
\begin{align}
P_{\chi=-1}(T,\ell,\ell')\sim\int db d\tau \int\bigg[\prod_{i=1}^4 e^{-S_0} e^{\ell_i} &d\ell_i\bigg]\;  P_{\chi=1}(T-T_1-T_2,\ell,\ell_4)\langle \ell_4|\ell_3,b\rangle 
\cr
&\times P_{\chi=1}(T_2,\ell_3,\ell_2) \langle \ell_2,b | \ell_1\rangle P_{\chi=1}(T_1,\ell_1,\ell').
\end{align}
The tilde indicates that in the actual expression for $P_{\chi=-1}$, the integral over $b$ and $\tau$ may not be naively performed as suggested by the formula. However, we may freely integrate over $b$ and $\tau$ if we use this formula in our expression \ref{OverlapFormula} for the state $|i_R\rangle_{\chi=-1}$ and include only one geodesic which goes through the baby universe in the matter propagator. In some sense the case of $T_1=0,\; T_2 = T$ is the simplest, as the matter geodesic stays on a purely Euclidean piece of the geometry. However, the more general case is interesting.

In this more general description of $|i_R\rangle_{\chi=-1}$, the baby universe is emitted at time $T_1$ after the particle is created, and is reabsorbed a time $T-T_1-T_2$ before we measure the state. For $T_1\gg 1$, the amplitude for the particle to be outside the black hole when the baby universe is emitted is very small, $\sim e^{-\Delta T_1}$. However, the amplitude for it to be absorbed by the baby universe and emitted out near the boundary does not depend on $T_1$ at all. This suggests that the particle may be emitted with a baby universe even if it is in the interior of the black hole. \textit{In other words, the particle can escape the interior of the black hole by going through the baby universe}.\footnote{One may think about this nonlocal effect more simply in the limit where the baby universe is small. One may approximately include the effects of small Euclidean wormholes by introducing a small bilocal coupling in the bulk theory, $\sim e^{-S_0} \int dx dy \mathcal{O}(x) \mathcal{O}(y)$, where $x$ and $y$ are integrated over all of spacetime.} This is a particularly interesting picture of the non-decaying behavior of the correlator.

\section{The ramp in the four-point function}\label{SectionRamp4pt}
In this section we will study the contribution to the out-of-time-ordered four-point function on geometries with the topology of a disk with two handles, and restrict our attention to a certain class of geodesics. We will make some brief remarks in Appendix \ref{AppendixGeodesics} about why we expect corrections from other geometries and geodesics to be small at late times but before the plateau time. However, we leave a systematic study of the corrections, as well as the calculation of higher point functions, to later work. 

\subsection{Calculation}\label{Subsection4ptCalculation}
We start by calculating a set of contributions to the Euclidean four-point function. As with the two-point function, the four-point function is given by an integral over geometries of the matter correlator on each geometry. For simplicity, we consider free matter, so the matter four-point function is simply the sum of products of two-point functions. Consider the matter four point function on a disk with two handles. We will restrict our attention to contributions to the four point function from geodesics of the form pictured below.
\begin{figure}[H]
\centering
\includegraphics[scale=1]{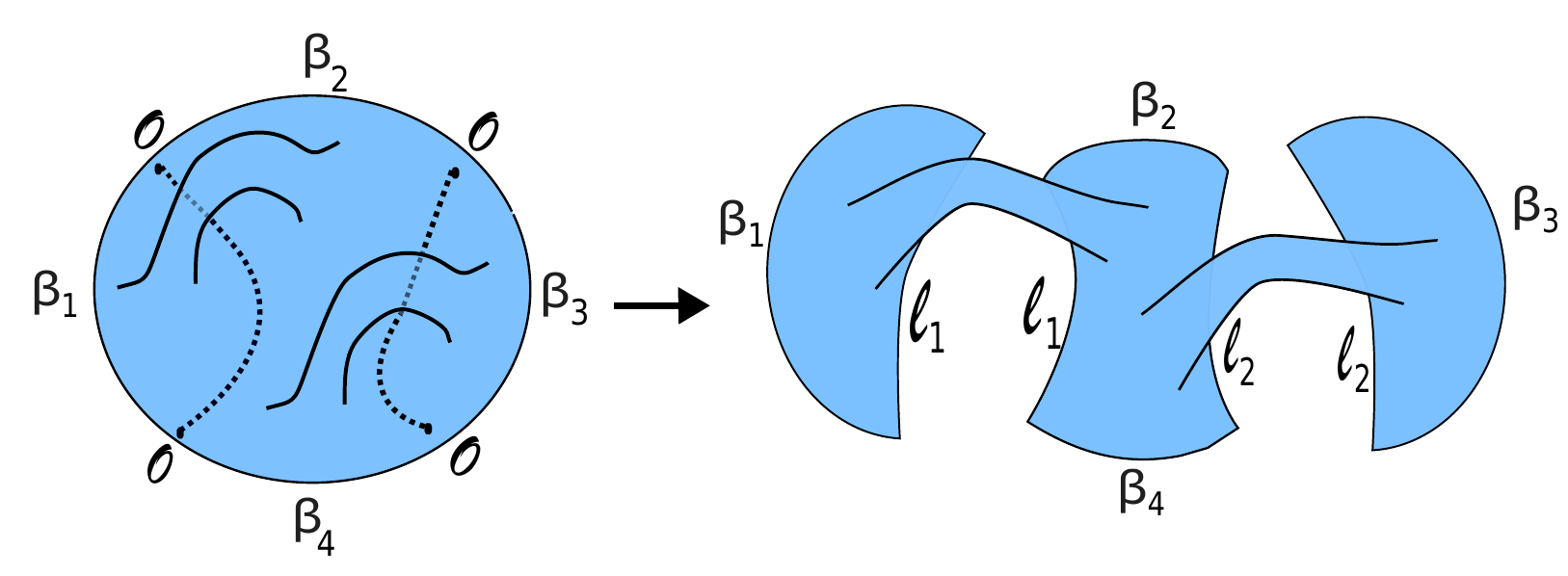}
\caption{\small Here we have pictured an example of a pair geodesics that we cut along. The geometry on the right is described by $\Psi(\ell_1,\ell_2)$.}
\label{fig:FigureFourPointCut}
\end{figure}
For each choice of geodesics, we may cut along the geodesics to produce the geometry pictured on the right, with $\ell_1$ and $\ell_2$ the lengths of the two geodesics. We will also have a similar contribution where we cut along geodesics connecting the operators separated by $\beta_2$ and $\beta_4$. For now we consider this configuration, adding the second contribution later.

 As with the single handle on a disk, the integral over geometries of the disk with two handles is complicated; we must not overcount geometries. However, by including the contributions of all geodesics that cut the two-handled disk as pictured above, we may arrive at a simple integral. 

Consider one of the cut geometries, on the right of Figure \ref{fig:FigureFourPointCut}. The integral over all geometries of this form is simple; we may decompose it into trumpets with geodesics cut out and integrate freely over the lengths and twists of the geodesics we cut along. However, two distinct geometries with this form may be glued together along the $\ell_1$ and $\ell_2$ geodesics to form the same two-handled disk geometry. We would like to use the sum over geodesics to allow us to integrate over all distinct cut geometries instead.

To see that we may do this, we must argue that each distinct cut geometry is counted once and only once by the integral over distinct two-handled disk geometries with a sum over geodesics. More precisely, we would like to argue that
\be
\int_{\text{Two-handled disks} }\mathcal{D} g \;e^{-I_{JT}} \sum_{\gamma,\gamma'} e^{-\Delta \ell_{\gamma}-\Delta \ell_{\gamma'}} = e^{-2 S_0} \int e^{\ell_1} d\ell_1 e^{\ell_2} d\ell_2 \; \Psi(\ell_1, \ell_2) e^{-\Delta \ell_1 - \Delta \ell_2}, \label{FourPointRearrange}
\ee
where $\Psi(\ell_1,\ell_2)$ is an integral over all geometries of the form pictured on the right of Figure \ref{fig:FigureFourPointCut}, and $\gamma, \;\gamma'$ are geodesics of the form described. 

First we establish that each geometry that contributes to the RHS of (\ref{FourPointRearrange}) is corresponds to a contribution on the LHS. This is obvious, as we can take each geometry and glue along the $\ell_1$ and $\ell_2$ geodesics to form a geometry that contributes to the LHS, where are we are weighting the geometry by the lengths of two geodesics $\gamma$ and $\gamma'$. 

Second, we argue that we are not overcounting; each geometry contributing to the RHS of (\ref{FourPointRearrange}) is counted no more than once by a contribution on the LHS. This could only happen when two geometries on the LHS form the same geometry when glued along $\ell_1$ and $\ell_2$. If we are not overcounting, then these two geometries on the RHS must correspond to cutting along two different choices of geodesics on the left. This is indeed the case, as cutting along geodesics $\gamma$ and $\gamma'$ yields a unique geometry of the form contributing to the RHS of (\ref{FourPointRearrange}). Unless there are two choices of pairs of geodesics $\gamma$, $\gamma'$ on the same geometry for the LHS which give the same cut geometry for the RHS, then each geometry on the RHS corresponds to only one contribution from the RHS. There are geometries for which two choices of pairs of geodesics $\gamma$, $\gamma'$ on the same geometry for the LHS which give the same cut geometry for the RHS, but these geometries have special symmetry (like a square torus) and are of measure zero in the integral.

Now we calculate $\Psi(\ell_1,\ell_2)$ by decomposing it into simpler pieces. We will choose a decomposition that is not the simplest possible, but will suggest a simple generalization for higher point OTOCs. This decomposition is pictured below
\begin{figure}[H]
\centering
\includegraphics[scale=1]{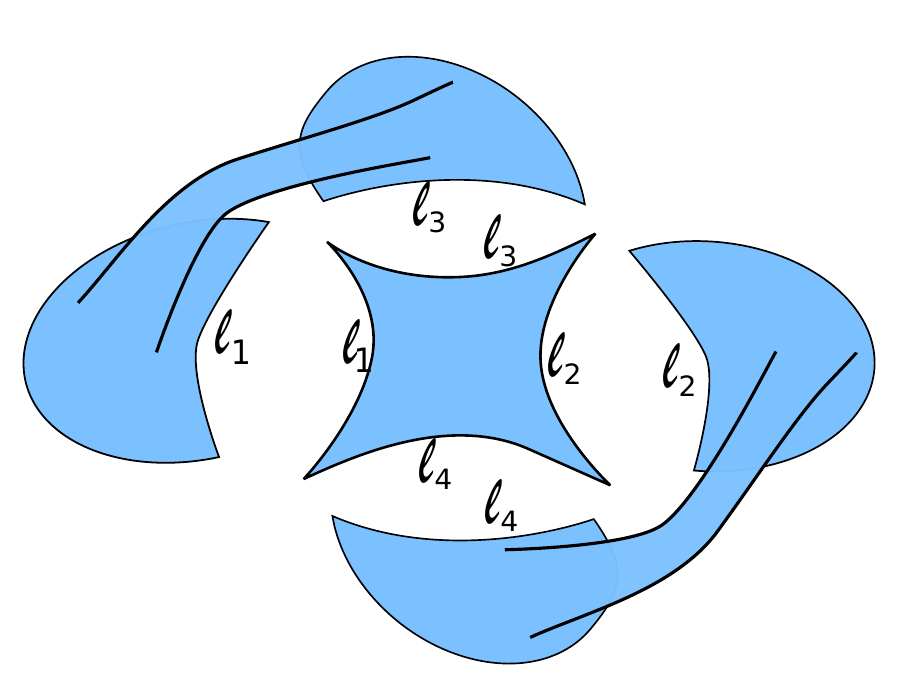}
\end{figure}
Here we have taken the geometry on the RHS of Figure \ref{fig:FigureFourPointCut} and cut along to geodesics $\ell_3$ and $\ell_4$. The double-trumpet like pieces are familiar from the previous sections. The quadrilateral with geodesic boundaries is described in \cite{Yang:2018gdb}. However, we will define it slightly differently, such that gluing along the geodesics $\ell_i$ is done with our usual measure. Then,
\be
\text{Q}(\ell_1 ,\ell_2,\ell_3,\ell_4) =\int dE \rho_0(E) \psi_E(\ell_1) \psi_E(\ell_2) \psi_E(\ell_3)\psi_E(\ell_4).
\ee
With this, we can write
\begin{align}
\Psi(\ell_1,\ell_2) = e^{-2 S_0} \int_{-\infty}^\infty e^{\ell_3} d\ell_3 \int_{-\infty}^\infty e^{\ell_4} d\ell_4 &\int_0^\infty b db \int_0^\infty b' db'\; \psi_{Tr,\beta_1}(\ell_1,b)\psi_{Tr,\beta_2}(\ell_3,b) 
\cr
&\times \psi_{Tr,\beta_3}(\ell_2,b')\psi_{Tr,\beta_4}(\ell_4,b') Q(\ell_1,\ell_2,\ell_3,\ell_4).
\end{align}
Plugging this into equation (\ref{FourPointRearrange}), using our expression (\ref{TrumpetWavefunctionFormula}) for the trumpet wavefunction in the energy basis, and performing the integrals over the $\ell_i$, $b$, and $b'$, we find an expression for the contribution to the four point function from these geodesics
\be
G_{4,\{\beta_i\}} \supset e^{-2 S_0} \int_0^\infty \prod_{i=1}^4 dE_i \frac{ \rho(E_1, E_2)_{Ramp}\rho(E_3, E_4)_{Ramp} }{\rho_0(E_2)} e^{-\sum_{i=1}^4 \beta_i E_i} |\mathcal{O}_{E_1, E_2}|^2|\mathcal{O}_{E_3, E_4}|^2 \delta(E_2-E_4).
\ee
From geodesics that connect the operators separated by $\beta_2$ and $\beta_4$, we find a similar contribution
\be
G_{4,\{\beta_i\}} \supset \int_0^\infty \prod_{i=1}^4 dE_i \frac{ \rho(E_1, E_2)_{Ramp}\rho(E_3, E_4)_{Ramp} }{\rho_0(E_1)} e^{-\sum_{i=1}^4 \beta_i E_i} |\mathcal{O}_{E_2, E_3}|^2|\mathcal{O}_{E_1, E_4}|^2 \delta(E_1-E_3).
\ee
Adding these and continuing the $\beta_i$, we match our predictions from Section \ref{SectionPredictions} for the late time four point function before the plateau time. In particular, we consider continuing $\beta_{1,3}\rightarrow \beta/2+i T$, $\beta_{2,4}\rightarrow \beta/2-i T$ to find the two-sided correlation function $G_{4,\beta/2}(T)_{JT}$. At long times we can approximate the energies as nearly equal to an average energy $E$ to find
\be
\boxed{G_{4,\beta/2}(T)_{JT} \approx 2 \;e^{-2 S_0} \int dE \frac{e^{-2\beta E}}{\rho_0(E)} |\mathcal{O}_E|^4 Y(E,T)^2, \hspace{20pt} 1\ll T \ll e^{S_0}.}
\ee
This precisely matches our prediction (\ref{FourPointLateTimePrediction}) for times before the plateau time. We will find continued agreement after the plateau time in the next section.

This four-point function is sensitive to correlations between different off-diagonal matrix elements. Our prediction (\ref{FourPointLateTimePrediction}) is based on the ETH expectation that these correlations are small, and do not affect the leading order behavior of the correlator; only contributions from paired matrix elements contribute to leading order.

We can interpret the Lorentzian correlator in a similar way to the two-point function. The geometries describing the out-of-time-ordered correlator involve multiple ``time-folds'', periods of Lorentzian evolution alternating forwards of backwards in time. During the periods of Lorentzian evolution, the parent JT universe can emit or absorb a baby universe. These baby universes can be ``traded'' between sections of forwards and backwards Lorentzian evolution. As there are two segments of forwards evolution, and two segments of backwards evolution, there are two ways in which the forwards and backwards segments can be paired up by trading baby universes. Each trade contributes a factor of $T$, for the reasons described in Section \ref{SectionRamp2pt}. And since there are two ways to trade two baby universes, the correlator goes as $2 T^2$.

\subsection{Higher point OTOCs}\label{SubsectionHigherptOTOCs}

We expect that a similar pattern holds for higher point OTOCs. Here we will briefly describe some expectations of how this might work. 

For higher point OTOCs, the random phases of the matrix elements force us to set some of the energies equal to each other. The leading energy pairings are described by planar chord diagrams. We can mimic this pattern in JT gravity. The energies between operators corresponds to the energy of the boundary particle in the formalism of \cite{Kitaev:2018wpr,Yang:2018gdb}. When we insert a bulk matter particle at the boundary, momentum is conserved; the bulk matter takes energy away from the boundary particle, and returns the energy when it reaches the boundary again. The pairings of boundary operators corresponding to planar chord diagrams is described in the JT gravity by matter propagating between the corresponding boundary points.

We expect that the the relevant geometry for the ramp in the $2k$ OTOC is a disk with $k$ handles. If we calculate the free $2k$ point correlator on such a geometry, restrict our attention to contractions of boundary operators described by planar chord diagrams, and further restrict our attention to geodesics between boundary points which ``cut through'' one handle each, then we expect that we may calculate the ramp in the OTOC via a procedure similar to the one we just followed for the four-point function. One may easily check that the $e^{-S_0}$ weight of these geometries matches our expectation (\ref{OTOClatetimeprediction}).

\section{Plateaus}\label{SectionPlateau}
In order to understand the plateaus in the correlation functions, we must appeal to the matrix integral description of JT gravity \cite{Saad:2019lba}. The effects that describe the plateau in the matrix integral description are related to the dynamics of individual eigenvalues \cite{Neuberger:1980qh,Ginsparg:1990as,David:1990sk,Shenker:1990uf}. To describe these effects in JT gravity, we allow a new boundary condition, the ``D-brane'' boundary condition.\footnote{These branes are also closely related to branes in topological string theories described by matrix models \cite{Dijkgraaf:2002fc,Aganagic:2003qj,Marino:2008ya,Marino:2012zq,Dijkgraaf:2018vnm}}\footnote{This boundary condition corresponds to an FZZT brane in minimal string theory \cite{Fateev:2000ik,Teschner:2000md,Seiberg:2003nm,Kutasov:2004fg,Maldacena:2004sn,Seiberg:2004at}} We include in our sum over geometries any number of disconnected geometries with any number of certain D-brane boundaries, in addition to asymptotically AdS boundaries which define partition functions or correlation functions. This mirrors Polchinski's calculation of D-brane effects in string theory \cite{Polchinski:1994fq}.

The rules for including D-brane boundaries do not have a clear origin in gravity; we must rely upon the matrix integral description of JT gravity to understand how to properly include these effects. In this section we will apply a natural generalization of these rules to our calculation of correlation functions. We start by briefly describing some aspects of the D-brane contributions to the partition function and spectral form factor in JT gravity. We find that the D-brane effects allow parent JT universes to emit baby universes in correlated ``D-brane'' states. This state, and the corresponding transition amplitude, are defined by the Euclidean path integral including appropriate D-brane boundaries and disconnected spacetimes. 

The Hilbert space description of the D-brane effects allows us to easily incorporate them into our calculations of correlation functions. We find that they give the expected plateau contributions to the two-point and four-point functions.

\subsection{The plateau in the spectral form factor}\label{SubsectionPlateauSFF}
We begin this section by briefly reviewing some aspects of the D-brane boundary conditions described in \cite{Saad:2019lba}. A helpful way of understanding these boundary conditions is to consider calculating the expectation of the resolvent $\Tr \frac{1}{E-H}$ with\footnote{Similar constructions are used extensively in the quantum chaos literature, see \cite{Efetov:1997fw,2005hep.ph....9286S,2009NJPh...11j3025M,haake2010quantum} for reviews.}
\be
R(E)\; \text{``=''}\; \lim_{E'\rightarrow E} \bigg\langle \Tr \frac{1}{E-H} \frac{\det(E-H)}{\det(E'-H)} \bigg\rangle.
\ee
The brackets denote an average over Hamiltonians $H$, $\langle \dots \rangle_H \equiv \int dH( \dots)P(H)$, and the quotation marks mean that the right hand side is divergent for $E\neq E'$.\footnote{This problem may be solved by regulating the determinants as in \cite{Saad:2019lba}. We refer the reader to \cite{Saad:2019lba} for a more detailed explanation.}

One may attempt to evaluate this semiclassically by writing the determinants as $\det(E-H)=\exp[\Tr \log(E-H)]$ and expanding the exponentials. Each trace corresponds to a boundary in JT gravity; the $\Tr \log(E-H)$ boundaries are ``brane'' or ``ghost-brane'' boundaries. Contributions from geometries with these boundary conditions are multivalued in $E$ and $E'$. The correct prescription for the sum over branches is determined by the matrix integral. With this prescription, one finds that there are contributions from an infinite number of brane or ghost-brane boundaries that do not cancel in the limit $E'\rightarrow E$. However, the disconnected contributions exponentiate and lead to a rapidly oscillating function.

The D-brane contributions which survive correspond to geometries with any number of disconnected spacetimes with D-brane boundaries, as well as a cylindrical geometry with a resolvent boundary and a D-brane boundary. For each of these geometries, we can cut this cylinder along a circular geodesic of length $b$ and separate out the piece of the geometry with the resolvent boundary. We define a fixed energy trumpet partition function, which has a resolvent type boundary, as $Z_{Tr}(E,b) = \int_0^\infty d\beta e^{-\beta E} Z_{Tr}(\beta,b)$. With $E_\pm\equiv E \pm i \epsilon$, the D-brane contribution to the resolvent may be described as
\be\label{DbraneResolvent}
R(E_\pm) \supset \int_0^\infty b db \; Z_{Tr}(E_\pm,b) \psi_{D-brane}(E_\pm;b).
\ee
$\psi_{D-brane}(E;b)$ includes the sum over all disconnected geometries as well as the other half of the cylindrical geometry with a D-brane boundary and geodesic boundary $b$.

Here we have defined $\psi_{D-brane}(E_\pm;b)$ to include only the connected contributions of opposite-branch branes and ghost-branes. These give the nonperturbative effects, while the contributions from disconnected D-brane boundaries and same-branch branes and ghost-branes lead to the perturbative contributions that we have already studied. The $\pm$ sign determines the behavior of the ghost-brane. 

The perturbative contributions to the resolvent, given by geometries with handles, translate into contributions to the return amplitude $Z(\beta+i T)_{JT}$ from the emission and reabsorption of baby universes. This D-brane contribution describes a process where a JT universe of fixed energy $E$ emits a baby universe in the D-brane state, without reabsorbing the baby universe. In other words, we allow external baby universe states determined by the D-brane wavefunction. The contribution (\ref{DbraneResolvent}) to the resolvent leads to an analogous contribution to the partition function. Using our interpretation of the partition function as the overlap of Hartle-Hawking states, and the expression for the Hartle-Hawking state in the energy basis $|HH_\beta\rangle = \int dE\rho_0(E) e^{-\beta E} |E\rangle$, we interpret this contribution to the partition function as
\be
\langle HH_{\beta'}, \text{D-brane}| HH_\beta\rangle \equiv \int dE dE' \rho_0(E)\rho_0(E') e^{-\beta E+ \beta' E'} \langle E', \text{D-brane}| E\rangle.
\ee
Our formula (\ref{DbraneResolvent}) tells us that the transition amplitude $\langle E', \text{D-brane}| E\rangle$ is diagonal in energy. Explicitly,
\begin{align}
\langle E', \text{D-brane}| E\rangle &= \frac{\delta(E-E') }{\rho_0(E)}  \langle E, \text{D-brane}| E\rangle
\cr
&=  \frac{\delta(E-E')}{\rho_0(E)}\sum_{\pm} \int b db  \;\frac{\mp Z_{Tr}(E_\pm,b)}{2\pi i \rho_0(E)} \psi_{D-brane}(E_\pm;b).
\end{align}

The double resolvent $\langle R(E_1) R(E_2)\rangle_H$, which we may use to calculate the spectral form factor, has similar contributions. First we briefly describe the perturbative corrections. There are disconnected contributions where handles are glued to two fixed energy trumpets $Z_{Tr}(E,b)$ and $Z_{Tr}(E',b')$. The ramp comes from gluing the two fixed energy trumpets together, $\int b db Z_{Tr}(E,b) Z_{Tr}(E',b)$. The nonperturbative contributions that lead to the plateau come from ending each trumpet in a D-brane state. However, the D-brane state of two baby universes is an entangled state, and leads to a non-factorized contribution to the double resolvent
\be
R(E_\pm)R(E_{\pm'}')\supset \int_0^\infty b db \int_0^\infty b' db' \; Z_{Tr}(E_\pm,b) Z_{Tr}(E'_{\pm'},b') \psi_{D-brane}(E_\pm, E'_{\pm'};b,b').
\ee
The D-brane state for two baby universes, $\psi_{D-brane}(E, E';b,b')$, is given by a sum over an infinite number of geometries with geodesic boundaries of lengths $b$ and $b'$, as well as any number of certain brane and ghost-brane boundaries. Contributions from connected geometries with multiple types of brane or ghost-brane boundaries, such as a cylinder with two different brane boundaries, are crucial for the plateau behavior and lead to an entangled state for the $b$ and $b'$ baby universes.
\begin{figure}[H]\label{DbraneWavefunction}
\centering
\includegraphics[scale=0.7]{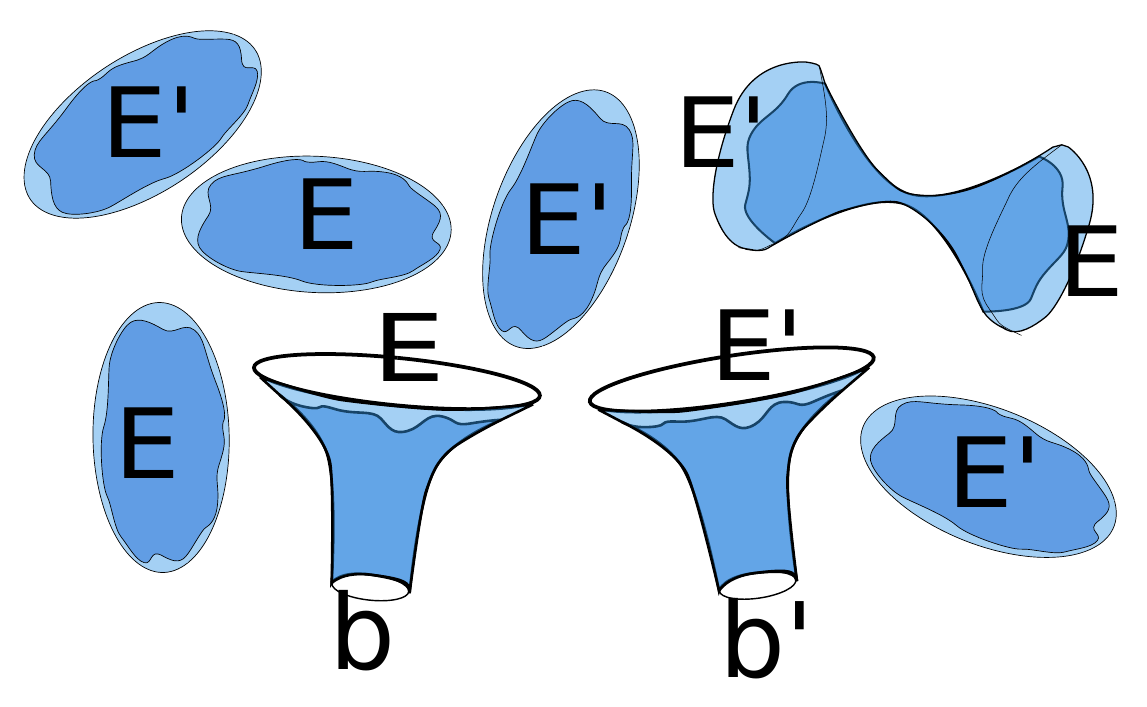}
\caption{\small Here we have pictured an example contribution to the wavefunction $\psi_{D-brane}(E,E';b,b')$. The asymptotically AdS boundaries pictured are brane or ghost-brane boundaries at energies $E$ and $E'$.}
\end{figure}
We may interpret the plateau in the spectral form factor as coming from a process where two JT gravity universes with very long ERBs emit very large baby universes. When these two baby universes are emitted in the same state, so that they are ``traded'', we find a ramp. However, we are now allowing them to be emitted in the entangled D-brane state. Explicitly, the transition amplitude describing this process is
\begin{align}\label{DbranePairAmplitude}
\langle& E_L' , E_R', \text{D-branes}| E_L, E_R\rangle = \frac{\delta(E_L-E_L') \delta(E_R-E_R')}{\rho_0(E_L) \rho_0(E_R)} 
\cr
&\times \sum_{\pm,\pm'} \int b db\int b' db' \frac{(\mp \mp')Z_{Tr}(E_{L,\pm},b)Z_{Tr}(E_{R,\pm'},b')}{(2\pi i)^2 \rho_0(E_L) \rho_0(E_R)} \psi_{D-brane}(E_{L,\pm} ,E_{R,\pm}; b,b').
\end{align}
Defining $\rho(E,E')_{Plateau}\equiv \rho(E,E')_{RMT}- \rho(E,E')_{Ramp}$, we may express the transition amplitude into the D-brane state as 
\be\label{DbraneAmplitudePairCorrelation}
\langle E_L, E_R, \text{D-branes}| E_L, E_R\rangle = \frac{\rho(E_L,E_R)_{Plateau}}{\rho_0(E_L)\rho_0(E_R)}.
\ee

\subsection{The plateau in the two-point function}\label{SubsectionPlateau2pt}
The ramp in the two-point function came from geometries with a handle, and geodesics which go through the handle. These geodesics measure the lengths of spatial slices after a baby universe is emitted, and since emitting a baby universe can dramatically shorten the length of the spatial slice, these geodesics give a non-decaying contribution at late times. 

Our discussion of the plateau in the spectral form factor suggests a similar calculation should give the plateau, but where we replace the handle with two circular geodesics boundaries glued onto a D-brane wavefunction. In order to do this correctly, we first use Laplace transforms to change the fixed length boundary segments between the operators to resolvent boundaries with energies to be above or below the real axis. With these resolvent boundaries, we simply adapt the rule derived from the matrix integral description of JT gravity and allow matching D-brane boundaries. This prescription amounts to including the transition amplitude (\ref{DbranePairAmplitude}) in the Lorentzian description of the correlator.

 Below we picture the relevant geometry and geodesic.
\begin{figure}[H]
\centering
\includegraphics[scale=0.8]{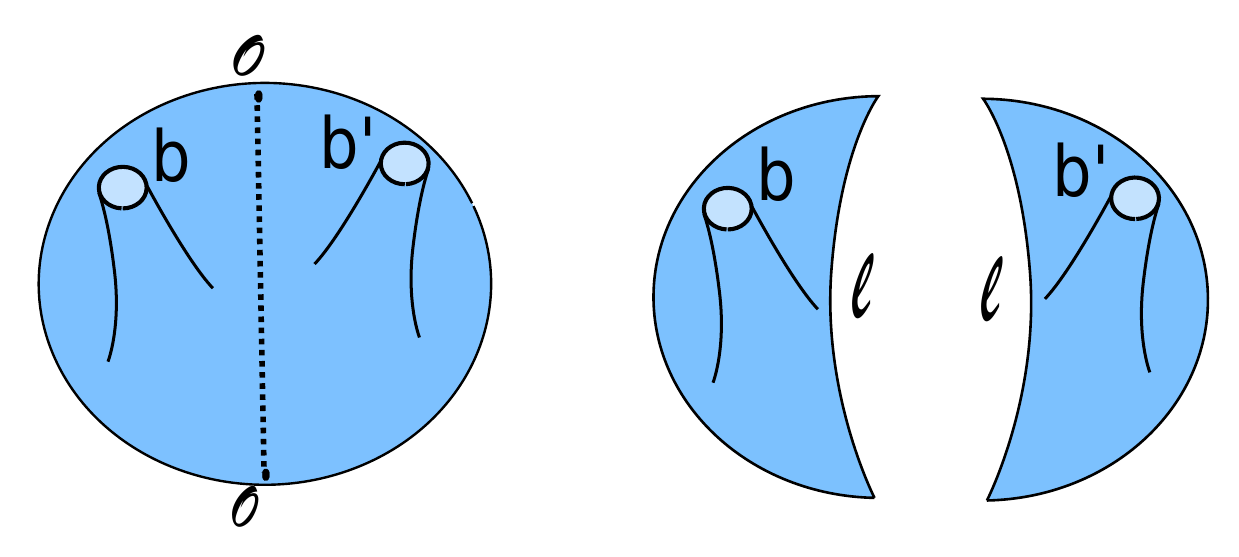}
\end{figure}
Fortunately, there is only one geodesic which cuts through the baby universes. The contribution of this geodesic to the correlator is
\begin{align}
G_{2,\beta}(-i\tau)_{JT} \supset &e^{-S_0}\int e^\ell d\ell \int b db \int b' db' \int dE dE' e^{-\tau E-(\beta-\tau)E'} 
\cr
& \times \sum_{\pm,\pm'}\frac{(\mp \mp')\psi_{Tr,E_\pm}(\ell,b)\psi_{Tr,E'_{\pm'}}(\ell,b')}{(2\pi i)^2}\psi_{D-brane}(E_\pm,E'_{\pm'};b,b') e^{-\Delta \ell}.
\end{align}
Here $\psi_{Tr,E_\pm}(\ell,b)$ is the Laplace transform of the trumpet wavefunction. The wavefunction $\psi_E(\ell)$ is single valued in $E$, so it factors out of the the discontinuity of $\psi_{Tr, E}(\ell,b)$ across the real axis. Using (\ref{DbranePairAmplitude}) and (\ref{DbraneAmplitudePairCorrelation}), performing the integral over $\ell$, and adding the ramp contribution (\ref{GJTRamp}), we find
\be\label{TwoPointPlateau}
\boxed{G_{2,\beta}(-i \tau)_{JT} \supset e^{-S_0} \int dE dE' \rho(E,E')_{RMT} e^{-\beta E'} e^{-\tau(E-E')} |\mathcal{O}_{E,E'}|^2, \hspace{20pt} T\gg 1.} 
\ee
Continuing to find real time correlators, we precisely match our prediction (\ref{GJTPrediction}). However, we must justify our statement that the squared diagonal matrix elements are indeed given by $|\mathcal{O}_{E,E}|^2$. 
We may calculate the averaged diagonal matrix elements by inverse Laplace transforming the thermal one-point functions. The one-point function for our scalar field on any geometry is zero, so the total one-point function is zero. Now we calculate the average of a product of one-point functions. This is equal to integral over all Euclidean geometries with one asymptotically AdS boundary of length $\beta$, with an operator $\mathcal{O}$ inserted along this boundary, and another asymptotically AdS boundary of length $\beta'$, with an operator $\mathcal{O}$ inserted somewhere along it.

To leading order, this is given by the integral over two disk geometries. The matter correlator on these disconnected geometries is zero, so the answer is zero to leading order. To next order, we include cylindrical geometries. Now the matter correlator is nonzero, it is given by the two-point function instead of just a product of one-point functions. On the cylinder geometry there are an infinite number of geodesics which contribute to this correlator. Cutting along any of these geodesics, we find a disk with two geodesics cut out. The integral over geometries of the sum over all geodesics is given by the unrestricted integral over these disk geometries, integrated over the length $\ell$ of the geodesic cut out, weighted by $e^{-\Delta \ell}$. The disk with two geodesics of length $\ell$ cut out is simply the Euclidean propagator, which we may obtain from \cite{Yang:2018gdb}. One asymptotically AdS segment of the disk's boundary has renormalized length $\beta$, and the other has renormalized length $\beta'$. However, the propagator is a function only of the sum $\beta+\beta'$,
\be
\int dE \rho_0(E) e^{-(\beta+\beta') E} \psi_E(\ell)^2.
\ee
The pair of one-point functions is then
\begin{align}
\langle \mathcal{O}\rangle_\beta\langle \mathcal{O}\rangle_{\beta'} \approx & e^{-S_0} \int e^\ell d\ell \int dE \rho_0(E) e^{-(\beta+\beta')E}   \psi_E(\ell)^2 e^{-\Delta \ell}
\cr
& = e^{-S_0} \int dE \rho_0(E) e^{-(\beta+\beta')E} |\mathcal{O}_{E,E}|^2
\cr
&= \int dE \rho_0(E) \int dE' \rho_0(E') \; e^{-\beta E} e^{-\beta' E'} \; \bigg[e^{-S_0}|\mathcal{O}_{E,E}|^2 \frac{\delta(E-E')}{\rho_0(E)} \bigg].
\end{align}
We can identifty the quantity in brackets as the prediction for $\langle E|\mathcal{O}|E\rangle\langle E'|\mathcal{O}|E'\rangle$, justifying (\ref{OnePointPrediction}).

Now we briefly discuss the physical interpretation of the plateau in the two-point function. First we focus on the ``shortening'' picture for the two-sided two-point function. The two-point function may be described by the expectation value of $e^{-\Delta \ell}$ in the density matrix
\be
\rho(\ell,\ell') = \psi_{D,\beta/2+i T}(\ell)\psi_{D,\beta/2-i T}(\ell') + \int dE dE' e^{- E (\frac{\beta}{2}+i T)}e^{-i E (\frac{\beta}{2}-i T)} \rho(E,E')_{RMT} \psi_E(\ell) \psi_{E'}(\ell').
\ee
To describe the piece of the density matrix which gives the plateau, we think of the density matrix as an entangled state of two copies of the system. Then we have
\begin{align}
\rho(\ell,\ell')_{Plateau}  = \langle \ell, \ell', \text{D-branes}|e^{-i \frac{T}{2}H_{Bulk,LR} }| HH_L, HH_R\rangle.
\end{align}
Since the D-brane state describes an entangled state of two baby universes, we interpret this contribution similarly to the ramp piece of the density matrix. The ramp contribution comes from simply tracing over the state of a baby universe that has been emitted. The D-brane piece behaves like a correction to this trace, modifying the inner product between baby universe states to give contributions from baby universes with $b\neq b'$.

We may also imagine an analog of the ``shortcut'' picture for the plateau. While this picture is not easily described by the worldline formalism for the matter correlation function, and requires us to imagine using fully quantum bulk matter as opposed to probe matter, we may roughly understand it as follows.

After a particle is created at the boundary of the Hartle-Hawking state, it may be emitted along with a baby universe, which ends in a D-brane state. At a time $T$ later, a different baby universe is absorbed, also in the D-brane state, so that it is entangled with the first baby universe. When the second baby universe is absorbed, matter may be absorbed along with it. The state of this emitted matter will be correlated with the state of the matter that was absorbed into the first baby universe; if the first baby universe contains a particle, the second is likely to also contain one. Since this matter may be reabsorbed into the parent universe near either one of the asymptotic boundaries with an amplitude which does not decay in time, this leads to a non-decaying contribution to the amplitude $\langle f_{L,R}| i_{R}\rangle$.

\subsection{The plateau in the four-point function}\label{SubsectionPlateau4pt}
Here we will only briefly discuss the plateau in the four-point function, as the calculation is very similar to the calculations we have performed so far. The geometry of interest is the disk with four circular geodesic holes and one asymptotically AdS boundary. There are only two choices of pairs of geodesics that contribute, each corresponding to a chord diagram. These geodesics cut through pairs of holes; if we imagine the geodesics, like our chord diagrams, as partitioning the disk into domains of equal energy, then there are two domains where no energies are paired, and just describe one of the boundary energies, and one domain with two energies paired, describing two paired boundary energies. In each domain we cut out a number of geodesic holes corresponding to the number of energies; one hole in the unpaired domains, and two holes in the paired domain. 

These four geodesic holes are glued to a D-brane wavefunction $\psi_{D-brane}(b_1,b_2,b_3,b_4)$. The plateau comes from factorized contributions to the D-brane wavefunctions,
\be
\psi_{D-brane}(b_1,b_2,b_3,b_4)\supset \psi_{D-brane}(b_1,b_2)\psi_{D-brane}(b_3,b_4).
\ee

\section{Discussion}\label{SectionDiscussion}
\subsection*{Higher point correlators and ETH}
In Section \ref{SubsectionOTOCPredictions}, we made precise predictions for the late-time behavior of a class of OTOCs. These OTOCs probe the statistics of the energy eigenstates; ETH predicts that these matrix elements average to zero unless they cancel in a particular way. Finding agreement in JT gravity with these predictions would be a step towards ``deriving'' ETH in JT gravity. 

However, these correlators only probe the behavior of ``chains'' of matrix elements, $\langle E_1|\mathcal{O}|E_2\rangle \langle E_2|\mathcal{O}|E_3\rangle \dots\langle E_{2k} |\mathcal{O}|E_1\rangle$.
To understand more general statistics of the eigenstates, we must look at more general combinations of matrix elements. One way to do this is to consider products of correlation functions. The average of general products of correlation functions probe general statistics of the matrix elements $\langle E_n|\mathcal{O}|E_m\rangle$, as well as general statistics of the energy levels. 

In JT gravity, these products of correlation functions will receive contributions from geometries that connect the asymptotic boundaries. We have already seen an example of this in Section \ref{SubsectionPlateau2pt}. 

Though we expect that the techniques used in this paper should be sufficient to answer this question, we leave this to future work.
\subsection*{Generalizations}
\subsubsection{Generalizations of JT gravity, SYK and other random matrix ensembles}
In this paper we have only considered JT gravity on orientable surfaces. Correspondingly, we found that the correlation functions obey our expectations for an ensemble-averaged theory in the GUE symmetry class. The work of \cite{Stanford:2019vob} found agreement between a class of generalizations of JT gravity and all of the known random matrix ensembles. It would be interesting to see if correlation functions in these generalizations of JT gravity also obey the expectations from the corresponding ensembles.

One might hope that these generalizations of JT gravity give good descriptions of the late-time behavior of correlation functions in the SYK model. In \cite{Cotler:2016fpe}, numerical evidence was found that suggests a ramp-plateau structure in SYK correlation functions in agreement with the predictions from ETH and RMT level statistics for the corresponding symmetry class. We expect some deviations from the RMT predictions due to large fluctuations in the density that are not present in a random matrix model or JT gravity.\footnote{See the discussion in \cite{Saad:2019lba} for more.} However, we expect that these effects do not have a large impact on the late-time behavior of the spectral form factor or correlation functions. 

It would be interesting to find a description of the ramps and plateaus in correlation functions in the $G-\Sigma$ functional integral description of the SYK model. This description makes close contact with the expected JT gravity description; perhaps the geometries discussed in this paper may be of use in searching for the relevant $G-\Sigma$ configurations.

\subsubsection{Higher dimensional generalizations}
A natural question to ask is whether the mechanisms described in this paper have analogs in higher dimensional theories of gravity. Euclidean wormholes in higher dimensional gravity have received a lot of study, see \cite{Lavrelashvili:1987jg,Hawking:1987mz,Giddings:1987cg,Coleman:1988cy,Coleman:1988tj,Giddings:1988wv,Klebanov:1988eh,Maldacena:2004rf,ArkaniHamed:2007js,Yin:2007at,Fu:2019oyc} for examples. However, it is generally unclear not only how we would include Euclidean wormholes in the path integral, but even whether or not we \textit{should} include them at all. Notably, contributions from Euclidean wormholes seem to be in tension with the expectation that many theories of higher dimensional gravity are dual to theories with a fixed Hamiltonian; partition functions for these theories should factorize, but naively, contributions from Euclidean wormholes seem to spoil factorization \cite{Maldacena:2004rf,ArkaniHamed:2007js,Yin:2007at,Saad:2018bqo}.

We will address the question of factorization in more detail in a later point of discussion; for now we will put it aside and discuss some hints that the mechanisms described in this paper do indeed have some analog in higher dimensions. 

In \cite{Saad:2018bqo}, a higher dimensional analog of the double-trumpet at long times and fixed energy, the ``double-cone'', was discussed as a possible origin of a ramp in higher dimensional gravity. Properly regulated, this geometry formally has the topology of a higher-dimensional cylinder, so it may be possible to interpret this geometry analogously to the double-trumpet in JT gravity, describing a process in which a closed baby universe is traded between two parent universes. The formula for the spectral form factor as a transition amplitude of Hartle-Hawking states also generalizes to higher dimensions. At late times, the higher-dimensional Hartle-Hawking state also evolves into a state with a large ERB, so presumably the double-cone is describing a process in which the ERBs of the pair of Hartle-Hawking states shrink. 

In JT gravity, the disk and double-trumpet geometries are the main ingredients in the D-brane mechanism for the plateau. With the Euclidean black hole as a higher-dimensional analog of the disk, and the double-cone as an analog of the double-trumpet, it seems plausible that an analog of the D-brane mechanism may explain the plateau in higher-dimensional gravity. 

If the double-cone and D-brane mechanism explain the ramp and plateau for the spectral form factor of higher-dimensional gravity, one might hope that similar mechanisms describe the ramp and plateau in correlation functions. In higher dimensional theories of gravity, thermal correlators should behave somewhat differently than in JT gravity due to contributions from the low-energy states without black holes \cite{Maldacena:2001kr,Barbon:2003aq}. However, by working in the microcanonical ensemble instead of the canonical ensemble we may avoid such contributions. The contributions to the correlator from high energy states should lead to a ramp-and plateau structure upon averaging (for example by time-averaging).
\subsection*{The D-brane mechanism in gravity}
In JT gravity, the rules for including D-brane boundaries are mysterious and do not have a clear direct justification from bulk gravity reasoning. In \cite{Saad:2019lba} and this paper, we have to rely on the correspondence between JT gravity and a matrix integral to explain the rules for the D-branes. Understanding the D-brane rules from a bulk perspective is an important open question. We will comment more on this issue in the next item of discussion. 
\subsection*{The information problem}
The effects described in this paper possibly have implications for the black hole information problem. Here we have seen that they are sufficient to resolve an analog of Maldacena's version of the information problem \cite{Maldacena:2001kr} for averaged theories.

The two-point function describes measurements around the eternal black hole perturbed by a simple operator. The lack of decay in the two-point function tells us that we may distinguish the eternal black hole from this perturbed state by careful measurements outside the black hole at arbitrarily long times later. The shortcut picture of the ramp gives a particularly vivid picture of how this happens; a particle in the interior of the black hole may escape to the boundary by going through a baby universe.

One might imagine asking a similar question about more general black hole states, formed by the insertion of many operators in the Euclidean evolution that defines the Hartle-Hawking state. It seems clear that the baby universe effects described in this paper will also lead to non-decaying values for measurements distinguishing these black holes from the eternal black hole at late times. While it is more difficult to find a precise prediction for these measurements based on general principles, and also to calculate the behavior of general correlators, one might hope that the contributions of Euclidean wormholes to general observables give answers consistent with the average of a unitary theory.

However, these effects are not sufficient to describe the unitary evolution of a conventional quantum system. Euclidean wormholes are in tension with basic properties of conventional quantum systems; for example, the contribution of a Euclidean wormhole to a pair of partition functions does not factorize. In order to understand the role of baby universes in the information problem, one must resolve these issues.

\subsection*{Non-disordered systems}
For a conventional quantum system with a fixed Hamiltonian, we expect the behavior of the spectral form factor and correlation functions to be very different than for an averaged system. Instead of a smooth ramp and plateau, we expect an erratic fluctuating behavior at late times. The size of these fluctuations is expected to be of order the size of the signal, so the average ramp and plateau behavior is barely distinguishable from the noise. An important challenge is to understand the origin of these large fluctuations in gravity theories dual to conventional quantum systems.

There are hints that the some modification of the rules for Euclidean wormholes/baby universe emission, may be enough to correctly capture this behavior. In JT gravity, Euclidean wormholes, along with the D-brane mechanism, are able to correctly describe all of the statistics of the fluctuations in the spectral form factor. During the ramp region, it seems possible that double-cone geometries also correctly describe the statistics of these fluctuations in higher-dimensional gravity, see \cite{Saad:2018bqo}. 

We now briefly describe a cartoon scenario of how a modification of the rules for baby universes may correctly describe the behavior of a conventional quantum system, inspired by the periodic orbit theory of semiclassical chaotic level statistics.\footnote{See \cite{2009NJPh...11j3025M,haake2010quantum} for reviews. Also see the discussion section in \cite{Saad:2019lba}.} This cartoon also shares properties with a description of the ramp and plateau in the spectral form factor for random unitary matrices \cite{Braun_2012,PhysRevE.87.052919}. We do not mean to take this cartoon as a serious proposal, instead we view it as merely suggestive.

In analogy to the Gutzwiller trace formula for $\Tr \big[e^{-i H T}\big]$, we imagine that we may write $Z(\beta+i T)$ for a putative two-dimensional theory of gravity described by a fixed Hamiltonian as
\be\label{OrbitSum}
Z(\beta+i T) \sim Z_0(\beta+i T) + \sum_a Z_a(\beta+i T).
\ee
$Z_0$ is the disk partition function, and the $Z_a$ are cylindrical geometries with an asymptotically AdS boundary and a boundary with some boundary condition labeled by $a$ which we do not specify. Such a sum might be appropriate if the cylindrical geometries described by the $Z_a$ are saddle points of $Z(\beta+i T)$ for the full bulk theory.\footnote{See \cite{Rey:1998yx,Dijkgraaf:2005bp} for examples in which baby universe effects have been related to instantons in a dual gauge theory.} We might also consider allowing similar contributions from geometries with more boundaries, but our point can be made without thinking about these.

The spectral form factor would then be described by a double sum over the $Z_a$. Below we represent a particular term in this sum, with the orange and purple boundaries representing two different boundary conditions.
\begin{figure}[H]
\centering
\includegraphics[scale=0.8]{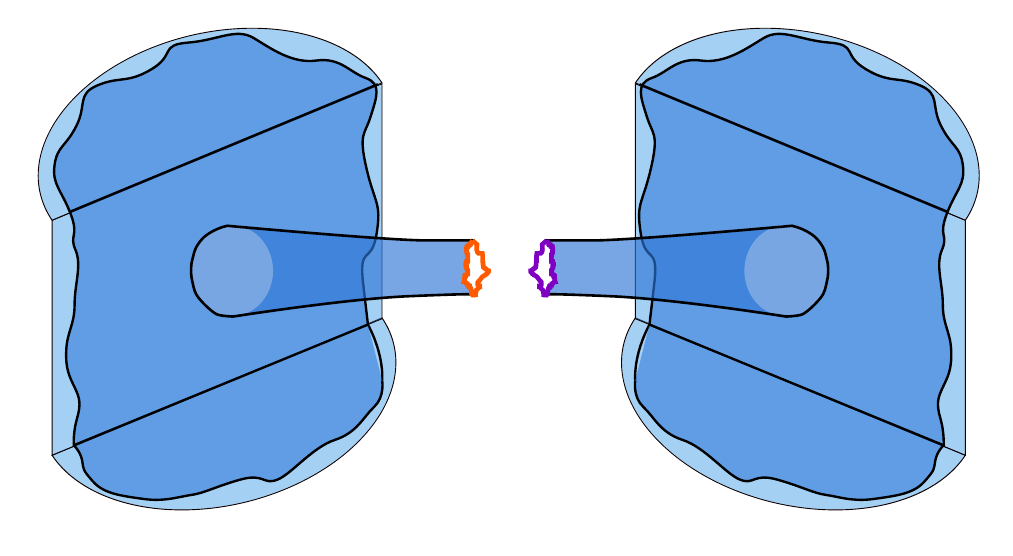}
\end{figure}
Now we assume that we can average over some parameter of the bulk theory, and upon averaging the phases of the $Z_a(\beta+i T)$ behave essentially as random phases at long times, so that
\be\label{DiagonalApprox}
\langle |Z(\beta+i T)|^2\rangle \sim \langle |Z_0(\beta+i T)|^2\rangle + \sum_a \langle |Z_a(\beta+i T)|^2\rangle, \hspace{20pt} 1\ll T\ll e^{S(\beta)}.
\ee
In periodic orbit theory, this is Berry's ``diagonal approximation'', and works because the actions of the long periodic orbits in a chaotic theory are essentially uncorrelated. In our cartoon, we might imagine that the boundary conditions of the cylindrical geometries corresponding to the $Z_a$ are complicated and depend sensitively on the details of the theory. 

If the boundaries at the end of the cylinder are spacelike, as we might expect in analogy to the geodesic boundaries on the trumpet, the boundary conditions at the end of the cylinder define a state $|a\rangle$ of a baby universe. These cylindrical geometries describe processes in which baby universes are allowed to be emitted in these special states. If we also imagine that the states $|a\rangle$ form a complete basis of the single baby universe Hilbert space, then we can replace the sum $\sum_a |Z_a(\beta+i T)|^2$ by the double-trumpet contribution to the spectral form factor, similarly to how we used the complete geodesic length $b$ basis to express the double-trumpet as the integral of two trumpets with $|b\rangle$ boundary conditions.\footnote{The potential for averaging to make connected geometries like these was pointed out in \cite{Maldacena:2004rf}.} 

In periodic orbit theory, this diagonal approximation breaks down at the plateau time $T \sim e^{S(\beta)}$, pointing to the importance of subtle correlations between the actions of different orbits. This timescale is the classical recurrence time; the orbits have started to completely cover phase space, and essentially start retracing other paths. Upon averaging, there are corrections to the diagonal form of the sum in (\ref{DiagonalApprox}), cancelling the linear growth of the ramp to form a plateau. These effects are reminiscent of the D-brane correction to the spectral form factor; while the ramp comes from ``matched'' baby universes, the D-brane wavefunction describes the contribution of baby universes which are not matched. This wavefunction $\psi_{D-brane}(E,E'; b,b')$ may be describing the corrections to the diagonal approximation from correlated baby universe amplitudes. Using methods similar to the determinant/D-brane method described in Section \ref{SectionPlateau}, the plateau has been described in terms of periodic orbits with correlated actions \cite{berry1990rule,keating1992semiclassical,keating2007resummation,2009NJPh...11j3025M,haake2010quantum}.

This cartoon scenario brings to mind Coleman's work on third quantization \cite{Coleman:1988cy}. Coleman's alpha parameters are analogous to the baby universe wavefunctions $|a\rangle$. Effects from baby universes give small corrections to expectation values of bulk operators, and we could determine the baby universe wavefunctions (like the alpha parameters) by careful measurements in the bulk theory. If we wish to describe our bulk theory without using baby universes, we could model these effects as small random-looking modifications to the coupling constants of the theory, but upon averaging over the baby universe wavefunctions, we may simply describe the theory by using Euclidean wormholes and D-branes.

\section*{Acknowledgements}

The author would like to give special thanks to S. Shenker for extensive discussions and guidance throughout this project, as well as to D. Stanford and Z. Yang for discussions and comments on the draft. We also want to thank A. Maloney and L. Susskind for discussions.
P.S. is partially supported by a Fellowship from the Ashok and Gita Vaish Charitable Trust and an ARCS Fellowship.

\appendix

\section{The trumpet wavefunction}\label{AppendixTrumpetWavefunction}
In this appendix we use the boundary particle formalism of \cite{Kitaev:2018wpr,Yang:2018gdb} to calculate the formula 
(\ref{TrumpetWavefunctionFormula}) for the trumpet wavefunction $\psi_{Tr,\beta/2}(\ell,b)$. 

We start by writing an expression for a contribution to the correlator on the trumpet geometry from a certain geodesic. This geodesic is homotopic to a segment of the boundary; cutting along this geodesic produces a disk geometry of the sort that contributes to the Hartle-Hawking wavefunction, and a similar geometry which also has a circular geodesic boundary, which contributes to the trumpet wavefunction. We write this contribution to the correlator as
\be\label{TrumpetCorrelator1}
\langle \mathcal{O}(u)  \mathcal{O}(u') \rangle_{Trumpet}= \frac{1}{2\pi}\int d\bold{x} d\bold{x}' \sqrt{g(\bold{x})} \sqrt{g(\bold{x}')} K_T(u,b, \bold{x}, \bold{x'}) K_D(u',\ell(\bold{x},\bold{x'}))e^{-\Delta \ell(\bold{x},\bold{x}')}.
\ee
$\ell$ is the regularized geodesic length between $\bold{x}$ and $\bold{x'}$ on the trumpet. $K_T(u,b,\bold{x},\bold{x'})$ is the propagator of the boundary particle (with the phase factor removed, as in \cite{Yang:2018gdb}) on the trumpet geometry between points $\bold{x}$ and $\bold{x'}$. This may be expressed as a sum over images of the disk propagator $K_D$, which depends on the coordinates $\bold{x}$ and $\bold{x'}$ only through the combination $\ell(\bold{x},\bold{x'})$. The factor of $2\pi$ out front accounts for the volume of the $U(1)$ gauge group which rotates $\bold{x}$ and $\bold{x'}$ by the same angle around the center of the trumpet.

We can describe the points $\bold{x}$ and $\bold{x}'$ in trumpet coordinates, $\bold{x}=(r,\theta)$, $\bold{x}'=(r',\theta')$. In these coordinates the metric is
\be\label{TrumpetMetric}
ds^2= dr^2 + \frac{b^2}{(2\pi)^2} \cosh^2(r)d\theta^2,\hspace{20pt} r>0,\;\; 0<\theta<2\pi.
\ee
 The points live out near the boundary where the metric simplifies so that
\be
\sqrt{g(\bold{x})}= \frac{b}{2\pi}\frac{e^{r}}{2}.
\ee
We now change variables to
\begin{align}
r_+&=r+r',
\cr
r_-&=r-r',
\cr
\theta_+&=\theta+\theta',
\cr
\theta_-&= \theta-\theta'.
\end{align}
Then the integration measure is
\be
 d\bold{x} d\bold{x}' \sqrt{g(\bold{x})} \sqrt{g(\bold{x}')}= \frac{1}{2} \bigg( \frac{b}{2\pi}\bigg)^2 e^{r_+} dr_+ dr_- d\theta_+ d\theta_-.
\ee
The integrand is independent of $\theta_+$; this is the $U(1)$ gauge symmetry. The integral over $\theta_+$ cancels the factor of $2\pi$ in (\ref{TrumpetCorrelator1}).

Now we do a further change of variables, exchanging $r_+$ for $\ell= r_+ + 2\log \sinh \frac{b \theta_-}{4\pi}- 2 \log \epsilon/2 $. $\ell$ is the renormalized length of the shortest geodesic between points with a given $r_+$ and $\theta_-$; we may compute this using the formulas (\ref{TrumpetGeodesicEquations}).
 
Now the measure is
 \be
 d\bold{x} d\bold{x}' \sqrt{g(\bold{x})} \sqrt{g(\bold{x}')}=  \frac{1}{2(\epsilon/2)^2} \bigg( \frac{b}{2\pi}\bigg)^2 \frac{e^\ell}{\sinh^2 \big(\frac{b\theta_-}{4\pi}\big)} d\ell dr_- d\theta_-.
 \ee
Absorbing a factor of $2/\epsilon$ into our definitions of the propagators $K_T$ and $K_D$, as in \cite{Yang:2018gdb}, our expression for the trumpet correlator becomes
\be
\langle \mathcal{O}(u)  \mathcal{O}(u') \rangle_{Trumpet} = \int e^\ell d\ell \bigg[\int dr_- d\theta_- \frac{ \big( \frac{b}{2\pi}\big)^2 }{2 \sinh^2 \big(\frac{b\theta_-}{4\pi}\big)} K_T(u,b, \ell, r_-,\theta_-)\bigg] K_D(u', \ell ) e^{-\Delta \ell}.
\ee
Up to a factor of $e^{S_0}$, $K_D$ is simply the disk Hartle-Hawking wavefunction $\psi_{D, 2 u'}(\ell)$. We define the quantity in the square brackets to be the trumpet wavefunction $\psi_{Tr, 2u}(b,\ell))$. We can see that the trumpet wavefunction is defined in a similar way to the disk Hartle-Hawking wavefunction; it is given by an integral over geometries with the topology of a cylinder, with a circular geodesic boundary of length $b$ and a boundary with an asymptotic portion of renormalized length $u$ and a geodesic portion of renormalized length $\ell$.

With this definition, may write
\be\label{TrumpetTrick}
\langle \mathcal{O}(u)  \mathcal{O}(u') \rangle_{Trumpet} =e^{-S_0} \int e^\ell d\ell\; \psi_{Tr,2u}( \ell,b) \psi_{D,2u'}( \ell) e^{-\Delta \ell}.
\ee
In particular, setting $\Delta=0$ we find an expression for the trumpet partition function. $\psi_{Tr,2u}(b,\ell)$ is an integral of trumpet propagators, which is a sum of the disk propagator at image points. The disk propagator may be written in the basis of orthogonal energy eigenfunctions $\psi_E(\ell)$, and thus the trumpet propagator may be as well. Exploiting this orthogonality, as well as the explicit form of the trumpet partition function from \cite{Saad:2019lba}, we may use (\ref{TrumpetTrick}) with $\Delta=0$ to find an integral expression for the trumpet wavefunction
\be
\psi_{Tr,2u}(\ell,b) = \int_0^\infty dE \frac{\cos( b \sqrt{2E})}{\pi \sqrt{2E}} e^{- u E} \psi_E(\ell).
\ee
Now we would like to confirm the measure $e^{-S_0}\int e^{\ell} d\ell $ for gluing two trumpet wavefunctions.
We start with an expression for a contribution the correlator on the disk with two geodesic boundaries, with lengths $b_1$ and $b_2$. This contribution only counts the geodesic between our operators that passes through the two geodesic boundaries.

In addition to integrating over the locations of the boundary points, we must integrate over the moduli of this space, $b$ and $\theta_b$. The full expression for the correlator is
\be
\langle \mathcal{O}(u)\mathcal{O}(u')\rangle_{b_1,b_2} =e^{-S_0}\frac{1}{2\pi} \int b db \frac{d\theta_b}{2\pi}d\bold{x} d\bold{x}' \sqrt{g(\bold{x})} \sqrt{g(\bold{x}')} K_T(u,b_1,\bold{x},\bold{x'})K_T(u',b_2,\bold{x},\bold{x'}) e^{-\ell(\bold{x},\bold{x'})}.
\ee
We can choose ``trumpet coordinates'' to describe the locations of the two boundary points. For a given location of the boundary points $\bold{x}$ and $\bold{x}'$, cutting along the geodesic connecting the two points that passes through the two geodesic boundaries disconnects the space into two trumpets with geodesics cut out. We can describe the locations of the two boundary points using the natural $r,\theta$ coordinates on either trumpet. First choose the $r,\theta$ coordinates on the trumpet with boundary $b_1$. We then find, after making appropriate changes of variables and integrating over an overall angular variable
\be\label{TrumpetGlueOneWay}
\langle \mathcal{O}(u)\mathcal{O}(u')\rangle_{b_1,b_2} =e^{-S_0} \int e^\ell d\ell \; \psi_{Tr,2u}(\ell, b_1) \bigg[\int b db \frac{d\theta_b}{2\pi} K_T(u', b_2, \ell, b, \theta_b)\bigg] e^{- \Delta\ell}.
\ee
The function
\be
\tilde{\psi}_{Tr,2u}(\ell,b_1)= \int b db \frac{d\theta_b}{2\pi} K_T(u', b_2, \ell, b, \theta_b)
\ee
is given by an integral over the same geometries that one integrates over to calculate the trumpet wavefunction. Note that $b$ and $\theta_b$ are independent of the parameters $r_-$, $\theta_-$, and $\ell$ which define the $b_1$ trumpet. By specifying $\ell$, $b$, and $\theta_b$, one may reverse the gluing procedure and recover the $b_2$ trumpet coordinates of the two boundary points. Integrating over $b$ and $\theta_b$ then can be traded for an integral over trumpet geometries for the $b_2$ trumpet. However, we would like to show that we in fact get the same integral that defines the trumpet wavefunction, so that $\tilde{\psi}=\psi$. 

To see that in fact $\tilde{\psi}=\psi$, we may derive an expression similar to (\ref{TrumpetGlueOneWay}) but with the two trumpets reversed. Then we find that for all $u$, $b_1$, $b_2$, and $\Delta>0$
\be
e^{S_0}\int e^\ell d\ell \;\psi_{Tr,2u}(\ell,b_1)\tilde{\psi}_{Tr,2u'}(\ell,b_2) e^{-\Delta\ell} = e^{-S_0}\int e^\ell d\ell\; \tilde{\psi}_{Tr,2u}(\ell,b_1)\psi_{Tr,2u'}(\ell,b_2) e^{-\Delta\ell}.
\ee
For this to be true we must have $\tilde{\psi}=\psi$. This establishes our rule for gluing trumpet wavefunctions used to derive (\ref{GNaiveContribution}).

\section{Corrections to the two-point function}\label{AppendixGeodesics}
In this appendix, we argue that the contributions to the two-point function from the geodesics not counted in Section \ref{Subsection2ptCalculation} and \ref{SubsectionPlateau2pt} give corrections which decay at long times. We leave a systematic study of these corrections, as well as the corrections to other higher-point functions, to future work.

There is a two-family parameter of geodesics between boundary points on the trumpet. Using trumpet coordinates (\ref{TrumpetMetric}), we label these geodesics by the average angle of the two boundary points $\theta_0$, and angle subtended by the geodesic $\Delta \theta$.
\begin{align}\label{TrumpetGeodesicEquations}
r(\tau) &= \sinh^{-1}  \frac{\cosh(\tau)}{\sinh \frac{b \Delta \theta}{4\pi}},
\cr
\theta(\tau)&= \theta_0+ \frac{2\pi}{b} \tanh^{-1}\bigg[ \tanh\bigg(\frac{b \Delta \theta}{4\pi}\bigg)\tanh(\tau)\bigg].
\end{align}
Consider two boundary points, where the shortest angle between the points is $\Delta \theta_0$, so that $0<\Delta \theta_0<\pi$. There are an infinite number of geodesics on the trumpet that end on these boundary points, with $\Delta \theta_k $ = $ 2k \pi - \Delta \theta_0$ for odd $k>0$ and $\Delta \theta_k = 2\pi k +\Delta \theta_0$ for even $k>0$. Only the geodesics for $k=0,1$ do not intersect themselves. The geodesics with $k>1$ all lie within the region bounded by the $k=0$ and $k=1$ geodesics. 

The above equations define unit-normalized geodesics, so if $\tau\in(\tau_{min},\tau_{max})$, the length is given by $\ell_b =\tau_{max}-\tau_{min}$. For these geodesics to have both ends at the boundary, we must have $\tau_{min}<0$ and $\tau_{max}>0$. For boundary-anchored geodesics, the formulas for the length simplifies, and we find a relation between the renormalized lengths $\ell_k$ of the geodesics between two boundary points
\be
\ell_k- \ell_j = 2\log \frac{\sinh \frac{b\Delta \theta_k}{4\pi}}{\sinh\frac{b\Delta \theta_j}{4\pi}}.
\ee
In particular, we have for $b\Delta\theta_0\gg 1$,
\be
\ell_k -\ell_0 = \begin{cases}bk- b\frac{\Delta\theta_0}{\pi},\hspace{20pt} &k>0\; \text{odd},
\cr
b k, \hspace{20pt} &k>0\; \text{even}.  \end{cases}
\ee
The self-intersecting geodesics are always longer than the shortest geodesic. Since these geodesics subtend an angle greater than $(k-1)2\pi$, we can see from (\ref{TrumpetMetric}) that their length grows with $b$ like $k b$ for large $k$.

Now we consider geodesics on the handle on a disk geometry. We may continue the geometry to find geometries for the Lorentzian two-sided correlator by using the ``shortening'' continuation, described in Section \ref{SubsectionPhysicalInterpretation}. On this geometry, all of the geodesics between the two boundary points of interest lie on a Euclidean portion of the geometry, which is a handle on a disk with a boundary consisting of two geodesics which meet at the two boundary points.
\begin{figure}[H]\label{GeodesicsCorrectionsFigure}
\centering
\includegraphics[scale=0.6]{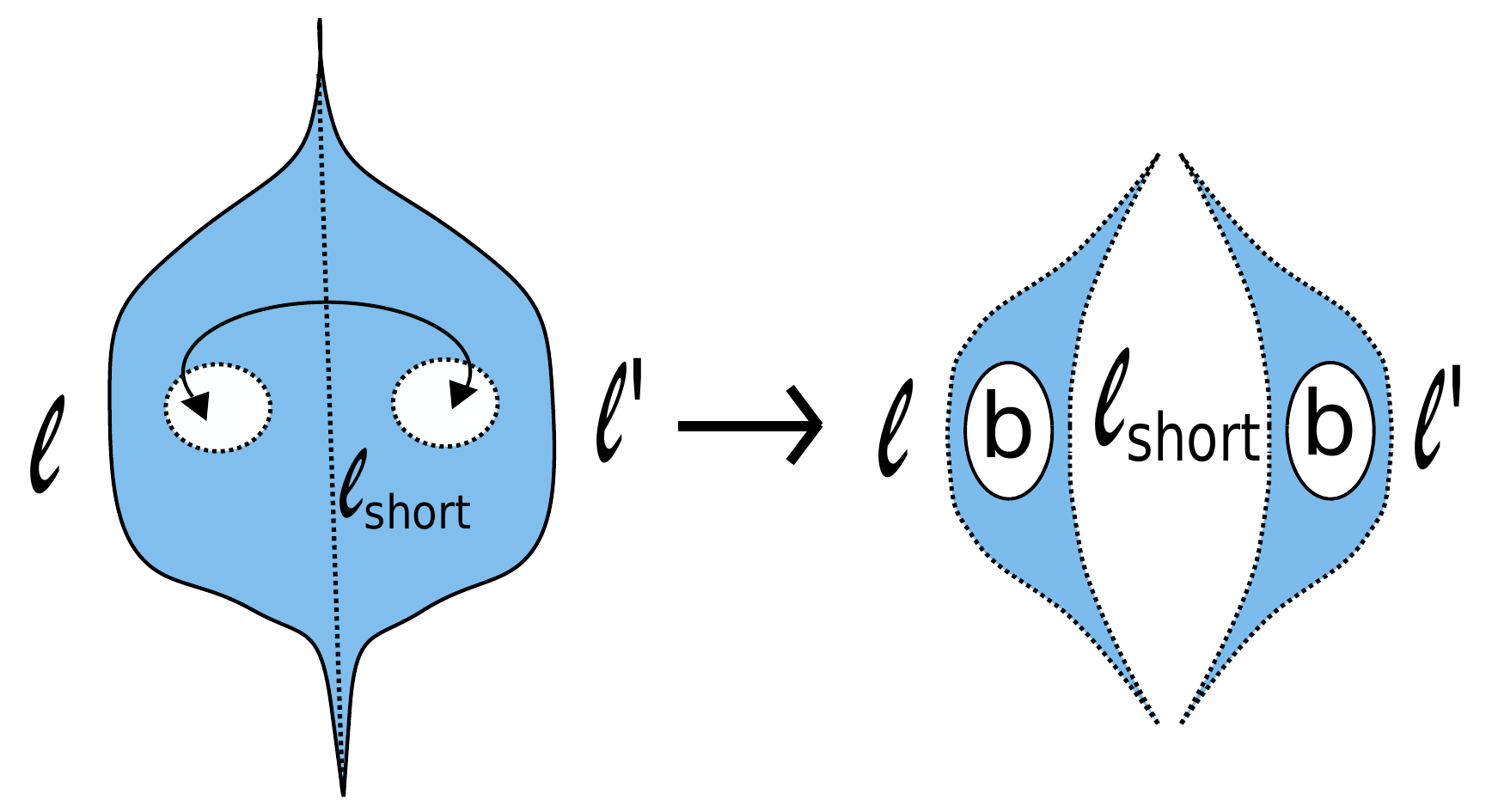}
\caption{\small Here we have pictured the Euclidean portion of the geometry that contributes to the ramp or plateau in the two-sided two-point function.}
\end{figure}
On the left, we have pictured the Euclidean portion of the geometry, which is a handle on a disk with a boundary formed by geodesics of lengths $\ell$ and $\ell'$. We have also pictured the shortest geodesic between the boundary points, with length $\ell_{short}$. Cutting along the short geodesic we find a geometry which is two pieces of the trumpet glued along a circular geodesic of length $b$. 

At long times, $\ell$ and $\ell'$ are forced to be long, of order $T$. For the geometry to contribute non-negligibly to the correlator, we must have $\ell_{short} \sim T^0$, which means that $b \sim T$. 

Now we consider all of the geodesics between the two boundary points. Two of these are the $\ell$ and $\ell'$ geodesics, which are long and give decaying contributions. Infinitely many of these intersect through the $b$ geodesic. In trumpet coordinates, the $r$ coordinate of these geodesics change monotonically from $r=-\infty$ at one boundary to $r=+\infty$ at the other, so that these geodesics do not intersect themselves. These are the geodesics counted in Section \ref{Subsection2ptCalculation}, which lead to the ramp. There are also infinitely many geodesics which do not intersect the $b$ geodesic and are not the shortest geodesic. These stay within one of the two shaded regions on the right of Figure \ref{GeodesicsCorrectionsFigure}. These geodesics intersect themselves and subtend angles greater than $2\pi$ on the trumpet geometry, with lengths that grow linearly with $b\sim T$. Their lengths grow with $b$ like $\ell_k \sim k b \sim k T$ for large $k$, so the sum over these geodesics converges and decays in time.

We can also consider corrections from other geodesics on the geometry considered in \ref{SubsectionPlateau2pt}. At long times, the geodesics stay on a Euclidean portion of the geometry pictured on the right of Figure \ref{GeodesicsCorrectionsFigure}. $b$ is also large, $b\sim T$, for the shortest geodesic to have a length of order one, so the argument that the corrections from other geodesics decay is the same as for the ramp geometry.

Now we briefly comment on the contributions of higher topologies. The work of \cite{Blommaert:2019hjr} hints that these contributions decay. Following their logic, we consider geodesics on these geometries that are not self-intersecting first. Cutting along these, we find geometries similar to those which contribute to the spectral form factor, but with geodesic portions on their boundaries. One might expect to argue that by combining the contributions from similar non-self-intersecting geodesics, the contributions from non-self-intersecting geodesics can be organized into contributions from geometries similar to those which contribute to the spectral form factor. Instead of two asymptotically AdS boundaries, these geometries would have mixed asymptotically AdS and geodesic boundaries. However, we expect that at long times, we can approximate their contribution to the correlator by their contribution to the spectral form factor. These corrections to the spectral form factor decay at long times.

{\footnotesize
\bibliography{references}

\providecommand{\href}[2]{#2}\begingroup\raggedright\begin{thebibliography}{10}

\bibitem{Maldacena:2001kr}
J.~M. Maldacena, ``{Eternal black holes in anti-de Sitter},''
  \href{http://dx.doi.org/10.1088/1126-6708/2003/04/021}{{\em JHEP} {\bfseries
  04} (2003) 021},
\href{http://arxiv.org/abs/hep-th/0106112}{{\ttfamily arXiv:hep-th/0106112
  [hep-th]}}.
%%CITATION = HEP-TH/0106112;%%.

\bibitem{Horowitz:1999jd}
G.~T. Horowitz and V.~E. Hubeny, ``{Quasinormal modes of AdS black holes and
  the approach to thermal equilibrium},''
  \href{http://dx.doi.org/10.1103/PhysRevD.62.024027}{{\em Phys. Rev.}
  {\bfseries D62} (2000) 024027},
\href{http://arxiv.org/abs/hep-th/9909056}{{\ttfamily arXiv:hep-th/9909056
  [hep-th]}}.
%%CITATION = HEP-TH/9909056;%%.

\bibitem{Goheer:2002vf}
N.~Goheer, M.~Kleban, and L.~Susskind, ``{The Trouble with de Sitter space},''
  \href{http://dx.doi.org/10.1088/1126-6708/2003/07/056}{{\em JHEP} {\bfseries
  07} (2003) 056},
\href{http://arxiv.org/abs/hep-th/0212209}{{\ttfamily arXiv:hep-th/0212209
  [hep-th]}}.
%%CITATION = HEP-TH/0212209;%%.

\bibitem{Dyson:2002pf}
L.~Dyson, M.~Kleban, and L.~Susskind, ``{Disturbing implications of a
  cosmological constant},''
  \href{http://dx.doi.org/10.1088/1126-6708/2002/10/011}{{\em JHEP} {\bfseries
  10} (2002) 011},
\href{http://arxiv.org/abs/hep-th/0208013}{{\ttfamily arXiv:hep-th/0208013
  [hep-th]}}.
%%CITATION = HEP-TH/0208013;%%.

\bibitem{Barbon:2003aq}
J.~L.~F. Barbon and E.~Rabinovici, ``{Very long time scales and black hole
  thermal equilibrium},''
  \href{http://dx.doi.org/10.1088/1126-6708/2003/11/047}{{\em JHEP} {\bfseries
  11} (2003) 047},
\href{http://arxiv.org/abs/hep-th/0308063}{{\ttfamily arXiv:hep-th/0308063
  [hep-th]}}.
%%CITATION = HEP-TH/0308063;%%.

\bibitem{prange1997spectral}
R.~Prange, ``The spectral form factor is not self-averaging,'' {\em Physical
  review letters} {\bfseries 78} no.~12, (1997) 2280.

\bibitem{Sachdev:1992fk}
S.~Sachdev and J.-w. Ye, ``{Gapless spin fluid ground state in a random,
  quantum Heisenberg magnet},''
  \href{http://dx.doi.org/10.1103/PhysRevLett.70.3339}{{\em Phys. Rev. Lett.}
  {\bfseries 70} (1993) 3339},
\href{http://arxiv.org/abs/cond-mat/9212030}{{\ttfamily arXiv:cond-mat/9212030
  [cond-mat]}}.
%%CITATION = COND-MAT/9212030;%%.

\bibitem{KitaevTalks}
A.~Kitaev, ``A simple model of quantum holography
  \href{http://online.kitp.ucsb.edu/online/entangled15/kitaev/}{talk1} and
  \href{http://online.kitp.ucsb.edu/online/entangled15/kitaev2/}{talk2}.''.
  Talks at KITP, April 7, 2015 and May 27, 2015.

\bibitem{Polchinski:2016xgd}
J.~Polchinski and V.~Rosenhaus, ``{The Spectrum in the Sachdev-Ye-Kitaev
  Model},'' \href{http://dx.doi.org/10.1007/JHEP04(2016)001}{{\em JHEP}
  {\bfseries 04} (2016) 001},
\href{http://arxiv.org/abs/1601.06768}{{\ttfamily arXiv:1601.06768 [hep-th]}}.
%%CITATION = ARXIV:1601.06768;%%.

\bibitem{Maldacena:2016hyu}
J.~Maldacena and D.~Stanford, ``{Remarks on the Sachdev-Ye-Kitaev model},''
  \href{http://dx.doi.org/10.1103/PhysRevD.94.106002}{{\em Phys. Rev.}
  {\bfseries D94} no.~10, (2016) 106002},
\href{http://arxiv.org/abs/1604.07818}{{\ttfamily arXiv:1604.07818 [hep-th]}}.
%%CITATION = ARXIV:1604.07818;%%.

\bibitem{Jensen:2016pah}
K.~Jensen, ``{Chaos in AdS$_2$ Holography},''
  \href{http://dx.doi.org/10.1103/PhysRevLett.117.111601}{{\em Phys. Rev.
  Lett.} {\bfseries 117} no.~11, (2016) 111601},
\href{http://arxiv.org/abs/1605.06098}{{\ttfamily arXiv:1605.06098 [hep-th]}}.
%%CITATION = ARXIV:1605.06098;%%.

\bibitem{Kitaev:2017awl}
A.~Kitaev and S.~J. Suh, ``{The soft mode in the Sachdev-Ye-Kitaev model and
  its gravity dual},'' \href{http://dx.doi.org/10.1007/JHEP05(2018)183}{{\em
  JHEP} {\bfseries 05} (2018) 183},
\href{http://arxiv.org/abs/1711.08467}{{\ttfamily arXiv:1711.08467 [hep-th]}}.
%%CITATION = ARXIV:1711.08467;%%.

\bibitem{Cotler:2016fpe}
J.~S. Cotler, G.~Gur-Ari, M.~Hanada, J.~Polchinski, P.~Saad, S.~H. Shenker,
  D.~Stanford, A.~Streicher, and M.~Tezuka, ``{Black Holes and Random
  Matrices},'' \href{http://dx.doi.org/10.1007/JHEP05(2017)118}{{\em JHEP}
  {\bfseries 05} (2017) 118},
\href{http://arxiv.org/abs/1611.04650}{{\ttfamily arXiv:1611.04650 [hep-th]}}.
%%CITATION = ARXIV:1611.04650;%%.

\bibitem{Saad:2018bqo}
P.~Saad, S.~H. Shenker, and D.~Stanford, ``{A semiclassical ramp in SYK and in
  gravity},''
\href{http://arxiv.org/abs/1806.06840}{{\ttfamily arXiv:1806.06840 [hep-th]}}.
%%CITATION = ARXIV:1806.06840;%%.

\bibitem{srednicki1994chaos}
M.~Srednicki, ``Chaos and quantum thermalization,'' {\em Physical Review E}
  {\bfseries 50} no.~2, (1994) 888.

\bibitem{deutsch1991quantum}
J.~Deutsch, ``Quantum statistical mechanics in a closed system,'' {\em Physical
  Review A} {\bfseries 43} no.~4, (1991) 2046.

\bibitem{Jackiw:1984je}
R.~Jackiw, ``{Lower Dimensional Gravity},''
\href{http://dx.doi.org/10.1016/0550-3213(85)90448-1}{{\em Nucl. Phys.}
  {\bfseries B252} (1985) 343--356}.
%%CITATION = NUPHA,B252,343;%%.

\bibitem{Teitelboim:1983ux}
C.~Teitelboim, ``{Gravitation and Hamiltonian Structure in Two Space-Time
  Dimensions},''
\href{http://dx.doi.org/10.1016/0370-2693(83)90012-6}{{\em Phys. Lett.}
  {\bfseries B126} (1983) 41--45}.
%%CITATION = PHLTA,B126,41;%%.

\bibitem{Almheiri:2014cka}
A.~Almheiri and J.~Polchinski, ``{Models of AdS$_{2}$ backreaction and
  holography},'' \href{http://dx.doi.org/10.1007/JHEP11(2015)014}{{\em JHEP}
  {\bfseries 11} (2015) 014},
\href{http://arxiv.org/abs/1402.6334}{{\ttfamily arXiv:1402.6334 [hep-th]}}.
%%CITATION = ARXIV:1402.6334;%%.

\bibitem{Maldacena:2016upp}
J.~Maldacena, D.~Stanford, and Z.~Yang, ``{Conformal symmetry and its breaking
  in two dimensional Nearly Anti-de-Sitter space},''
  \href{http://dx.doi.org/10.1093/ptep/ptw124}{{\em PTEP} {\bfseries 2016}
  no.~12, (2016) 12C104},
\href{http://arxiv.org/abs/1606.01857}{{\ttfamily arXiv:1606.01857 [hep-th]}}.
%%CITATION = ARXIV:1606.01857;%%.

\bibitem{Yang:2018gdb}
Z.~Yang, ``{The Quantum Gravity Dynamics of Near Extremal Black Holes},''
\href{http://arxiv.org/abs/1809.08647}{{\ttfamily arXiv:1809.08647 [hep-th]}}.
%%CITATION = ARXIV:1809.08647;%%.

\bibitem{Gross:2017aos}
D.~J. Gross and V.~Rosenhaus, ``{All point correlation functions in SYK},''
  \href{http://dx.doi.org/10.1007/JHEP12(2017)148}{{\em JHEP} {\bfseries 12}
  (2017) 148},
\href{http://arxiv.org/abs/1710.08113}{{\ttfamily arXiv:1710.08113 [hep-th]}}.
%%CITATION = ARXIV:1710.08113;%%.

\bibitem{Lam:2018pvp}
H.~T. Lam, T.~G. Mertens, G.~J. Turiaci, and H.~Verlinde, ``{Shockwave S-matrix
  from Schwarzian Quantum Mechanics},''
  \href{http://dx.doi.org/10.1007/JHEP11(2018)182}{{\em JHEP} {\bfseries 11}
  (2018) 182},
\href{http://arxiv.org/abs/1804.09834}{{\ttfamily arXiv:1804.09834 [hep-th]}}.
%%CITATION = ARXIV:1804.09834;%%.

\bibitem{Mertens:2017mtv}
T.~G. Mertens, G.~J. Turiaci, and H.~L. Verlinde, ``{Solving the Schwarzian via
  the Conformal Bootstrap},''
  \href{http://dx.doi.org/10.1007/JHEP08(2017)136}{{\em JHEP} {\bfseries 08}
  (2017) 136},
\href{http://arxiv.org/abs/1705.08408}{{\ttfamily arXiv:1705.08408 [hep-th]}}.
%%CITATION = ARXIV:1705.08408;%%.

\bibitem{Blommaert:2018oro}
A.~Blommaert, T.~G. Mertens, and H.~Verschelde, ``{The Schwarzian Theory - A
  Wilson Line Perspective},''
  \href{http://dx.doi.org/10.1007/JHEP12(2018)022}{{\em JHEP} {\bfseries 12}
  (2018) 022},
\href{http://arxiv.org/abs/1806.07765}{{\ttfamily arXiv:1806.07765 [hep-th]}}.
%%CITATION = ARXIV:1806.07765;%%.

\bibitem{Blommaert:2019hjr}
A.~Blommaert, T.~G. Mertens, and H.~Verschelde, ``{Clocks and Rods in
  Jackiw-Teitelboim Quantum Gravity},''
\href{http://arxiv.org/abs/1902.11194}{{\ttfamily arXiv:1902.11194 [hep-th]}}.
%%CITATION = ARXIV:1902.11194;%%.

\bibitem{Bulycheva:2019naf}
K.~Bulycheva, ``{Semiclassical correlators in Jackiw-Teitelboim gravity},''
\href{http://arxiv.org/abs/1905.05692}{{\ttfamily arXiv:1905.05692 [hep-th]}}.
%%CITATION = ARXIV:1905.05692;%%.

\bibitem{Iliesiu:2019xuh}
L.~V. Iliesiu, S.~S. Pufu, H.~Verlinde, and Y.~Wang, ``{An exact quantization
  of Jackiw-Teitelboim gravity},''
\href{http://arxiv.org/abs/1905.02726}{{\ttfamily arXiv:1905.02726 [hep-th]}}.
%%CITATION = ARXIV:1905.02726;%%.

\bibitem{Saad:2019lba}
P.~Saad, S.~H. Shenker, and D.~Stanford, ``{JT gravity as a matrix integral},''
\href{http://arxiv.org/abs/1903.11115}{{\ttfamily arXiv:1903.11115 [hep-th]}}.
%%CITATION = ARXIV:1903.11115;%%.

\bibitem{Cotler:2017jue}
J.~Cotler, N.~Hunter-Jones, J.~Liu, and B.~Yoshida, ``{Chaos, Complexity, and
  Random Matrices},'' \href{http://dx.doi.org/10.1007/JHEP11(2017)048}{{\em
  JHEP} {\bfseries 11} (2017) 048},
\href{http://arxiv.org/abs/1706.05400}{{\ttfamily arXiv:1706.05400 [hep-th]}}.
%%CITATION = ARXIV:1706.05400;%%.

\bibitem{Lavrelashvili:1987jg}
G.~V. Lavrelashvili, V.~A. Rubakov, and P.~G. Tinyakov, ``{Disruption of
  Quantum Coherence upon a Change in Spatial Topology in Quantum Gravity},''
  {\em JETP Lett.} {\bfseries 46} (1987) 167--169.
[Pisma Zh. Eksp. Teor. Fiz.46,134(1987)].
%%CITATION = JTPLA,46,167;%%.

\bibitem{Hawking:1987mz}
S.~W. Hawking, ``{Quantum Coherence Down the Wormhole},''
\href{http://dx.doi.org/10.1016/0370-2693(87)90028-1}{{\em Phys. Lett.}
  {\bfseries B195} (1987) 337}.
%%CITATION = PHLTA,B195,337;%%.

\bibitem{Giddings:1987cg}
S.~B. Giddings and A.~Strominger, ``{Axion Induced Topology Change in Quantum
  Gravity and String Theory},''
\href{http://dx.doi.org/10.1016/0550-3213(88)90446-4}{{\em Nucl. Phys.}
  {\bfseries B306} (1988) 890--907}.
%%CITATION = NUPHA,B306,890;%%.

\bibitem{Coleman:1988cy}
S.~R. Coleman, ``{Black Holes as Red Herrings: Topological Fluctuations and the
  Loss of Quantum Coherence},''
\href{http://dx.doi.org/10.1016/0550-3213(88)90110-1}{{\em Nucl. Phys.}
  {\bfseries B307} (1988) 867--882}.
%%CITATION = NUPHA,B307,867;%%.

\bibitem{Coleman:1988tj}
S.~R. Coleman, ``{Why There Is Nothing Rather Than Something: A Theory of the
  Cosmological Constant},''
\href{http://dx.doi.org/10.1016/0550-3213(88)90097-1}{{\em Nucl. Phys.}
  {\bfseries B310} (1988) 643--668}.
%%CITATION = NUPHA,B310,643;%%.

\bibitem{Giddings:1988wv}
S.~B. Giddings and A.~Strominger, ``{Baby Universes, Third Quantization and the
  Cosmological Constant},''
\href{http://dx.doi.org/10.1016/0550-3213(89)90353-2}{{\em Nucl. Phys.}
  {\bfseries B321} (1989) 481--508}.
%%CITATION = NUPHA,B321,481;%%.

\bibitem{Klebanov:1988eh}
I.~R. Klebanov, L.~Susskind, and T.~Banks, ``{Wormholes and the Cosmological
  Constant},''
\href{http://dx.doi.org/10.1016/0550-3213(89)90538-5}{{\em Nucl. Phys.}
  {\bfseries B317} (1989) 665--692}.
%%CITATION = NUPHA,B317,665;%%.

\bibitem{Maldacena:2004rf}
J.~M. Maldacena and L.~Maoz, ``{Wormholes in AdS},''
  \href{http://dx.doi.org/10.1088/1126-6708/2004/02/053}{{\em JHEP} {\bfseries
  02} (2004) 053},
\href{http://arxiv.org/abs/hep-th/0401024}{{\ttfamily arXiv:hep-th/0401024
  [hep-th]}}.
%%CITATION = HEP-TH/0401024;%%.

\bibitem{ArkaniHamed:2007js}
N.~Arkani-Hamed, J.~Orgera, and J.~Polchinski, ``{Euclidean wormholes in string
  theory},'' \href{http://dx.doi.org/10.1088/1126-6708/2007/12/018}{{\em JHEP}
  {\bfseries 12} (2007) 018},
\href{http://arxiv.org/abs/0705.2768}{{\ttfamily arXiv:0705.2768 [hep-th]}}.
%%CITATION = ARXIV:0705.2768;%%.

\bibitem{Jafferis:2017tiu}
D.~L. Jafferis, ``{Bulk reconstruction and the Hartle-Hawking wavefunction},''
\href{http://arxiv.org/abs/1703.01519}{{\ttfamily arXiv:1703.01519 [hep-th]}}.
%%CITATION = ARXIV:1703.01519;%%.

\bibitem{Harlow:2018tqv}
D.~Harlow and D.~Jafferis, ``{The Factorization Problem in Jackiw-Teitelboim
  Gravity},''
\href{http://arxiv.org/abs/1804.01081}{{\ttfamily arXiv:1804.01081 [hep-th]}}.
%%CITATION = ARXIV:1804.01081;%%.

\bibitem{Polchinski:1985zf}
J.~Polchinski, ``{Evaluation of the One Loop String Path Integral},''
\href{http://dx.doi.org/10.1007/BF01210791}{{\em Commun. Math. Phys.}
  {\bfseries 104} (1986) 37}.
%%CITATION = CMPHA,104,37;%%.

\bibitem{Bagrets:2017pwq}
D.~Bagrets, A.~Altland, and A.~Kamenev, ``{Power-law out of time order
  correlation functions in the SYK model},''
  \href{http://dx.doi.org/10.1016/j.nuclphysb.2017.06.012}{{\em Nucl. Phys.}
  {\bfseries B921} (2017) 727--752},
\href{http://arxiv.org/abs/1702.08902}{{\ttfamily arXiv:1702.08902
  [cond-mat.str-el]}}.
%%CITATION = ARXIV:1702.08902;%%.

\bibitem{Stanford:2017thb}
D.~Stanford and E.~Witten, ``{Fermionic Localization of the Schwarzian
  Theory},'' \href{http://dx.doi.org/10.1007/JHEP10(2017)008}{{\em JHEP}
  {\bfseries 10} (2017) 008},
\href{http://arxiv.org/abs/1703.04612}{{\ttfamily arXiv:1703.04612 [hep-th]}}.
%%CITATION = ARXIV:1703.04612;%%.

\bibitem{Kitaev:2018wpr}
A.~Kitaev and S.~J. Suh, ``{Statistical mechanics of a two-dimensional black
  hole},''
\href{http://arxiv.org/abs/1808.07032}{{\ttfamily arXiv:1808.07032 [hep-th]}}.
%%CITATION = ARXIV:1808.07032;%%.

\bibitem{Stanford:2019vob}
D.~Stanford and E.~Witten, ``{JT Gravity and the Ensembles of Random Matrix
  Theory},''
\href{http://arxiv.org/abs/1907.03363}{{\ttfamily arXiv:1907.03363 [hep-th]}}.
%%CITATION = ARXIV:1907.03363;%%.

\bibitem{Iliesiu:2019lfc}
L.~V. Iliesiu, ``{On 2D gauge theories in Jackiw-Teitelboim gravity},''
\href{http://arxiv.org/abs/1909.05253}{{\ttfamily arXiv:1909.05253 [hep-th]}}.
%%CITATION = ARXIV:1909.05253;%%.

\bibitem{Foini:2018sdb}
L.~Foini and J.~Kurchan, ``{Eigenstate thermalization hypothesis and out of
  time order correlators},''
  \href{http://dx.doi.org/10.1103/PhysRevE.99.042139}{{\em Phys. Rev.}
  {\bfseries E99} no.~4, (2019) 042139},
\href{http://arxiv.org/abs/1803.10658}{{\ttfamily arXiv:1803.10658
  [cond-mat.stat-mech]}}.
%%CITATION = ARXIV:1803.10658;%%.

\bibitem{HalfPipeWiki}
{Wikipedia contributors}, ``Half-pipe,''.
  \url{https://en.wikipedia.org/wiki/Half-pipe}.

\bibitem{Huang:2017fng}
Y.~Huang, F.~G. S.~L. Brandão, and Y.-L. Zhang, ``{Finite-size scaling of
  out-of-time-ordered correlators at late times},''
  \href{http://dx.doi.org/10.1103/PhysRevLett.123.010601}{{\em Phys. Rev.
  Lett.} {\bfseries 123} no.~1, (2019) 010601},
\href{http://arxiv.org/abs/1705.07597}{{\ttfamily arXiv:1705.07597
  [quant-ph]}}.
%%CITATION = ARXIV:1705.07597;%%.

\bibitem{Maldacena:2019cbz}
J.~Maldacena, G.~J. Turiaci, and Z.~Yang, ``{Two dimensional Nearly de Sitter
  gravity},''
\href{http://arxiv.org/abs/1904.01911}{{\ttfamily arXiv:1904.01911 [hep-th]}}.
%%CITATION = ARXIV:1904.01911;%%.

\bibitem{Cotler:2019nbi}
J.~Cotler, K.~Jensen, and A.~Maloney, ``{Low-dimensional de Sitter quantum
  gravity},''
\href{http://arxiv.org/abs/1905.03780}{{\ttfamily arXiv:1905.03780 [hep-th]}}.
%%CITATION = ARXIV:1905.03780;%%.

\bibitem{Papadodimas:2015xma}
K.~Papadodimas and S.~Raju, ``{Local Operators in the Eternal Black Hole},''
  \href{http://dx.doi.org/10.1103/PhysRevLett.115.211601}{{\em Phys. Rev.
  Lett.} {\bfseries 115} no.~21, (2015) 211601},
\href{http://arxiv.org/abs/1502.06692}{{\ttfamily arXiv:1502.06692 [hep-th]}}.
%%CITATION = ARXIV:1502.06692;%%.

\bibitem{Numasawa:2019gnl}
T.~Numasawa, ``{Late Time Quantum Chaos of pure states in the SYK model},''
\href{http://arxiv.org/abs/1901.02025}{{\ttfamily arXiv:1901.02025 [hep-th]}}.
%%CITATION = ARXIV:1901.02025;%%.

\bibitem{Susskind:2014rva}
L.~Susskind, ``{Computational Complexity and Black Hole Horizons},''
  \href{http://dx.doi.org/10.1002/prop.201500093, 10.1002/prop.201500092}{{\em
  Fortsch. Phys.} {\bfseries 64} (2016) 44--48},
  \href{http://arxiv.org/abs/1403.5695}{{\ttfamily arXiv:1403.5695 [hep-th]}}.
[Fortsch. Phys.64,24(2016)].
%%CITATION = ARXIV:1403.5695;%%.

\bibitem{Brown:2018bms}
A.~R. Brown, H.~Gharibyan, H.~W. Lin, L.~Susskind, L.~Thorlacius, and Y.~Zhao,
  ``{Complexity of Jackiw-Teitelboim gravity},''
  \href{http://dx.doi.org/10.1103/PhysRevD.99.046016}{{\em Phys. Rev.}
  {\bfseries D99} no.~4, (2019) 046016},
\href{http://arxiv.org/abs/1810.08741}{{\ttfamily arXiv:1810.08741 [hep-th]}}.
%%CITATION = ARXIV:1810.08741;%%.

\bibitem{wolpert1985weil}
S.~Wolpert, ``On the {Weil-Petersson} geometry of the moduli space of curves,''
  {\em American Journal of Mathematics} {\bfseries 107} no.~4, (1985) 969--997.

\bibitem{Maldacena:2018lmt}
J.~Maldacena and X.-L. Qi, ``{Eternal traversable wormhole},''
\href{http://arxiv.org/abs/1804.00491}{{\ttfamily arXiv:1804.00491 [hep-th]}}.
%%CITATION = ARXIV:1804.00491;%%.

\bibitem{Kim:2015qoa}
J.~Kim and M.~Porrati, ``{On a Canonical Quantization of 3D Anti de Sitter Pure
  Gravity},'' \href{http://dx.doi.org/10.1007/JHEP10(2015)096}{{\em JHEP}
  {\bfseries 10} (2015) 096},
\href{http://arxiv.org/abs/1508.03638}{{\ttfamily arXiv:1508.03638 [hep-th]}}.
%%CITATION = ARXIV:1508.03638;%%.

\bibitem{Neuberger:1980qh}
H.~Neuberger, ``{Nonperturbative Contributions in Models With a Nonanalytic
  Behavior at Infinite $N$},''
\href{http://dx.doi.org/10.1016/0550-3213(81)90238-8}{{\em Nucl. Phys.}
  {\bfseries B179} (1981) 253--282}.
%%CITATION = NUPHA,B179,253;%%.

\bibitem{Ginsparg:1990as}
P.~H. Ginsparg and J.~Zinn-Justin, ``{2-d Gravity + 1-d Matter},''
\href{http://dx.doi.org/10.1016/0370-2693(90)91108-N}{{\em Phys. Lett.}
  {\bfseries B240} (1990) 333--340}.
%%CITATION = PHLTA,B240,333;%%.

\bibitem{David:1990sk}
F.~David, ``{Phases of the large N matrix model and nonperturbative effects in
  2-d gravity},''
\href{http://dx.doi.org/10.1016/0550-3213(91)90202-9}{{\em Nucl. Phys.}
  {\bfseries B348} (1991) 507--524}.
%%CITATION = NUPHA,B348,507;%%.

\bibitem{Shenker:1990uf}
S.~H. Shenker, \href{http://dx.doi.org/10.1142/9789814365802_0057}{``{The
  strength of nonperturbative effects in string theory},''} in {\em {The Large
  N expansion in quantum field theory and statistical physics: From spin
  systems to two-dimensional gravity}}, pp.~809--819.
\newblock World Scientific, [Brezin, E., Wadia, S. R. eds., ], 1993, [preprint
  RU-90-47].
\newblock \url{https://lib-extopc.kek.jp/preprints/PDF/2000/0035/0035186.pdf}.

\bibitem{Dijkgraaf:2002fc}
R.~Dijkgraaf and C.~Vafa, ``{Matrix models, topological strings, and
  supersymmetric gauge theories},''
  \href{http://dx.doi.org/10.1016/S0550-3213(02)00766-6}{{\em Nucl. Phys.}
  {\bfseries B644} (2002) 3--20},
\href{http://arxiv.org/abs/hep-th/0206255}{{\ttfamily arXiv:hep-th/0206255
  [hep-th]}}.
%%CITATION = HEP-TH/0206255;%%.

\bibitem{Aganagic:2003qj}
M.~Aganagic, R.~Dijkgraaf, A.~Klemm, M.~Mari\~{n}o, and C.~Vafa, ``{Topological
  strings and integrable hierarchies},''
  \href{http://dx.doi.org/10.1007/s00220-005-1448-9}{{\em Commun. Math. Phys.}
  {\bfseries 261} (2006) 451--516},
\href{http://arxiv.org/abs/hep-th/0312085}{{\ttfamily arXiv:hep-th/0312085
  [hep-th]}}.
%%CITATION = HEP-TH/0312085;%%.

\bibitem{Marino:2008ya}
M.~Mari\~{n}o, ``{Nonperturbative effects and nonperturbative definitions in
  matrix models and topological strings},''
  \href{http://dx.doi.org/10.1088/1126-6708/2008/12/114}{{\em JHEP} {\bfseries
  12} (2008) 114},
\href{http://arxiv.org/abs/0805.3033}{{\ttfamily arXiv:0805.3033 [hep-th]}}.
%%CITATION = ARXIV:0805.3033;%%.

\bibitem{Marino:2012zq}
M.~Mari\~{n}o, ``{Lectures on non-perturbative effects in large $N$ gauge
  theories, matrix models and strings},''
  \href{http://dx.doi.org/10.1002/prop.201400005}{{\em Fortsch. Phys.}
  {\bfseries 62} (2014) 455--540},
\href{http://arxiv.org/abs/1206.6272}{{\ttfamily arXiv:1206.6272 [hep-th]}}.
%%CITATION = ARXIV:1206.6272;%%.

\bibitem{Dijkgraaf:2018vnm}
R.~Dijkgraaf and E.~Witten, ``{Developments in Topological Gravity},''
\href{http://arxiv.org/abs/1804.03275}{{\ttfamily arXiv:1804.03275 [hep-th]}}.
%%CITATION = ARXIV:1804.03275;%%.

\bibitem{Fateev:2000ik}
V.~Fateev, A.~B. Zamolodchikov, and A.~B. Zamolodchikov, ``{Boundary Liouville
  field theory. 1. Boundary state and boundary two point function},''
\href{http://arxiv.org/abs/hep-th/0001012}{{\ttfamily arXiv:hep-th/0001012
  [hep-th]}}.
%%CITATION = HEP-TH/0001012;%%.

\bibitem{Teschner:2000md}
J.~Teschner, ``{Remarks on Liouville theory with boundary},''
  \href{http://dx.doi.org/10.22323/1.006.0041}{{\em PoS} {\bfseries tmr2000}
  (2000) 041},
\href{http://arxiv.org/abs/hep-th/0009138}{{\ttfamily arXiv:hep-th/0009138
  [hep-th]}}.
%%CITATION = HEP-TH/0009138;%%.

\bibitem{Seiberg:2003nm}
N.~Seiberg and D.~Shih, ``{Branes, rings and matrix models in minimal
  (super)string theory},''
  \href{http://dx.doi.org/10.1088/1126-6708/2004/02/021}{{\em JHEP} {\bfseries
  02} (2004) 021},
\href{http://arxiv.org/abs/hep-th/0312170}{{\ttfamily arXiv:hep-th/0312170
  [hep-th]}}.
%%CITATION = HEP-TH/0312170;%%.

\bibitem{Kutasov:2004fg}
D.~Kutasov, K.~Okuyama, J.-w. Park, N.~Seiberg, and D.~Shih, ``{Annulus
  amplitudes and ZZ branes in minimal string theory},''
  \href{http://dx.doi.org/10.1088/1126-6708/2004/08/026}{{\em JHEP} {\bfseries
  08} (2004) 026},
\href{http://arxiv.org/abs/hep-th/0406030}{{\ttfamily arXiv:hep-th/0406030
  [hep-th]}}.
%%CITATION = HEP-TH/0406030;%%.

\bibitem{Maldacena:2004sn}
J.~M. Maldacena, G.~W. Moore, N.~Seiberg, and D.~Shih, ``{Exact vs.
  semiclassical target space of the minimal string},''
  \href{http://dx.doi.org/10.1088/1126-6708/2004/10/020}{{\em JHEP} {\bfseries
  10} (2004) 020},
\href{http://arxiv.org/abs/hep-th/0408039}{{\ttfamily arXiv:hep-th/0408039
  [hep-th]}}.
%%CITATION = HEP-TH/0408039;%%.

\bibitem{Seiberg:2004at}
N.~Seiberg and D.~Shih, ``{Minimal string theory},''
  \href{http://dx.doi.org/10.1016/j.crhy.2004.12.007}{{\em Comptes Rendus
  Physique} {\bfseries 6} (2005) 165--174},
\href{http://arxiv.org/abs/hep-th/0409306}{{\ttfamily arXiv:hep-th/0409306
  [hep-th]}}.
%%CITATION = HEP-TH/0409306;%%.

\bibitem{Polchinski:1994fq}
J.~Polchinski, ``{Combinatorics of boundaries in string theory},''
  \href{http://dx.doi.org/10.1103/PhysRevD.50.R6041}{{\em Phys. Rev.}
  {\bfseries D50} (1994) R6041--R6045},
\href{http://arxiv.org/abs/hep-th/9407031}{{\ttfamily arXiv:hep-th/9407031
  [hep-th]}}.
%%CITATION = HEP-TH/9407031;%%.

\bibitem{Efetov:1997fw}
K.~Efetov, {\em {Supersymmetry in disorder and chaos}}.
\newblock Cambridge Univ. Press, Cambridge, UK, 2012.
\newblock
\url{http://www.cambridge.org/mw/academic/subjects/physics/condensed-matter-physics-nanoscience-and-mesoscopic-physics/supersymmetry-disorder-and-chaos?format=AR}.
\newblock
%%CITATION = INSPIRE-460245;%%.

\bibitem{2005hep.ph....9286S}
M.~A. {Stephanov}, J.~J.~M. {Verbaarschot}, and T.~{Wettig}, ``{Random
  Matrices},'' {\em arXiv e-prints} (Sep, 2005) hep--ph/0509286,
  \href{http://arxiv.org/abs/hep-ph/0509286}{{\ttfamily arXiv:hep-ph/0509286
  [hep-ph]}}.

\bibitem{2009NJPh...11j3025M}
S.~{M{\"u}ller}, S.~{Heusler}, A.~{Altland}, P.~{Braun}, and F.~{Haake},
  ``{Periodic-orbit theory of universal level correlations in quantum chaos},''
  \href{http://dx.doi.org/10.1088/1367-2630/11/10/103025}{{\em New Journal of
  Physics} {\bfseries 11} (Oct, 2009) 103025},
  \href{http://arxiv.org/abs/0906.1960}{{\ttfamily arXiv:0906.1960 [nlin.CD]}}.

\bibitem{haake2010quantum}
F.~Haake, {\em Quantum Signatures of Chaos}.
\newblock Springer Series in Synergetics. Springer Berlin Heidelberg, 2010.
\newblock \url{https://books.google.com/books?id=hxFkUIqmYC8C}.

\bibitem{Yin:2007at}
X.~Yin, ``{On Non-handlebody Instantons in 3D Gravity},''
  \href{http://dx.doi.org/10.1088/1126-6708/2008/09/120}{{\em JHEP} {\bfseries
  09} (2008) 120},
\href{http://arxiv.org/abs/0711.2803}{{\ttfamily arXiv:0711.2803 [hep-th]}}.
%%CITATION = ARXIV:0711.2803;%%.

\bibitem{Fu:2019oyc}
Z.~Fu and D.~Marolf, ``{Bag-of-gold spacetimes, Euclidean wormholes, and
  inflation from domain walls in AdS/CFT},''
\href{http://arxiv.org/abs/1909.02505}{{\ttfamily arXiv:1909.02505 [hep-th]}}.
%%CITATION = ARXIV:1909.02505;%%.

\bibitem{Braun_2012}
P.~Braun and F.~Haake, ``Chaotic maps and flows: exact
  riemann{\textendash}siegel lookalike for spectral fluctuations,''
  \href{http://dx.doi.org/10.1088/1751-8113/45/42/425101}{{\em Journal of
  Physics A: Mathematical and Theoretical} {\bfseries 45} no.~42, (Oct, 2012)
  425101}. \url{https://doi.org/10.1088%2F1751-8113%2F45%2F42%2F425101}.

\bibitem{PhysRevE.87.052919}
D.~Waltner, S.~Gnutzmann, G.~Tanner, and K.~Richter, ``Subdeterminant approach
  for pseudo-orbit expansions of spectral determinants in quantum maps and
  quantum graphs,'' \href{http://dx.doi.org/10.1103/PhysRevE.87.052919}{{\em
  Phys. Rev. E} {\bfseries 87} (May, 2013) 052919}.
  \url{https://link.aps.org/doi/10.1103/PhysRevE.87.052919}.

\bibitem{Rey:1998yx}
S.-J. Rey, ``{Holographic principle and topology change in string theory},''
  \href{http://dx.doi.org/10.1088/0264-9381/16/7/102}{{\em Class. Quant. Grav.}
  {\bfseries 16} (1999) L37--L43},
\href{http://arxiv.org/abs/hep-th/9807241}{{\ttfamily arXiv:hep-th/9807241
  [hep-th]}}.
%%CITATION = HEP-TH/9807241;%%.

\bibitem{Dijkgraaf:2005bp}
R.~Dijkgraaf, R.~Gopakumar, H.~Ooguri, and C.~Vafa, ``{Baby universes in string
  theory},'' \href{http://dx.doi.org/10.1103/PhysRevD.73.066002}{{\em Phys.
  Rev.} {\bfseries D73} (2006) 066002},
\href{http://arxiv.org/abs/hep-th/0504221}{{\ttfamily arXiv:hep-th/0504221
  [hep-th]}}.
%%CITATION = HEP-TH/0504221;%%.

\bibitem{berry1990rule}
M.~V. Berry and J.~P. Keating, ``A rule for quantizing chaos?,'' {\em Journal
  of Physics A: Mathematical and General} {\bfseries 23} no.~21, (1990) 4839.

\bibitem{keating1992semiclassical}
J.~Keating, ``The semiclassical functional equation,'' {\em Chaos: An
  Interdisciplinary Journal of Nonlinear Science} {\bfseries 2} no.~1, (1992)
  15--17.

\bibitem{keating2007resummation}
J.~P. Keating and S.~M{\"u}ller, ``Resummation and the semiclassical theory of
  spectral statistics,'' in {\em Proceedings of the Royal Society of London A:
  Mathematical, Physical and Engineering Sciences}, vol.~463, pp.~3241--3250,
  The Royal Society.
\newblock 2007.

\end{thebibliography}\endgroup
}

\bibliographystyle{utphys}

\end{document}